\documentclass[10pt,journal,compsoc]{IEEEtran}
\usepackage[ruled,vlined]{algorithm2e}
\usepackage{setspace}
\usepackage{enumitem}
\usepackage{multirow}
\usepackage{stackengine}
\usepackage[nocompress]{cite}
\usepackage{graphicx}
\usepackage{amsthm}
\usepackage[colorlinks, urlcolor=black, linkcolor=black, anchorcolor=black, citecolor=black]{hyperref}
\usepackage{ragged2e}
\usepackage{tablefootnote}
\newtheorem{myDef}{Definition}
\newtheorem{myExa}{Example}

\begin{document}

\title{Navigable Proximity Graph-Driven Native Hybrid Queries with Structured and Unstructured Constraints}

\author{Mengzhao Wang,
        Lingwei Lv,
        Xiaoliang Xu,
        Yuxiang Wang,~\IEEEmembership{Member,~IEEE,}
        Qiang Yue,
        and~Jiongkang Ni
\IEEEcompsocitemizethanks{\IEEEcompsocthanksitem M. Wang, L. Lv, X. Xu, Y. Wang, Q. Yue, and J. Ni are with the School of Computer Science and Technology, Hangzhou Dianzi University, Hangzhou 310018, China.\protect\\
E-mail: \{mzwang, llw, xxl, lsswyx, yq, hananoyuuki\}@hdu.edu.cn
}

\thanks{Corresponding author: Xiaoliang Xu}}


\IEEEtitleabstractindextext{
\begin{abstract}
\justifying{As research interest surges, vector similarity search is applied in multiple fields, including data mining, computer vision, and information retrieval. {Given a set of objects (e.g., a set of images) and a query object, we can easily transform each object into a feature vector and apply the vector similarity search to retrieve the most similar objects. However, the original vector similarity search cannot well support \textit{hybrid queries}, where users not only input unstructured query constraint (i.e., the feature vector of query object) but also structured query constraint (i.e., the desired attributes of interest). Hybrid query processing aims at identifying these objects with similar feature vectors to query object and satisfying the given attribute constraints. Recent efforts have attempted to answer a hybrid query by performing attribute filtering and vector similarity search separately and then merging the results later, which limits efficiency and accuracy because they are not purpose-built for hybrid queries.} In this paper, we propose a native hybrid query (NHQ) framework based on proximity graph (PG), which provides the specialized \textit{composite index and joint pruning} modules for hybrid queries. We easily deploy existing various PGs on this framework to process hybrid queries efficiently. Moreover, we present two novel navigable PGs (NPGs) with optimized edge selection and routing strategies, which obtain better overall performance than existing PGs. After that, we deploy the proposed NPGs in NHQ to form two hybrid query methods, which significantly outperform the state-of-the-art competitors on all experimental datasets (10$\times$ faster under the same \textit{Recall}), including eight public and one in-house real-world datasets. Our code and datasets have been released at \url{https://github.com/AshenOn3/NHQ}.}

\end{abstract}
}

\maketitle
\IEEEdisplaynontitleabstractindextext
\IEEEpeerreviewmaketitle

\begin{figure*}[!tb]
  \centering
  \setlength{\abovecaptionskip}{0cm}
  \includegraphics[width=1.0\linewidth]{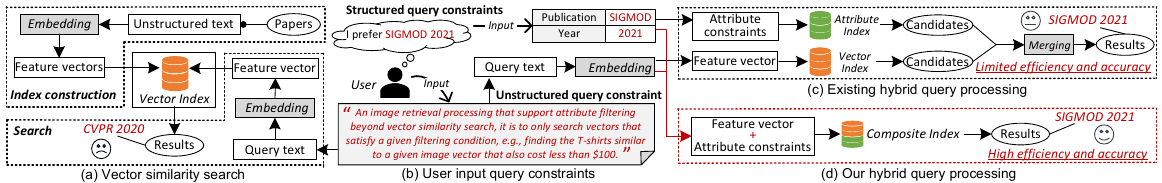}
  \caption{A running example of different paper-retrieval schemes, including (a) vector similarity search and (c–d) hybrid query. Vector similarity search retrieves results with semantically similar content, but it cannot ensure the structured query constraints. A hybrid query processing obtains results that satisfy the structured and unstructured query constraints.}\label{fig:hybrid_query_intro}
  \vspace{-0.5cm}
\end{figure*}

\IEEEraisesectionheading{\section{Introduction}\label{sec:intro}}
\IEEEPARstart{W}{ith} smart devices and applications on the rise, a tremendous amount of unstructured data (such as video, text, image) is being produced \cite{EchihabiZPB19,ValkanasLG17}. International Data Corporation (IDC) projects that by 2025, 80\% of the digital universe will be unstructured \cite{unstructured_data}, but thanks to the rapid development of deep neural networks (DNNs), we can describe unstructured data accurately via feature vectors. Then we can process unstructured data through vector similarity search \cite{Milvus_sigmod2021} to support a multitude of downstream applications, e.g., image retrieval \cite{SPTAG1}, paper retrieval \cite{liujun}, and recommendation systems \cite{Milvus_sigmod2021}. As {Fig. \ref{fig:hybrid_query_intro}}(a) shows, {in a paper-retrieval system based on vector similarity search, a user aims to find the most semantically similar papers to her input query text. Specifically, each paper's unstructured text can be embedded in a high-dimensional space to establish a feature vector; then the query text is also transformed into a feature vector in the same space. Next, the system performs a vector similarity search using a vector index to obtain papers with semantically similar content. In this case, the feature vector of query text can be viewed as an unstructured query constraint that all returned papers should satisfy (measured by vector similarity).} Many effective vector similarity search methods have been proposed to achieve better trade-offs between query efficiency and accuracy \cite{NSSG,NSG,HNSW,Graph_theory}, which we often see in AI scenarios.

Even though vector similarity search is widely studied, it does not support many important real-world scenarios {where users not only expect the unstructured query constraint to be satisfied but also some structured query constraints can be matched as well \cite{Milvus_sigmod2021,Wangwei_tutorial}. For example, in many applications, e.g., visual products search and movie recommendation, it is very common that a user can come up with a \textit{hybrid query} by inputting unstructured and structured constraints simultaneously \cite{ADBV,Jingdong_paper,Milvus_sigmod2021,XuLWX20}. Similar to \cite{ADBV,Milvus_sigmod2021}, the unstructured query constraint is provided as a feature vector of the query object (e.g., a paper) and the structured query constraint is given as a set of attributes of interest (e.g., {\sf topic}, {\sf venue}, and {\sf publish\_year} of a paper). We call such a query \textit{a hybrid query that identifies objects with similar feature vectors to the query object and satisfying the given attribute query constraints} \cite{Milvus_sigmod2021,ADBV,PASE,Jingdong_paper,Vo0S0L0H19,Wangwei_tutorial,Qin000W21}.} For example, in {Fig. \ref{fig:hybrid_query_intro}}(b), a user wants to find some of the latest top-tier conference papers related to her research interest. She can form a hybrid query by providing a detailed descriptive text of her research interest as the unstructured constraint and paper's two attributes (e.g., {\sf SIGMOD} and {\sf 2021}) as the structured constraints. Unfortunately, the traditional vector similarity search shown in {Fig. \ref{fig:hybrid_query_intro}}(a) only obtains papers with semantically similar content, and it is difficult to ensure matching attribute constraints. In response, many have tried to expand attribute filtering on top of vector similarity search to answer a hybrid query.

Vearch (a library provided by the famous e-commerce company Jingdong in China) \cite{Jingdong_lib} first obtains candidates that satisfy the unstructured constraint through vector similarity search, and then performs attribute filtering to return the final results \cite{Jingdong_paper}. This strategy extends easily to other dedicated vector similarity search libraries, e.g., SPTAG \cite{SPTAG}, NGT \cite{NGT}, Faiss \cite{Faiss}. In contrast, Alibaba AnalyticDB-V (ADBV) \cite{ADBV} adds multiple query plans based on product quantization (PQ)\footnote{PQ is a quantization-based vector similarity search method.}\cite{PQ}, including performing attribute filtering first and then vector similarity search to return the final results. This method has good scalability and is suitable for most use cases \cite{Milvus_sigmod2021}. {Milvus \cite{Milvus} pre-partitions the dataset into several subsets based on frequently used attributes, so that 
it can quickly select the subsets whose attribute values overlap with the query's attribute constraints, then it only need to perform the vector similarity search on these subsets rather than the entire dataset, thereby improving the hybrid query processing's performance \cite{Milvus_sigmod2021}.}

{In a nutshell, the basic idea behind each of these hybrid query solutions is to execute attribute filtering and vector similarity search separately in a different order (we illustrate the details in \textbf{{Sec. \ref{Implementation Strategies}}}). Regardless of the execution order of attribute filtering and vector similarity search, the entire query pipeline can be abstracted into the logic shown in {Fig. \ref{fig:hybrid_query_intro}}(c), where two subquery systems (for attribute filtering and vector similarity search) are established independently. Although they can answer a hybrid query, each subquery system was not originally tailored for hybrid queries, which limits overall efficiency and effectiveness.} We next analyze their limitations, and then propose our solution.

\vspace{-0.3cm}
\subsection{Existing Methods' Limitations}
\label{sec:limitations}

\noindent\textbf{L1: Two indexes must be maintained simultaneously.} {Fig. \ref{fig:hybrid_query_intro}}(c) shows that existing solutions require maintaining both the attribute index and vector index simultaneously for two separate subquery systems. Not only does this increase memory overhead, it also introduces additional logic and a post-processing step (i.e., merging) to ensure two indexes' consistency and query results' correctness \cite{ADBV}. With the advent of frequently updated data, updating different indexes synchronously is now an onerous task.

\vspace{0.1cm}
\noindent\textbf{L2: Unnecessary computational overhead caused by two separate pruning strategies.} For existing solutions, pruning is performed based on attribute index and vector index separately, which increases computational overhead. {For example, suppose that the vector index in {Fig. \ref{fig:hybrid_query_intro}}(c) is implemented as a proximity graph (PG, we will give the formal definition of PG in \textbf{Sec. \ref{sec:motivation}}) \cite{graph_survey_vldb2021}, where each vertex in a PG corresponds to an object's feature vector. Then, some vertices close to the query object's feature vector but inconsistent with the attribute constraints causing the search to follow the wrong paths and return incorrect answers to a hybrid query. Because we eventually prune these incorrect answers in the attribute-filtering phase, and the computational cost spent in the vector similarity search phase for returning them is unnecessary and inefficient, we should avoid it.}

\vspace{0.1cm}
{\noindent\textbf{L3: Query results depend on the merging of the candidates obtained from both subquery systems.} Given a hybrid query, we usually expect to return the top-$k$ results that are similar to the query object's feature vector and match the attribute constraints. As shown in {Fig. \ref{fig:hybrid_query_intro}}(c), a post-processing step, namely merging, is required to combine the top-$k$ candidates $C_1$ and $C_2$ obtained from both subquery systems to generate the final results $R=C_1 \cap C_2$. However, usually we have $|R|<k$. This is because existing solutions perform vector similarity search and attribute filtering separately, so that each subquery system can only guarantee it will return candidates satisfying one type of query constraint. To ensure $|R|= k$, we need to enlarge $C_1$ and $C_2$, which seriously increases both subquery systems' latency \cite{PASE}.}

\vspace{0.1cm}
\noindent\textbf{L4: Existing solutions are not friendly to PGs.} {Recent solutions, e.g., ADBV \cite{ADBV} and Milvus \cite{Milvus_sigmod2021}, adopt a ``first attribute filtering, then vector similarity search'' strategy to answer a hybrid query; they only deploy PQ \cite{PQ} for vector similarity search. Because many studies have shown PG is $10\times$ faster than PQ for vector similarity search \cite{annbenchmark_lib,DPG}, it is necessary to exploit a way to integrate PG into hybrid query methods. A straightforward way is to replace PQ-based vector index with another PG-based one, unfortunately, it is problematic for the following reasons: (1) In the ``first attribute filtering, then vector similarity search'' strategy
, we expect to find similar vectors from a dynamic space of the vectors filtered by attribute constraints. To be more precise, different user input attribute constraints cause different filtered vectors (this is a dynamic process in runtime regarding user input). It is not worth building a vector index for these filtered vectors dynamically in runtime because of the efficiency issue. (2) Worse still, if we enumerate all possible conjunctions of attribute values and prebuild a vector index for each one offline (the same as \cite{XuLWX20}), it would suffer from exorbitant memory overhead, e.g., $m^n$ different indexes for $n$ attributes where each attribute has $m$ values.}


{To sum up, most the aforementioned limitations occur because existing solutions are committed to answer hybrid queries in a ``decomposition-assembly'' model, where a hybrid query is decomposed into two standalone subquery systems that process separately and then assemble the final results. This motivates us to present a novel generalized framework tailored for hybrid queries that works well with existing PGs, offering the well-designed \textit{composite index} and \textit{joint pruning} modules to support unstructured and structured query constraints simultaneously, rather than maintaining two independent indexes and performing prune operations separately.}  

\vspace{-0.3cm}
\subsection{Our Solution and Contributions}

To the best of our knowledge, {we are the first to present a \textbf{N}ative \textbf{H}ybrid \textbf{Q}uery framework (NHQ) for answering hybrid queries with unstructured and structured constraints in a fused way, rather than the ``decomposition-assembly'' model adopted by existing solutions. As {Fig. \ref{fig:hybrid_query_intro}}(d) shows, essentially the framework first embeds feature vectors and attributes in a well-designed {composite index} (for \textbf{L1}); then it prunes the search space collectively on the composite index by considering both the given unstructured and structured query constraints (for \textbf{L2}); and finally, it obtains the final top-$k$ results directly rather than resorting to a merging operation (for \textbf{L3}). It is worth noting that our NHQ is a generalized framework, where the composite index is built based on PG, therefore existing PGs deploy easily into NHQ (for \textbf{L4}) and it also supports custom-optimized PGs (e.g., we present two navigable PGs, discussed in \textbf{Sec. \ref{NPG}}).}

Specifically, our solution includes three aspects. (1) \textbf{NHQ framework} (\textbf{Sec. \ref{framework}}). {The framework contains two modules: \textit{composite index} and \textit{joint pruning}. The former is built based on a PG via the presented \textit{fusion distance} that can measure the similarity of two objects by considering both their feature vectors and attributes simultaneously. The latter concurrently prunes unpromising objects with dissimilar feature vectors and mismatched attributes by evaluating the fusion distance when searching on the composite index, thereby returning hybrid query results in one step. We emphasize that our NHQ is a generalized framework, and thus we can deploy current PGs (such as HNSW \cite{HNSW}) on it to serve as a composite index with a minor modification (\textbf{Sec. \ref{sec: NHQ validation}}).} (2) \textbf{Navigable PGs (NPGs)} (\textbf{Sec. \ref{NPG}}). In NHQ, PG plays a key role and a PG's \textit{edge selection} and \textit{routing} strategies have an important impact on the hybrid query processing's performance. Although we can implement different existing PGs on NHQ, the inherent limitations of existing PGs still leave us room for optimization. Hence, we present two NPGs with novel edge selection and routing strategies, which offer the state-of-the-art performance of PG. (3) \textbf{NPG-based hybrid query methods} (\textbf{Sec. \ref{NPG-NHQ}}). We obtain two hybrid query methods by integrating the proposed NPGs into our NHQ framework. Specifically, we build two NPG-based composite indexes with our fusion distance measure. Meanwhile, we implement an optimized joint pruning on a NPG-based composite index. We experimentally prove that our NPG-based hybrid query methods significantly outperform the state-of-the-art competitors (e.g., Vearch, ADBV, and Milvus) on all experimental datasets.

\setlength{\textfloatsep}{0cm}
\setlength{\floatsep}{0cm}
\begin{table}[!tb]
  \centering
  \setlength{\abovecaptionskip}{0.05cm}
  \fontsize{8pt}{3.3mm}\selectfont
  \caption{Frequently Used Notations}
  \label{notations}
  \begin{tabular}{p{45pt}|p{175pt}}
    \hline
    \textbf{Notations} & \textbf{Descriptions}\\
    \hline
    \hline
    $\mathcal{S}$ & An object set\\
    \hline
    $\mathcal{X}$ & {The feature vector space of all objects in $\mathcal{S}$}\\
    \hline
    $\mathcal{Y}$ & {The attribute vector space of all objects in $\mathcal{S}$}\\
    \hline
    $\nu(e)$ & {$\nu(e)\in \mathcal{X}$ is the feature vector of object $e$}\\
    \hline
    $\ell(e)$ & {$\ell(e)\in \mathcal{Y}$ is the attribute vector of object $e$}\\
    \hline
    ${q}$ & A query object\\
    \hline
    $\delta (,)$ & The Euclidean distance between feature vectors \\
    \hline
    $G=(V,E)$ & A PG $G$ with a vertex set $V$ and an edge set $E$\\
    \hline
  \end{tabular}
\end{table}

Thus, our main contributions are as follows:

\begin{itemize}[leftmargin=*]
  \item {We present NHQ, a generalized framework tailored for answering hybrid queries with unstructured and structured constraints in a fused way. Our NHQ is friendly to existing popular PGs, and we can offer them the powerful ability to handle hybrid queries by deploying them in NHQ with a lightweight modification.}
  \item {We propose two NPGs by designing novel edge selection and routing strategies, which yield better performance regarding efficiency, accuracy, and memory usage compared to existing PGs.}
  \item {We deploy two NPGs in our NHQ framework, thus obtaining two new hybrid query methods that significantly outperform the state-of-the-art competitors on all datasets} (10$\times$ faster under the same \textit{Recall}).
  \item We verify our approaches' superiority of effectiveness and efficiency by various metrics on nine real-world datasets, compared with six popular PGs and six hybrid query methods, respectively.
\end{itemize}

The remainder of this paper is organized as follows. Preliminaries are presented in \textbf{Sec. \ref{Preliminaries}}. \textbf{Sec. \ref{framework}} outlines the NHQ framework. \textbf{Sec. \ref{NPG}} presents two proposed NPGs with the optimized edge selection and routing strategies. The NPG-based hybrid query methods are discussed in \textbf{Sec. \ref{NPG-NHQ}}. \textbf{Sec. \ref{experiments}} shows the experimental evaluation. \textbf{Sec. \ref{related_work}} reviews the related work, and \textbf{Sec. \ref{conclusion}} concludes this paper.

\section{Preliminaries}
\label{Preliminaries}
We first define the hybrid query in \textbf{Sec. \ref{Problem Definition}} and then briefly overview different types of solutions to hybrid queries in \textbf{Sec. \ref{Implementation Strategies}}. For ease of understanding, we summarize frequently used notations in {Tab. \ref{notations}}.

\subsection{Problem Definition}
\label{Problem Definition}

{Similar to \cite{aoyama2011fast,houle2014rank}, we use the term \textit{object set} to refer to the data that we deal with in this paper.}

\begin{myDef}
  \label{object_set_definition}
  {\textbf{Object Set.} An object set is defined as a set $\mathcal{S}=\{e_0,\dots,e_{n-1}\}$ of size $n$. For each object $e\in \mathcal{S}$, (1) its features are represented as a high-dimensional vector, denoted by $\nu(e)$, (2) $e$ has a set of attributes denoted by $\{a_0,\dots,a_{m-1}\}$ and $e.a_i$ indicates the value of attribute $a_i$ of $e$. Moreover, we define the feature vector space of all objects in $\mathcal{S}$ as $\mathcal{X}=\{\nu(e)|e\in \mathcal{S}\}$.}
\end{myDef}

\begin{myExa}
  \label{object_set_example}
  {An object set could refer to different types of data, e.g., a set of images or papers. When we specify each object $e$ as a paper, it carries two types of information: one is the implicit semantics behind the text, which is usually represented as a feature vector $\nu(e)$ through deep neural network (DNN), e.g., BERT \cite{devlin2018bert}; and another is the explicit attributes, e.g., $\{{\sf publish\_year}, {\sf venue}, {\sf topic}\}$, such as $e.{\sf venue}=$``${\sf SIGMOD}$''.}
\end{myExa}

Given an object set $\mathcal{S}$ and a user-input query object $q$, many approaches have been studied to retrieve the most similar objects to $q$ from $\mathcal{S}$ by considering feature vectors' distance \cite{Echihabi20,HNSW,EchihabiZP21}. In the following, we first introduce vector similarity search, then formally define the hybrid query that we focus in this paper.

{
Given an object $e\in \mathcal{S}$ and its feature vector $\nu(e) \in \mathcal{X}$, we have $\nu(e)=[\nu(e)^{0},\nu(e)^{1},\dots,\nu(e)^{d-1}]$, where $\nu(e)^{i}$ denotes the value of $\nu(e)$ on the $i$-th dimension. In particular, we are interested in the high-dimensional case where $d$ ranges from hundreds to thousands. For any two objects $e,o\in \mathcal{S}$ with feature vectors $\nu(e)$, $\nu(o)$ $\in$ $\mathcal{X}$, we can measure their vector similarity through various methods, e.g., Euclidean distance \cite{NSG} and Cosine similarity \cite{HNSW}. Among these methods, Euclidean distance is the most commonly used in the literature \cite{DPG}, which is given in Eq. \ref{euclidean_distance}.
\begin{equation}
  \label{euclidean_distance}
  \delta({\nu(e)},{\nu(o)})= \sqrt {\sum_{i=0}^{d-1} (\nu(e)^{i}-\nu(o)^{i})^2}
\end{equation}
}

We define the exact vector similarity search as follows \cite{NSSG,NNS_definition,skopal2011nonmetric,shimomura2018performance}:

\begin{myDef}
  \label{definition: exact vector similarity search}
  \textbf{Exact Vector Similarity Search.} Given an object set $\mathcal{S}$, a query object $q$ with the feature vector $\nu(q)$, and an integer $k$, the exact vector similarity search aims at obtaining the exact top-$k$ objects from $\mathcal{S}$ whose feature vectors are closest to $\nu(q)$.
\end{myDef}

In {Def. \ref{definition: exact vector similarity search}}, the exact top-$k$ objects (denoted by $\mathcal{G}$) hold that
\begin{equation}
  \label{exact top-k}
  \mathcal{G}=\arg \min_{\mathcal{G} \subseteq \mathcal{S}, \vert \mathcal{G} \vert = k} \sum_{e \in \mathcal{G}}\delta({\nu(e)},{\nu(q)})\quad.
\end{equation}
Applying exact vector similarity search on a large $\mathcal{S}$ is impractical because of the high computational cost \cite{graph_survey_vldb2021}. So, an approximate vector similarity search is of realistic significance, as it balances a trade-off between accuracy and efficiency using a vector index \cite{graph_survey_vldb2021} and can be defined as follows \cite{NSSG,NNS_definition,skopal2011nonmetric,shimomura2018performance}:

\begin{myDef}
  \label{definition: approximate vector similarity search}
  \textbf{Approximate Vector Similarity Search.} Given an object set $\mathcal{S}$, a query object $q$ with the feature vector $\nu(q)$, and parameters $\epsilon$ ($>$0) and $k$, the approximate vector similarity search aims at obtaining the approximate top-$k$ objects $\mathcal{D}$ from $\mathcal{S}$, such that $\forall i=0,1,\cdots,k-1$, $\delta(\nu(e_i),\nu(q)) \leq (1+\epsilon)\cdot \delta(\nu(o), \nu(q))$, where $e_i \in \mathcal{D}$, and $o \in \mathcal{S}$ is the exact top-$1$ object whose feature vector is closest to $\nu(q)$.
\end{myDef}

Given the exact top-$k$ objects $\mathcal{G}$ from {Def. \ref{definition: exact vector similarity search}} and the approximate top-$k$ objects $\mathcal{D}$ from {Def. \ref{definition: approximate vector similarity search}}, we can evaluate the approximate vector similarity search's accuracy by recall rate $Recall@k$:

\begin{equation}
  \label{recall}
  Recall@k=\frac{|\mathcal{D}\cap\mathcal{G}|}{k}\quad.
\end{equation}
A larger $Recall@k$ indicates that more accurate results are returned for approximate vector similarity search. Hereafter, we refer to \textit{approximate vector similarity search} simply as \textit{vector similarity search} unless otherwise specified.

To efficiently return the query results, vector similarity search preprocess $\mathcal{S}$ by building an vector index $\mathcal{I(\mathcal{S})}$ based on the feature vector space $\mathcal{X}$ of $\mathcal{S}$. According to how the index $\mathcal{I}(\mathcal{S})$ is built, we divide vector similarity search methods into four common genres: quantization-based \cite{PQ,ScaNN}; tree-based \cite{Silpa-AnanH08,AroraSK018}; hashing-based \cite{GongWOX20,HuangFZFN15}; and proximity graph (PG)-based groups \cite{HNSW,NSG}. Many works in the literature \cite{DPG,NSG} have demonstrated that the PG-based methods achieve better speedup vs recall rate trade-off.

Recently, a rich spectrum of emerging applications—e.g., finding similar products in e-commerce \cite{Milvus_sigmod2021,ADBV}—not only expect that the query results and given query object will have similar feature vectors, but also have the same attributes \cite{PASE,ADBV,Milvus_sigmod2021,Jingdong_paper}. For queries in this context, the query object's feature vector and attributes of interests can be viewed as the unstructured and structured query constraints, respectively \cite{ADBV}. We aim to identify similar objects that satisfy both the given feature vector and attribute constraints. We define such a hybrid query as follows:

\begin{myDef}
  \label{definition: hybrid query}
  {\textbf{Hybrid Query.} Given an object set $\mathcal{S}$ and a query object $q$ with the feature vector $\nu(q)$ and a set of attributes $\{a_0,\dots,a_{m-1}\}$ of size $m$, a hybrid query returns approximate top-$k$ objects from $\mathcal{S}$, denoted by $\mathcal{D}$. The relationship between the feature vectors of objects in $\mathcal{D}$ and $\nu(q)$ satisfies {Def. \ref{definition: approximate vector similarity search}} and objects' attributes are the same as those of query object $q$—that is, for $e \in \mathcal{D}$, $\forall i=0,1,\cdots,m-1, e.a_i=q.a_i$.}
\end{myDef}

Although the hybrid query can be viewed as an extended vector similarity search with attribute constraints \cite{Milvus_sigmod2021}, it is a non-trivial problem to solve, because all existing solutions suffer from effectiveness and efficiency issues (\textbf{L1-L4 in Sec. \ref{sec:limitations}}), which is what inspired our research. We next briefly introduce essential differences in the implementation strategies of hybrid query processing between existing solutions and ours, to highlight our solution's novelty.

\subsection{Implementation Strategies}
\label{Implementation Strategies}
\begin{figure}[!tb]
  \centering
  \setlength{\abovecaptionskip}{0cm}
  \includegraphics[width=\linewidth]{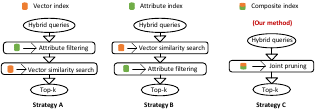}
  \caption{{Three different hybrid query implementation strategies. Our framework is based on Strategy C.}}
  \label{fig:implementation_strategies}
\end{figure}

{
As \textbf{{Sec. \ref{sec:limitations}}} mentions, existing solutions to hybrid queries are designed in a ``decomposition-assembly'' model, where a hybrid query is decomposed to two standalone subqueries for processing separately and final results assembling—that is, the vector similarity search is based on a vector index and attribute filtering is based on an attribute index. There are two implementation strategies regarding the execution order of these two subqueries, i.e., Strategies A and B in {Fig. \ref{fig:implementation_strategies}}. Current work mainly falls in these categories and cannot achieve the best performance. By contrast, our solution works by using a new fusion strategy (Strategy C).
}

\vspace{0.1cm}
\noindent\textbf{Strategy A: ``first attribute filtering, then vector similarity search.''} {In general, attribute filtering shows higher unit efficiency than vector similarity search, e.g., in our evaluation, the average overhead time is about 0.02$\mu$s and 0.50$\mu$s for checking whether two attributes are equal and computing the distance of a pair of feature vectors, respectively. Therefore, when attribute filtering runs first, vector similarity search can be performed on the obtained candidate subset rather than the entire object set, which will boost query efficiency to some extent. As a side effect, this strategy suffers from memory and efficiency issues regarding vector index construction for searching similar vectors, as we analyzed in \textbf{Sec \ref{sec:limitations}} (\textbf{L4}).} To address this problem, existing efforts \cite{ADBV,Milvus_sigmod2021} encode the feature vectors by PQ \cite{PQ} for approximating Euclidean distance ({Eq. \ref{euclidean_distance}}). Yet, they yield remarkably low query accuracy, especially compared to vector similarity search on top of the PG-based vector index \cite{DPG}.

\vspace{0.1cm}
\noindent\textbf{Strategy B: ``first vector similarity search, then attribute filtering.''} {This strategy can be classified further into two different categories: (1) One is to perform attribute filtering after completing vector similarity search; that is, we only need to check the attributes of the returned approximate top-$k$ objects. (2) The other is to perform attribute filtering after each step of vector similarity search; that is, we need to check the attributes for each object explored during the vector similarity search (i.e., the entire search space). In our evaluation (see \textbf{Sec. \ref{sec: NHQ validation}}), (1) is more efficient than (2) in most cases, so that (1) is adopted more widely in most existing work \cite{Jingdong_paper,Jingdong_lib}. One possible explanation is that, because current vector indexes are built with feature vectors rather than attributes, prematurely filtering the object that mismatches the attributes in each search step may impair vector similarity search's performance \cite{graph_survey_vldb2021}. Nevertheless, a hybrid query working off (1) usually requires vector similarity search to return more than $k$ candidates for subsequent attribute filtering (e.g., it may need to process 300 candidates for obtaining top-10 nearest objects), which limits vector similarity search's efficiency (\textbf{Sec. \ref{sec:limitations}, L3}).}

In addition, Strategies A and B must maintain the attribute index and vector index simultaneously ({Fig. \ref{fig:implementation_strategies}}), and separately searching on the two indexes would increase computational overhead (we discussed these in \textbf{Sec. \ref{sec:limitations}, L1–2}). To overcome these limitations, we design and implement the following novel solution.

\vspace{0.1cm}
\noindent\textbf{Strategy C: ``joint attribute filtering and vector similarity search''.} {This strategy carries out vector similarity search and attribute filtering concurrently on a {composite index} ({Fig. \ref{fig:implementation_strategies}}) that contains both the feature vectors' and attributes' information, which obviously differs from the vector index or attribute index in Strategies A and B. For Strategy C, given a query object and a {composite index}, we jointly prune unpromising objects with dissimilar vectors and mismatched attributes during query processing, so that we return query results in one step without intermediate candidates. To the best of our knowledge, this is the first work to present a hybrid query solution in a fused way. With this essential difference, compared to existing solutions we achieve 10$\times$ improvement in effectiveness and efficiency.
}

\section{Native Hybrid Query Framework}
\label{framework}


Rather than answering a hybrid query by two separate subqueries on different indexes, we propose a native hybrid query (NHQ) framework, with a well-designed {composite index} to support unstructured and structured query constraints simultaneously. We first show the intuition of our solution (\textbf{Sec. \ref{sec:motivation}}), and then describe the NHQ framework working off Strategy C in detail (\textbf{Sec. \ref{sec:NHQ}}).

\subsection{Intuition}
\label{sec:motivation}
To build a purpose-built index for hybrid queries, it is natural to assemble both unstructured and structured information in this index (i.e., a composite index). Research suggests that the graph has inherent advantages in embedding complex information \cite{zhang2020relational,huang2017label,choudhary2021survey}. For example, the multimodal knowledge graph embeds various modal data (e.g., image, text) into a graph, thereby benefiting a multitude of downstream tasks \cite{ZhuZRF15,KannanFAKCRYMF20}. Driven by this, we explore how to integrate various types of information into a graph as our {composite index}. {In the literature of vector similarity search, the mainstream vector index is implemented based on PG (see Def. \ref{def: PG}), which has been proven to offer state-of-the-art performance compared to other index types \cite{NSSG,SONG,Zhao_graph_construction}. However, existing PGs only contain feature vectors' neighborhood relationship while excluding attributes, so they cannot be used directly for building the composite index.}

\begin{myDef}
\label{def: PG}
  {\textbf{Proximity graph (PG).} Given an object set $\mathcal{S}$, we define the PG of $\mathcal{S}$ w.r.t. an distance threshold $\mathit{\Theta}$ as a graph $G=(V,E)$ with the vertex set $V$ and edge set $E$. (1) For each vertex $u \in V$, it corresponds to an object $e\in \mathcal{S}$. (2) For any two vertices $u_i$ and $u_j$ from $V$, we have an edge $u_iu_j\in E$, iff $\delta (\nu(u_i),\nu(u_j)) \leq \mathit{\Theta}$, where $\nu(u_i)$ and $\nu(u_j)$ is the feature vectors of objects $u_i$ and $u_j$, respectively.}
\end{myDef}

For a vertex $u_i$, we denote by $\mathcal{N}_{G}(u_i)$ the set of all neighbors of $u_i$ in $G$, that is, $\{u_i \mid u_j \in V, u_i u_j \in E\}$; we omit the subscript $G$ when the graph is clear from the context. In {Def. \ref{def: PG}}, $\mathit{\Theta}$ controls the vertex $u_i$'s neighbors. The higher the value of $\mathit{\Theta}$, the more neighbors $u_i$ has. Specifically, a larger $\mathit{\Theta}$ yields more neighbors of $u_i$, thus increasing the search space expanded from $u_i$, in turn, reducing search efficiency \cite{HNSW}. Meanwhile, a smaller $\mathit{\Theta}$ may break a PG's connectivity, which impacts search accuracy \cite{NSG}. In general, some specific PGs' differences mainly fall into the implementation of (2) in {Def. \ref{def: PG}} (e.g., DPG \cite{DPG} and NSG \cite{NSG}).

Instead of building an ordinary PG only based on objects' feature vectors, our insight is to build a new PG based on both the feature vectors and attributes to serve as the {composite index}, so that we can answer hybrid queries directly with structured and unstructured constraints. For example, consider an ordinary PG $G=(V,E)$ built for an object set $\mathcal{S}$. For any vertex $u \in V$, it only links neighbors (i.e., $\mathcal{N}(u)$) with similar feature vectors. In contrast, for our new PG, we not only expect the feature vector of each object in $\mathcal{N}(u)$ to be similar to $u$'s feature vector, but also to have the matched attributes. On this basis, we can jointly prune the unpromising vertices with dissimilar feature vectors and mismatched attributes by evaluating the fusion distance (defined in \textbf{Sec. \ref{sec:NHQ}}).

\begin{figure}[!tb]
  \centering
  \setlength{\abovecaptionskip}{0cm}
  \includegraphics[width=\linewidth]{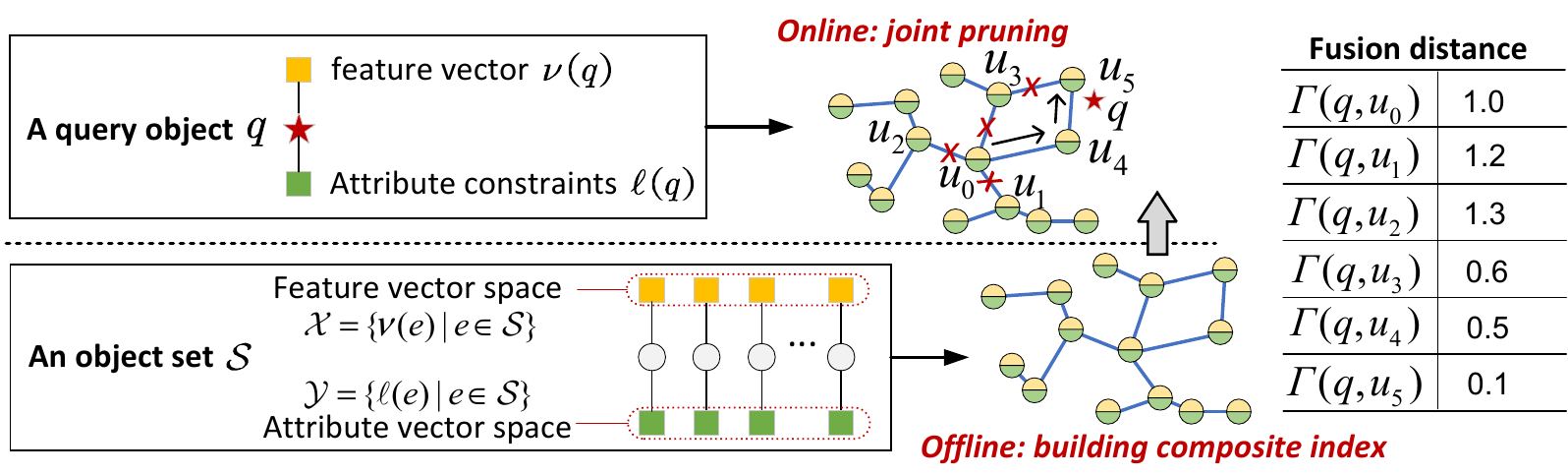}
  \caption{{NHQ framework working off Strategy C.}}
  \label{fig:framework}
\end{figure}

\subsection{NHQ Framework}
\label{sec:NHQ}
{Fig. \ref{fig:framework}} overviews our NHQ framework, which consists of two modules: \textit{composite index} and \textit{joint pruning}.

\vspace{0.1cm}
\noindent\textbf{Overview.} {To build a composite index, the key challenge is to measure the distance between objects—or rather, the fusion distance of feature vectors and attributes. We can evaluate the distance of feature vectors by {Eq. \ref{euclidean_distance}}, while for the distance of attributes, we must quantify a set of attributes for each object and then define a distance metric function on it \cite{Encoding3}. Given the distance of feature vectors and distance of attributes, fusing them into a unified distance is the basic premise of calculating the objects' distance. Next, we discuss how to compute the fusion distance between two objects, and show the procedure of a composite index construction based on this fusion distance. Then we present a joint pruning strategy based on the built composite index.}

\vspace{0.1cm}
\noindent\textbf{Fusion distance.} {Given an object set $\mathcal{S}$ with the feature vector space $\mathcal{X}$ shown in {Fig. \ref{fig:framework}} (bottom left), we first apply ordinal encoding \cite{sun2008ordinal} to encode the attribute values for each object $e\in \mathcal{S}$ as an attribute vector ${\ell(e)}=[{\ell(e)}^0,\cdots,{\ell(e)}^{m-1}]$, where ${\ell(e)}^i$ is the encoded value of attribute value $e.a_i$. Then, we have the attribute vector space ${\mathcal{Y}}=\{{\ell(e)}|e \in \mathcal{S}\}$.}

For $ e_i, e_j \in \mathcal{S}$, the distance between $\nu(e_i) \in \mathcal{X}$ and $\nu(e_j) \in \mathcal{X}$ can be evaluated by $\delta (\nu(e_i), \nu(e_j))$ (Eq. \ref{euclidean_distance}). The smaller $\delta (\nu(e_i), \nu(e_j))$ is, the more similar $\nu(e_i)$ and $\nu(e_j)$ are. We measure the distance between ${\ell(e_i)} \in {\mathcal{Y}}$ and ${\ell(e_j)} \in {\mathcal{Y}}$ as follows:
\vspace{-0.1cm}
\begin{equation}
  \label{label_distance}
  \chi ({\ell(e_i)},{\ell(e_j)})=\sum _{k=0} ^{m-1} \phi ({\ell(e_i)}^k, {\ell(e_j)}^k)\quad ,
\end{equation}
where $\phi ({\ell(e_i)}^k, {\ell(e_j)}^k)$ satisfies
\begin{equation}
  \label{label_value}
  \setlength{\nulldelimiterspace}{0pt}
  \phi ({\ell(e_i)}^k, {\ell(e_j)}^k) = \left\{\begin{IEEEeqnarraybox}[\relax][c]{l's}
    0, &${\ell(e_i)}^k= {\ell(e_j)}^k$ \\
    1, &${\ell(e_i)}^k\neq {\ell(e_j)}^k$
  \end{IEEEeqnarraybox}\right.\quad .
\end{equation}

In Eq. \ref{label_distance}–\ref{label_value}, $m$ is the number of dimensions of ${\ell(e_i)}$ and ${\ell(e_i)}^{k}$ denotes the values of ${\ell(e_i)}$ on the $k$-th dimension. The smaller the $\chi ({\ell(e_i)}, {\ell(e_j)})$, the higher the similarity between ${\ell(e_i)}$ and ${\ell(e_j)}$.

Given $\delta (\nu(e_i), \nu(e_j))$ and $\chi ({\ell(e_i)},{\ell(e_j)})$, we deal with the fusion distance $\mathit{\Gamma} (e_i,e_j)$ of objects $e_i$ and $e_j$ as follows:
\begin{equation}
  \label{hybrid_distance}
  \mathit{\Gamma} (e_i,e_j)= \omega _{\nu} \cdot \delta(\nu(e_i),\nu(e_j))+\omega _{\ell} \cdot \chi ({\ell(e_i)},{\ell(e_j)})\quad ,
\end{equation}
where $\omega _{\nu}$, $\omega _{\ell}$ are the weights of feature vectors' and attribute vectors' distances, respectively; and the smaller $\mathit{\Gamma} (e_i,e_j)$ is, the more $e_i$ and $e_j$ are similar in both the feature vectors and attributes. 

{{Eq. \ref{hybrid_distance}} shows a simple and practically feasible way to fuse two completely different distances. It is especially friendly for building a composite index on top of existing PGs, because we easily extend an existing PG to a composite index simply by replacing the distance measure from {Eq. \ref{euclidean_distance}} to {Eq. \ref{hybrid_distance}}. For example, if we set $\omega _{\nu}=1$ and $\omega _{\ell}=0$, then we have $\mathit{\Gamma} (e_i, e_j)=\delta(\nu(e_i),\nu(e_j))$, which is exactly the case of building an original PG based on the feature vector distance. Meanwhile, if we set $\omega _{\nu}=0$ and $\omega _{\ell}=1$—that is, $\mathit{\Gamma} (e_i, e_j)=\chi ({\ell(e_i)},{\ell(e_j)})$—we then get a PG based on the distance of attribute vectors. So, we can obtain an optimal compromise between feature vector distance and attribute vector distance by adjusting $\omega _{\nu}$ and $\omega _{\ell}$.}

\begin{algorithm}[t]
\label{alg: composite index}
  \caption{Building Composite Index ($\mathcal{S}$)}
  \LinesNumbered
  \KwIn{Object set $\mathcal{S}$}
  \KwOut{\textit{Composite Index} ${G}=({V},{E})$}
  
  ${V} \gets \mathcal{S}$, ${E} \gets \emptyset $
  
  \ForAll{$u_i \in {V}$}{
    \ForAll{$u_j \in {V} \setminus \{ u_i \}$}{
      \If{$\mathit{\Gamma} (u_i, u_j) \leq {\mathit{\Theta}}^{\prime}$}{
        ${E} = {E} \cup \{ u_i u_j  \}$
      }
    }
  }
  return ${G}=({V}, {E})$
\end{algorithm}

\begin{algorithm}[t]
\label{alg: joint pruning}
  \caption{Joint Pruning (${G}$, $q$, $\mathcal{B}$)}
  \LinesNumbered
  \KwIn{\textit{Composite index} ${G}$, query object $q$, seed set $\mathcal{B}$}
  \KwOut{result set $\mathcal{R}$}
  
  candidate set $\mathcal{C} \gets \mathcal{B}$; result set $\mathcal{R} \gets \mathcal{B}$
  
  \While{$\mathcal{R}$ is updated}{
    {$u_i \gets \arg\, \min _{u_i \in \mathcal{C}}{\mathit{\Gamma} (q,u_i)}$; $\mathcal{C}=\mathcal{C} \setminus \left \{u_i \right \}$}
    
    {$\mathcal{N}(u_i) \gets$ the neighbors of $u_i$; $\mathcal{ C}=\mathcal{ C} \cup \mathcal{N}(u_i)$}
    
    \ForAll{{$u_j \in \mathcal{N}(u_i)$}}{	
            
      $u_r \gets \arg \, \max _{u_r \in \mathcal{R}} {\mathit{\Gamma} (q,u_r)}$

      \If{$\mathit{\Gamma} (q,u_j) <\mathit{\Gamma} (q,u_r)$}{
        {$\mathcal{R}=\mathcal{R} \setminus \left \{u_r \right \}$; $\mathcal{R}=\mathcal {R} \cup \left \{u_j \right \}$}
      }
      
    }

  }
  return $\mathcal{R}$
\end{algorithm}

\vspace{0.1cm}
\noindent{\textbf{Optimal weight configuration.}} {In our experimental study, $\omega _{\nu}=1$ and $\omega _{\ell}=\delta(\nu(e_i),\nu(e_j))/m$ ($m$ is the dimension of an attribute vector) offer the best hybrid querying performance and are recommended for most datasets and algorithms (see \textbf{Sec. \ref{sec: parameter sensitivity}}). In other words, this configuration of $\omega _{\nu}$ and $\omega _{\ell}$ is dataset independent—that is, they only relate to the feature vector distance $\delta(\nu(e_i),\nu(e_j))$ of two specific objects and the attribute vector dimension $m$—so we do not need to configure them specifically for different datasets. The basic idea behind this configuration is that we obtain a fusion distance by fine-tuning the feature vector distance via the attribute distance. More precisely, for two objects $e_i$ and $e_j$, we first obtain their feature vector distance $\delta (\nu(e_i), \nu(e_j))$, then the attribute distance $\chi (\ell(e_i), \ell(e_j))$ serves as a fine-tuning effect for the fusion distance. For example, if $e_i$ and $e_j$ have exactly matched attributes, that is, $\chi (\ell(e_i), \ell(e_j))=0$, we make no changes on the original feature vector distance (i.e., $\mathit{\Gamma} (e_i,e_j)=\delta (\nu(e_i), \nu(e_j))$). While if the attributes of $e_i$ and $e_j$ do not match at all—that is, $\chi (\ell(e_i), \ell(e_j)) =m$—we impose a penalty on the original feature vector distance and achieve a fusion distance of $\mathit{\Gamma} (e_i,e_j)=2 \cdot \delta (\nu(e_i), \nu(e_j))$. Finally, we get that $\mathit{\Gamma} (e_i,e_j)$ satisfies $\delta (\nu(e_i), \nu(e_j)) \leq \mathit{\Gamma} (e_i,e_j) \leq 2 \cdot \delta (\nu(e_i), \nu(e_j))$ for different attribute distances.}

\vspace{0.1cm}
\noindent\textbf{Composite index.} The original PGs construct the neighborhood relationship between two objects based on their feature vector distance (Eq. \ref{euclidean_distance}), so that we say only the objects' unstructured information is preserved in the PG's topological structure. By contrast, we take into consideration both the distances of feature vectors and attribute vectors by {Eq. \ref{hybrid_distance}}. So, we preserve both the unstructured and structured information in the neighborhood relationship between objects, thereby generating a new PG and we can take it as the composite index (as {Fig. \ref{fig:framework}} (bottom right) shows).

{Specifically, given an object set $\mathcal{S}$, we build a composite index ${G} = ({V}, {E})$ on $\mathcal{S}$ through \hyperref[alg: composite index]{Alg. 1}. We first initialize the vertex set ${V}$ by $\mathcal{S}$ to ensure that each object $e\in \mathcal{S}$ corresponds to a vertex $u\in V$ in $G$, as well as an empty edge set $E$ (line 1); and then we connect two objects $u_i, u_j \in {V}$ with an edge $u_iu_j\in E$, iff $\mathit{\Gamma} (u_i, u_j)\leq {\mathit{\Theta}}^{\prime}$, ${\mathit{\Theta}}^{\prime}$ is a predefined fusion distance threshold (lines 2–5). In our experimental study, ${\mathit{\Theta}}^{\prime}$ has an optimal value to achieve a robust query performance. We found that ${\mathit{\Theta}}^{\prime}$ is strongly related to the degree of vertex, and the upper bound on each vertex's degree is always 20 for an optimal ${\mathit{\Theta}}^{\prime}$ on most datasets, e.g., the optimal ${\mathit{\Theta}}^{\prime}$ is around 230 on SIFT1M dataset\textsuperscript{\ref{texmex}}. We can determine this value via a grid search \cite{huang2012improved}.}

\vspace{0.1cm}
\noindent\textbf{Joint pruning.} Different from the separate pruning for unstructured and structured query constraints in the existing work \cite{ADBV,Milvus_sigmod2021,Jingdong_paper} (Strategies A and B in {Fig. \ref{fig:implementation_strategies}}), we simultaneously prune the unpromising objects with dissimilar feature vectors and mismatched attributes (measured by the fusion distance, {Eq. \ref{hybrid_distance}}) based on the aforementioned composite index (top right in {Fig. \ref{fig:framework}}).

{Given a composite index ${G}=(V,E)$ built for an object set $\mathcal{S}$, a query object $q$, and a seed set $\mathcal{B} \subseteq V$ (usually these are selected randomly from $V$ \cite{graph_survey_vldb2021}), we obtain the approximate top-$k$ objects via joint pruning using the following steps (\hyperref[alg: joint pruning]{Alg. 2}): (1) \textit{Initialization}. We use a visited vertex set $\mathcal{C}$ to record the vertices for further search expansion and use a result set $\mathcal{R}$ of size $k$ to record current query results, both of which are initialized by $\mathcal{B}$ (line 1). (2) \textit{Search expansion}. We take out the vertex $u_i$ with the smallest $\mathit{\Gamma} (q,u_i)$ from $\mathcal{C}$ as the next visited vertex for search expansion (line 3), and then expand $\mathcal{C}$ by $\mathcal{C}\cup \mathcal{N}(u_i)$ (lines 3–4). (3) \textit{Query results update}. We update $\mathcal{R}$ by the better vertices in $\mathcal{N}(u_i)$ (lines 5–8). To be more precise, for any vertex $u_j \in \mathcal{N}(u_i)$ and a vertex $u_r\in \mathcal{R}$ that has the farthest distance to $q$, we replace $u_r$ in $\mathcal{R}$ by $u_j$, if $u_j$ is closer to $q$ than $u_r$ in terms of our fusion distance, i.e., $\mathit{\Gamma} (q,u_j) < \mathit{\Gamma} (q,u_r)$. This is because $u_j$ is more similar to $q$ in both the feature vectors and attributes, compared with $u_r$. Finally, we repeat (2) and (3) until $\mathcal{R}$ cannot be further updated, then we terminate the query and return $\mathcal{R}$ as the approximate top-$k$ objects.}

\begin{myExa}
  \label{example: joint pruning}
  {In {Fig. \ref{fig:framework}} (top right), we show the joint pruning of returning the nearest object (i.e., top-1) to the query object $q$. First, we initialize the seed vertex set with a randomly selected vertex, i.e., $\mathcal{B}=\{u_0\}$, then we have $\mathcal{C}=\{u_0\}$, $\mathcal{R}=\{u_0\}$. Second, we expand the search space with $u_0$'s neighbors and update $\mathcal{C}=\{ u_1,u_2,u_3,u_4 \}$. Third, we replace $u_0$ in $\mathcal{R}$ by $u_4$ because $\mathit{\Gamma} (q,u_4) = 0.5$ is smaller than $\mathit{\Gamma} (q,u_0) = 1.0$. We then keep expanding the search space by updating $\mathcal{C}$ with the neighbors of $u_4$, $u_5$, and finally we have $R=\{u_5\}$. Because each vertex $u_j \in \mathcal{N}(u_5)$ satisfies $\mathit{\Gamma} (q,u_j) > \mathit{\Gamma} (q,u_5)$, we return $u_5$ as the approximate top-1 nearest object. In the process, we prune the search space from vertices $u_1$, $u_2$, and $u_3$.}
\end{myExa}

\noindent\textbf{Remarks.} (1) It is worth emphasizing that joint pruning can obtain the hybrid query results in one step, thereby avoiding the post-processing operation (merging in {Fig. \ref{fig:hybrid_query_intro}}(c)) of solutions following the ``decomposition-assembly'' model. (2) Although the fusion distance weights' optimal configuration ($\omega _{\nu}$ and $\omega _{\ell}$ in {Eq. \ref{hybrid_distance}}) is dataset independent, users still can set them flexibly according to their query preference, thereby getting personalized query results (e.g., only considering the attributes by setting $\omega _{\nu}$$=$$0$ and $\omega _{\ell}=1$).

\section{Navigable Proximity Graph}
\label{NPG}
{NHQ is a generalized framework works well with existing PGs. We easily can deploy a specific PG in NHQ to form the composite index by changing its original distance measure to our fusion distance (Eq. \ref{hybrid_distance}), so that we can offer this PG the ability to handle hybrid queries effectively. However, current PGs' limitations would have a side-effect on NHQ's overall performance with the PG-based composite index. This motivates us to present two new Navigable PGs (NPGs) in this section, by optimizing the \textit{edge selection} and \textit{routing} strategies widely used for building a PG and searching on a PG, respectively.} We discuss the details of how to deploy our NPGs in NHQ in \textbf{Sec. \ref{NPG-NHQ}}.

\subsection{Edge Selection}\label{sec: edge selection}
{Edge selection is a key step in building a PG \cite{graph_survey_vldb2021}. Given an object set $\mathcal{S}$, we obtain the neighbors of each object $e$ in $\mathcal{S}$ based on an edge selection strategy. Different strategies produce noteworthy index structure discrepancies for a PG, thereby impacting search performance on the PG \cite{NSG,DPG}.}

\vspace{0.1cm}
\noindent{\textbf{Intuition.}} In general, existing PGs focus on two factors during edge selection: \textit{distance between two vertices} (D1) and \textit{distribution of all vertices} (D2) \cite{graph_survey_vldb2021}. Early PGs such as NSW \cite{NSW} and KGraph \cite{NNDescent} only consider D1 when selecting edges—that is, each vertex is connected with a certain number of its nearest neighbors. \cite{lin2019graph} argued that only considering D1 would lead to redundant computations and impair search efficiency, as illustrated in {Example \ref{example:distance_factor}}.

\begin{figure}[!tb]
  \centering
  \setlength{\abovecaptionskip}{0cm}
  \includegraphics[width=\linewidth]{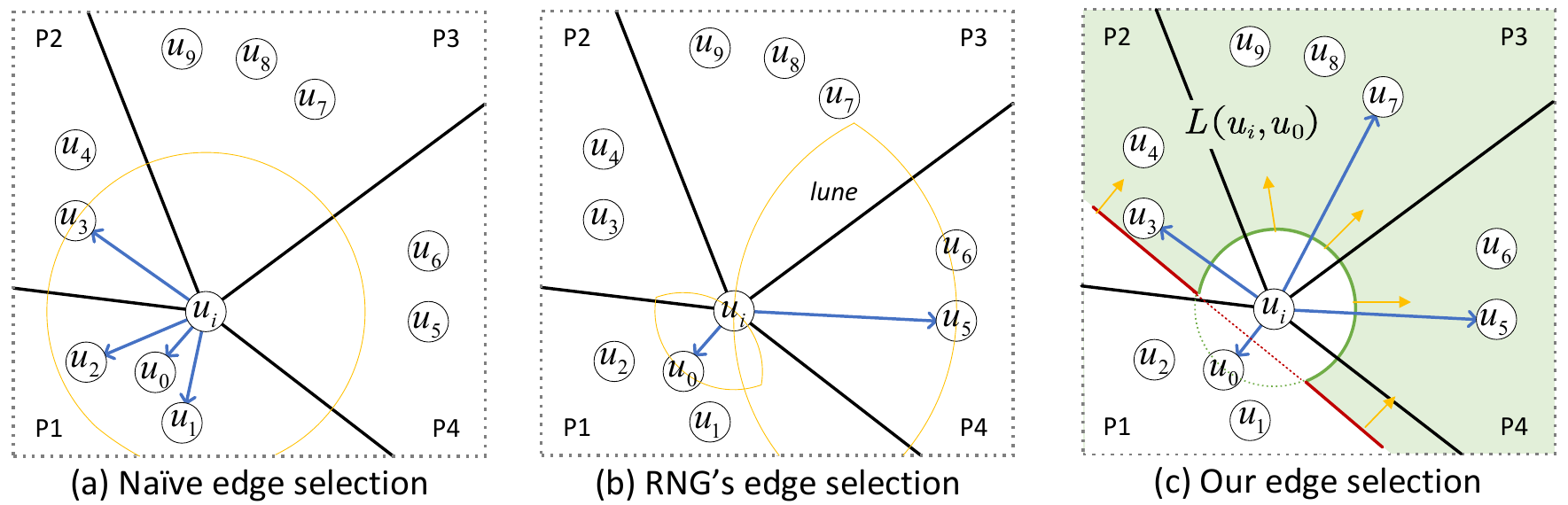}
  \caption{{Neighbors of $u_{i}$ obtained by different edge selection strategies (we assume $u_i$ has a maximum of four neighbors).}}
  \label{fig:edge_selection}
\end{figure}

\begin{myExa}
  \label{example:distance_factor}
  {Fig. \ref{fig:edge_selection}}(a) illustrates a situation where four nearest neighbors are connected to $u_{i}$. We divide $u_{i}$'s neighborhood into four parts (P1–P4) by a black solid line, and the neighbors located in each part will guide the search to approach the query object $q$ along this orientation. Suppose we are going to expand the search space from $u_i$. Because most of $u_i$'s neighbors are in the same area (such as $u_{0}$–$u_{2}$ in P1), we cannot use these neighbors to guide a search toward $q$ that is located in different areas (e.g., P3). Instead, we need extra computation to find a new route to $q$'s area, which adds computational overhead and reduces query efficiency.
\end{myExa}

Recent PGs add D2 to edge selection, which diversifies each vertex's neighbors under the premise of ensuring a similarity of neighbors \cite{DPG}, thereby significantly improving the search performance. \cite{graph_survey_vldb2021} concluded that state-of-the-art PGs connect neighbors in more directions through the edge selection strategy of Relative Neighborhood Graph (RNG), which we define next.

\begin{myDef}
  \label{definition:RNG}
  {\textbf{RNG \cite{RNG1}.} A RNG $G=(V,E)$ w.r.t a given object set $\mathcal{S}$ holds that {Def. \ref{def: PG}} (PG's definition), and for any two vertices $u_{i}$, $ u_{j}$ from $V$, we have an edge $u_iu_{j} \in E$, iff $\delta (\nu(u_i)$, $\nu(u_j))$$<$$\delta (\nu(u_i), \nu(u_k))$, or $\delta (\nu(u_i),\nu(u_j))$$<$$\delta (\nu(u_k), \nu(u_j))$, where $u_k\in V$ is an arbitrary vertex that satisfies $u_k \neq u_i \neq u_j$.}
\end{myDef}

\begin{myExa}
  \label{example:RNG_edge_selecton}
  As {Fig. \ref{fig:edge_selection}}(b) shows, the yellow lune is formed by two circles' intersection. For example, the small lune is formed by two circles with vertices $u_{i}$ and $u_{0}$ as the centers and distance $\delta (\nu(u_{i}),\nu(u_{0}))$ as the radius. {We add an edge between $u_i$ and $u_0$, iff the lune formed by $u_i$ and $u_0$ does not include any other vertices, which is equivalent to the condition of RNG's edge selection in {Def. \ref{definition:RNG}} (refer to \cite{bose2012proximity} for proof).} Under this rule, RNG's edge selection prevents $u_{i}$'s neighbors from congregating in the same area, so that there is more opportunity to link neighbors in different areas (such as $u_{5}$ in P4) \cite{DPG}. However, RNG's edge selection is still problematic, e.g., the vertices in P2 and P3 cannot be connected to $u_{i}$ in {Fig. \ref{fig:edge_selection}}(b) because they violate the above rule. For instance, the lune formed by $u_3$ and $u_i$ already includes a vertex $u_2$, so we cannot add an edge between $u_3$ and $u_i$. As a result, for the case that query object $q$ in the area of P2 or P3, we still need more computations for finding a new route to $q$'s area.
\end{myExa}

\noindent{\textbf{Our edge selection.}} Based on the above observations, we design a new edge selection strategy considering D1 and D2, which is committed to link one nearest neighbor to $u_i$ in each area of $u_i$ ({Fig. \ref{fig:edge_selection}}(c)). {We first give the definition of the \textit{landing zone} formed by $u_i$ and $u_i$'s one neighbor $u_j\in\mathcal{N}(u_i)$, which depicts the area that only the vertices in it could be added to $\mathcal{N}(u_i)$. We then give our edge selection strategy working off the landing zone.}

\begin{myDef}
  \label{def: landing zone}
  {\textbf{Landing zone.} Given a vertex $u_i$ and $u_i$'s one neighbor $u_j\in\mathcal{N}(u_i)$, the landing zone $L(u_i, u_j)$ formed by $u_i$ and $u_j$ is an area defined by ${H}(u_{i},u_{j})\setminus{B} (u_{i},\delta (\nu(u_i),\nu(u_j)))$, where ${H}(u_{i},u_{j})$ is the half space containing $u_{i}$, it is divided by the perpendicular bisectionplane $\mathcal{U}(u_{i},u_{j})$ of the line connecting $u_{i}$ and $u_{j}$ \cite{onishi1996construction}, and ${B} (u_{i}, \delta(\nu(u_i),\nu(u_j)))$ is the hypersphere with $u_{i}$ as the center and $\delta (\nu(u_i),\nu(u_j))$ as the radius \cite{NSG}.}
\end{myDef}

\begin{myExa}
  \label{exa: landing zone}
  {As {Fig. \ref{fig:edge_selection}}(c) shows, in a two dimensional space, $\mathcal{U}(u_{i},u_{0})$ is a perpendicular bisector of the line connecting $u_i$ and $u_0$ (i.e., the red line), ${H}(u_{i}, u_{0})$ is located on the upper side of $\mathcal{U}(u_{i},u_{0})$, and ${B}(u_{i}, \delta (\nu(u_i), \nu(u_0)))$ is the area enclosed by the green circle. Therefore, the landing zone $L(u_i, u_0)$ formed by $u_i$ and $u_0$ is the green shaded region (i.e., ${H}(u_{i},u_{0})\setminus{B} (u_{i},\delta (\nu(u_i),\nu(u_0)))$).}
\end{myExa}

{By defining a landing zone $L(u_i, u_j)$ for each neighbor $u_j\in \mathcal{N}(u_i)$, we easily locate the areas without any neighbor of $u_i$ by the intersection of the landing zones of $u_i$'s all neighbors (i.e., $\bigcap _{u_j \in \mathcal{N}(u_i)}L(u_i, u_j)$). Then, we add a nearest vertex in this intersection area to $\mathcal{N}(u_i)$, which facilitates the diversity of neighbors regarding areas.} We now introduce the core steps of our edge selection strategy for building a PG $G=(V,E)$; more specifically, for $\forall u_{i} \in V$, we aim to obtain its neighbors $\mathcal{N}(u_{i})$ as follows.

\noindent (1){ \textit{Candidates initialization}. We acquire a candidate neighbor set $\mathcal{C}(u_{i})$ (a subset of $(V \setminus \{ u_{i} \})$) of size $l$ (by randomly sampling \cite{NNDescent} or an additional index \cite{NSW}); it holds that $l \geq k$, where $k$ is the upper bound of the number of $u_{i}$'s neighbors (i.e., $|\mathcal{N}(u_i)| \leq k$).}

\noindent (2){ \textit{Neighbors initialization}. We sort the vertices in $\mathcal{C}(u_{i})$ in ascending order w.r.t their distances to $u_{i}$. We then initialize $\mathcal{N}(u_{i})$ with $u_i$'s nearest candidate neighbor $u_t$ from $\mathcal{C}(u_i)$ and add $u_t$ to $\mathcal{N}(u_i)$, as well as removing $u_t$ from $\mathcal{C}(u_i)$.}

\noindent (3){ \textit{Neighbors update}. We remove the nearest vertex $u_p$ of $u_i$ from $\mathcal{C}(u_i)$, and add $u_p$ to $\mathcal{N}(u_i)$, iff $u_p$ is in the new intersection area (i.e., $\bigcap _{u_j \in \mathcal{N}(u_i)}L(u_i, u_j)$) of the landing zones formed by $u_i$ and all its neighbors. We repeat this process until $\mathcal{C}(u_{i})=\emptyset$.}

\vspace{0.1cm}
\noindent\textbf{Remarks.} {Since each $u_p$ added to $\mathcal{N}(u_i)$ is located in a new area, our strategy ensures the diversity of $u_i$'s neighbors regarding areas where $\mathcal{C}(u_i)$ is located.}

\begin{myExa}
  \label{example:ORNG}
  {In {Fig. \ref{fig:edge_selection}}(c), suppose we have $\mathcal{N}(u_{i})=\{ u_{0} \}$ and $\mathcal{C}(u_{i})=\{ u_{3}, u_{4}, \cdots, u_{9}  \}$, we can add $u_{3}$ to $\mathcal{N}(u_{i})$ because $u_{3}$ is in the landing zone $L(u_i, u_0)$ (i.e., the green shaded region); then we can add $u_5$ to $\mathcal{N}(u_{i})$ because $u_5$ is in the intersection of $L(u_i, u_0)$ and $L(u_i, u_3)$ (i.e., $L(u_i, u_0)\cap L(u_i, u_3)$).} We repeat our edge selection until each new area contains one neighbor of $u_i$ (e.g., $u_7$ in P3 and $u_5$ in P4). Compared to RNG's edge selection, the neighbors acquired by ours are more diverse. Hence, we can route to the query object $q$ based on $u_i$'s neighbor that is located in the same area of $q$, thereby improving the query efficiency.
\end{myExa}

\noindent{\textbf{Complexity.}} For $u_{i} \in V$, the time complexity of sorting $\mathcal{C}(u_{i})$ is $O(l\cdot \log(l))$. For each $u_p \in \mathcal{C}(u_i)$, {we must check $|\mathcal{N}(u_i)|$ ($\leq k$) times to see if $u_p$ is in the landing zone $L(u_i, u_j)$ for each $u_j\in\mathcal{N}(u_i)$. Thus, the check times of getting the final $\mathcal{N} (u_{i})$ via our edge selection is $l \cdot k$ ($\ll |V|$).} So, the time complexity of performing our edge selection on $V$ is $O(l \cdot (k + \log (l)) \cdot |V|)$.

\subsection{Routing}
\label{sec:routing}
\begin{figure}[!tb]
  \centering
  \setlength{\abovecaptionskip}{0cm}
  \includegraphics[width=\linewidth]{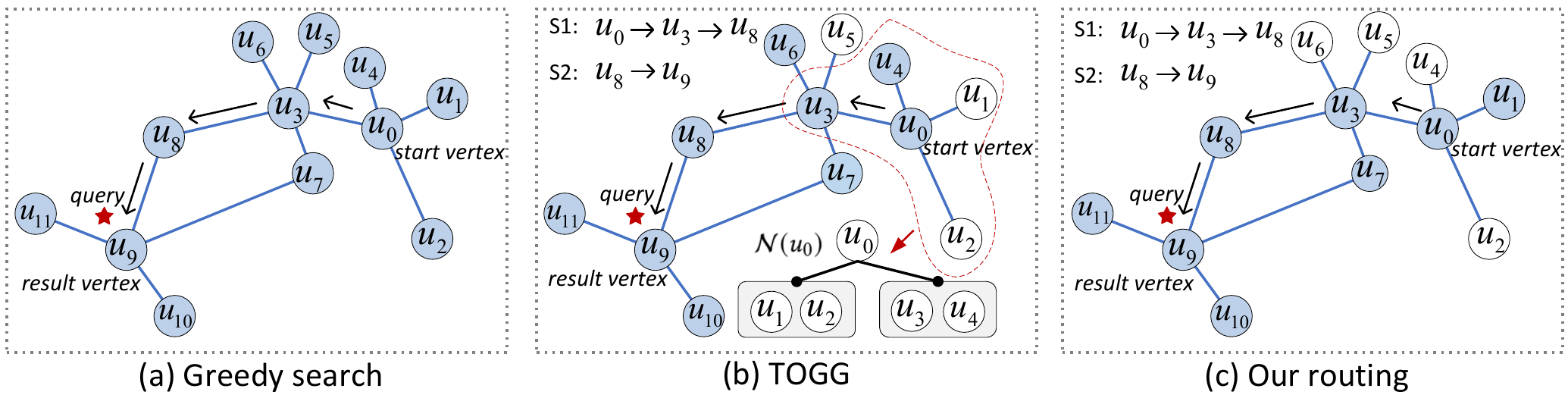}
  \caption{{Different routing strategies on PG. The white vertices do not need to, but the blue ones must calculate the distance from the query.}}
  \label{fig:routing_strategies}
\end{figure}

{Routing is a key step of searching on a PG \cite{TOGG}. Given a PG and a query object, the process of finding the query results is implemented using a proper routing strategy that determines a routing path from the start vertex to the result vertex \cite{TOGG}, e.g., the path indicated by the black arrows in {Fig. \ref{fig:routing_strategies}}. Obviously, the routing strategy directly affects the efficiency and accuracy of searching on the PG \cite{HCNNG}.}

\vspace{0.1cm}
\noindent\textbf{Intuition.} Most existing PGs adopt a greedy search as their routing strategy, which leads to some redundant computations (see {Example \ref{example:routing}}) \cite{BaranchukPSB19,LiZAH20}. Some machine learning optimizations \cite{LiZAH20} are used to mitigate this problem, which achieve better \textit{Speedup} vs \textit{Recall} trade-off at the expense of more index processing time and memory \cite{graph_survey_vldb2021}. \cite{TOGG} specifies two search stages on which routing has different requirements, i.e., \textit{the stage far from the query object} (S1) and \textit{the stage closer to the query object} (S2). In S1, \cite{TOGG} claims that we should quickly locate the query object's neighborhood (for efficiency requirement), while in S2, we should focus on comprehensively visiting vertices nearest to the query object (for accuracy requirement). Therefore, \cite{TOGG} designs a tailored routing strategy for each stage, to form a two-stage routing strategy, called TOGG. {However, TOGG attaches a tree-based index (e.g., KD-tree is used) to organize the neighbors of each vertex (see {Example \ref{example:routing}}), which increases the index construction time and memory overhead.}

\vspace{0.1cm}
\noindent\textbf{Our routing.} As a simple and effective solution, we design a random TOGG. In S1, {for each visited vertex $u_{i}$, rather than obtaining partial neighbors of $u_i$ for distance calculation by an additional tree (as TOGG did), we randomly select $u_i$'s $\lceil {k}/{h} \rceil$ ($1\leq h \leq k$, where $k$ is the upper bound of the vertex degree) neighbors from $\mathcal{N}(u_{i})$ for distance calculation and quickly approach the query object's neighborhood.} While in S2, we evaluate the distance of all neighbors in $\mathcal{N}(u_{i})$ to the query object to ensure query accuracy, which is the same as the greedy search. Different from TOGG, our routing can achieve the comparable accuracy but does not introduce additional index management overhead.

\begin{myExa}
  \label{example:routing}
  {Fig. \ref{fig:routing_strategies}} shows the routing process (indicated by black arrows) from the start vertex $u_{0}$ to the result vertex $u_{9}$ working off different routing strategies. When we use greedy search as the routing strategy ({Fig \ref{fig:routing_strategies}}(a)), neighbors of each visiting vertex are explored fully; e.g., for $u_0$, the distances from all vertices in $\mathcal{N}(u_0)$=$\{u_1, u_2, u_3\}$ to the query object are computed. As a result, the greedy search has 12 distance calculations. In {Fig. \ref{fig:routing_strategies}(b)}, TOGG organizes each vertex's neighbors ($\mathcal{N}(u_{0})$) in a tree-based index. In S1 ($u_0$$\rightarrow$$u_3$$\rightarrow$$u_8$), TOGG selects the next hop based on each visited vertex's neighbor index. In S2 ($u_8$$\rightarrow$$u_9$), TOGG uses a similar routing to the greedy search. Finally, TOGG only needs 9 distance calculations. Instead of an additional tree-based index for each vertex's neighbors, our routing randomly selects $\lceil k/h \rceil$=2 neighbors (when we set $k$=4 and $h$=2) for computing the distance to the query object in S1 and performs the greedy search in S2. Similar to TOGG, our strategy requires only 8 distance calculations in {Fig. \ref{fig:routing_strategies}(c)}. It is worth noting that our routing avoids some unnecessary calculations and improves query efficiency without extra index processing and memory overhead.
\end{myExa}

\noindent{\textbf{Complexity.}} According to \cite{HNSW,NSG,TOGG}, the greedy search's time complexity on a PG is about $O(k \cdot \log (|V|))$, where $k$ $\ll$ $|V|$ is the maximum number of neighbors of each vertex and $\log (|V|)$ is approximately the length of the routing path on average. {In our routing, we set the length of the routing path in S1 as $l_1$ and the length of the routing path in S2 as $l_2$ on average; then we have $l_1$+$l_2$=$\log (|V|)$. Therefore, our routing's time complexity is $O((\lceil {k}/{h} \rceil) \cdot l_1 + k \cdot l_2)$.}

\subsection{NPG with Our Edge Selection and Routing}
\label{sec: algorithm optimization} 

We present two NPGs, i.e., NPG\_nsw and NPG\_kgraph, based on two mainstream PGs, NSW \cite{NSW} and KGraph \cite{NNDescent}, by using our edge selection and routing strategies. More precisely, we construct two NPGs through our edge selection and conduct the search on them through our routing.

\vspace{0.1cm}
\noindent{\textbf{NPG\_nsw.}} NPG\_nsw is constructed by inserting an object incrementally. Specifically, a newly inserted object $e_{i}$ $\in$ $\mathcal{S}$ (corresponding to a vertex $u_i$) is regarded as a query object, {so we conduct a greedy search \cite{NSG} to obtain $l$ vertices closest to $u_{i}$ as $u_{i}$'s candidate neighbors $\mathcal{C}(u_{i})$ from the NPG\_nsw built on previously inserted objects \cite{NSW}}. Then we apply our edge selection to form $u_i$'s neighbors $\mathcal{N}(u_{i})$. We repeat these operations for each object in $\mathcal{S}$, yielding a NPG\_nsw. 

\vspace{0.1cm}
\noindent{\textbf{NPG\_kgraph.}} This is built by iteratively updating neighbors. For each $e_{i} \in \mathcal{S}$ (corresponding to a vertex $u_i$), we randomly generate its $l$ initial candidate neighbors $\mathcal{C}(u_{i})$. {To make the vertices in $\mathcal{C}(u_{i})$ are closer to $u_i$, we refine $\mathcal{C}(u_{i})$ through the neighbors of $u_i$'s neighbors, as \textit{the neighbors of a vertex's neighbors are likely to be neighbors of the vertex} \cite{NNDescent}. Then we obtain $\mathcal{N}(u_i)$ by using our edge selection in the final $\mathcal{C}(u_{i})$.} Specifically, we execute the following process until the graph quality (see {Def. \ref{definition:graph_quality}}) \cite{BoutetKMT16} reaches a preset threshold (in our experiment, 0.8 is enough to achieve a good performance): for any $u_{j} \in \mathcal{C}(u_{i})$ and $u_k \in \mathcal{C}(u_j)$, we add $u_k$ to $\mathcal{C}(u_{i})$, if $\delta (u_{i},u_{k}) < \delta (u_{i},u_{t})$, where $u_t$ is the farthest vertex to $u_i$ in $\mathcal{C}(u_{i})$. After the iteration completes, for each $ u_{i} \in V$, its neighbors $\mathcal{N}(u_{i})$ are produced from $\mathcal{C}(u_{i})$ via our edge selection.

\begin{myDef}
  \label{definition:graph_quality}
  {\textbf{Graph quality \cite{graph_survey_vldb2021}.} Given a PG $G=(V,E)$, we define the graph quality $Q_G$ of $G$ as the mean ratio of the number of $u$'s neighbors (i.e., $\mathcal{N}(u)$) in $\mathcal{M}(u)$ over $|\mathcal{M}(u)|$ for all $u\in V$ ({Eq. \ref{eq: graph quality}}), where $\mathcal{M}(u)$ is the $k$ nearest vertices of $u$ in $(V\setminus \{u\})$.}
\end{myDef}

\begin{equation}
\label{eq: graph quality}
{Q_G=\frac{1}{|V|}\sum _{u\in V}\frac{|\mathcal{N}(u)\cap \mathcal{M}(u)|}{k}}
\end{equation}

Given the two constructed NPGs, we can apply our routing strategy on them directly to return query results efficiently, starting from a randomly acquired seed vertex.

\vspace{0.1cm}
\noindent\textbf{Remarks.} In addition to NSW and KGraph, our edge selection and routing also apply to many other PGs, e.g., SPTAG \cite{SPTAG}. So, we easily obtain a new user customized PG by improving the original PG with our edge selection and routing (just like we built NPG\_nsw and NPG\_kgraph). Our experiment shows that our proposed NPGs yield significant performance improvement, comparing to the original PGs.

\section{NPG-Based Hybrid Query Methods}
\label{NPG-NHQ}
We integrate the proposed NPGs (i.e., NPG\_nsw and NPG\_kgraph) into our NHQ framework, thereby yielding two practical hybrid query methods; namely, NHQ-NPG\_nsw and NHQ-NPG\_kgraph. According to the evaluation (\textbf{Sec. \ref{sec: hybrid query}}), our methods far exceed the state-of-the-art competitors (e.g., Vearch, ADBV, and Milvus).

\subsection{NPG-Based Composite Index}\label{sec: Composite index optimization}
{We modify the build process of NPGs by changing its original Euclidean distance measure to our fusion distance ({Eq. \ref{hybrid_distance}}). Then we deploy NPG\_nsw (NPG\_kgraph) in NHQ to serve as a composite index, so generating a hybrid query method called NHQ-NPG\_nsw (NHQ-NPG\_kgraph).}

\vspace{0.1cm}
\noindent\textbf{Remarks.} We emphasize that our NHQ is practically flexible, so it is friendly to existing PGs as well as custom-optimized PGs, by simply replacing their distance measures by our fusion distance.

\subsection{{Joint Pruning} Optimization}\label{sec: Joint pruning optimization}
{Driven by our routing strategy, we optimize the joint pruning of NHQ by performing a two-stage search (stages S1 and S2 mentioned in {Sec. \ref{sec:routing}}) on a composite index.} Specifically, in S1, our search only visits $\lceil k/h \rceil$ neighbors of each vertex in the search path for efficiency; thus we modify line 5 of \hyperref[alg: joint pruning]{Alg. 2} (joint pruning) to randomly visit each vertex's $\lceil {k}/{h} \rceil$ neighbors, and execute the loop in lines 2–8 until it falls into the local optimum (i.e., it reaches the vicinity of the result vertices). In S2, we set $\mathcal{C}=\mathcal{C} \cup \mathcal{R}$, and continue to perform lines 2–8 (without modifying line 5) to visit all neighbors of the vertices in the search path comprehensively. Finally, $\mathcal{R}$ is returned as the query results.

\vspace{0.1cm}
\noindent\textbf{Remarks.} (1) {In our optimized joint pruning, different stages have specific requirements when searching on a {composite index}. In S1, it quickly reaches a small area of objects that are similar to the given query object's feature vector and attributes; while in S2, it accurately obtains top-$k$ objects with similar feature vectors and matched attributes.} (2) Given the NHQ-NPG\_kgraph and NHQ-NPG\_nsw built in {Sec. \ref{sec: Composite index optimization}}, we can apply the optimized joint pruning directly on them to return hybrid query results efficiently.

\section{Experiments}
\label{experiments}
{In this section, we demonstrate our methods' effectiveness and efficiency from the following aspects. First, we deploy several mainstream PGs into our NHQ framework to verify the universality of NHQ in \textbf{Sec. \ref{sec: NHQ validation}}. We then compare the proposed NPGs with six existing PGs in \textbf{Sec. \ref{sec: NPG evaluation}} to demonstrate the superiority our NPGs. In \textbf{Sec. \ref{sec: hybrid query}}, we deploy our NPGs in NHQ to form two NPG-based hybrid query methods and we verify their state-of-the-art performance. In addition, we study the parameter sensitivity of fusion distance weights ($\omega _{\nu}$ and $\omega _{\ell}$ in {Eq. \ref{hybrid_distance}}) in \textbf{Sec. \ref{sec: parameter sensitivity}}.}

\subsection{Experimental Setting}
\label{sec: experimental setting}

\setlength{\textfloatsep}{0cm}
\setlength{\floatsep}{0cm}
\begin{table}[!tb]
  \centering
  \setlength{\abovecaptionskip}{0.05cm}
  \setstretch{0.8}
  \fontsize{6.5pt}{3.3mm}\selectfont
  \caption{Statistics of real-world datasets.}
  \label{tab: Dataset}
  \setlength{\tabcolsep}{.013\linewidth}{
  \begin{tabular}{l|l|l|l|l|l}
    \hline
    \textbf{Dataset} & \textbf{Dimension} & \textbf{\# Base} & \textbf{\# Query} & \textbf{LID}~\cite{DPG,NSSG} & \textbf{Type}\\
    \hline
    \hline
    UQ-V\tablefootnote{\url{http://staff.itee.uq.edu.au/shenht/UQVIDEO/}} & 256 & 1,000,000 & 10,000 & 7.2 & Video + Attributes \\
    \hline
    Msong\tablefootnote{\url{http://www.ifs.tuwien.ac.at/mir/msd/}} & 420 & 992,272 & 200 & 9.5 & Audio + Attributes \\
    \hline
    Audio\tablefootnote{\url{https: //www.cs.princeton.edu/cass/demos.htm}} & 192 & 53,387 & 200 & 5.6 & Audio + Attributes \\
    \hline
    SIFT1M\tablefootnote{\url{http://corpus-texmex.irisa.fr/}\label{texmex}} & 128 & 1,000,000 & 10,000 & 9.3 & Image + Attributes \\
    \hline
    GIST1M\textsuperscript{\ref{texmex}} & 960 & 1,000,000 & 1,000 & 18.9 & Image + Attributes \\
    \hline
    Crawl\tablefootnote{\url{https://commoncrawl.org/}} & 300 & 1,989,995 & 10,000 & 15.7 & Text + Attributes \\
    \hline
    GloVe\tablefootnote{\url{https://nlp.stanford.edu/projects/glove/}} & 100 & 1,183,514 & 10,000 & 20.0 & Text + Attributes \\
    \hline
    Enron\tablefootnote{\url{https://www.cs.cmu.edu/~./enron/}} & 1,369 & 94,987 & 200 & 11.7 & Text + Attributes \\
    \hline
    Paper\tablefootnote{\url{https://github.com/AshenOn3/NHQ}\label{code_dataset}} & 200 & 2,029,997 & 10,000 & - & Text + Attributes \\
    \hline
  \end{tabular}
  }
\end{table}

\noindent{\textbf{Datasets.}} We used eight publicly available real-world datasets and one in-house dataset\textsuperscript{\ref{code_dataset}}; they cover various modes, such as video, image, audio, and text. We summarize their main characteristics in {Tab. \ref{tab: Dataset}}. Among them, the first eight public datasets are composed of high-dimensional feature vectors extracted from different unstructured information, 
which do not originally contain structured attributes; so, we generate attributes for each object in the eight public datasets following the same method in \cite{Milvus_sigmod2021,XuLWX20}. For example, we add attributes such as {\sf date, location, size} to each image on SIFT1M to form an object set having both feature vectors and attributes. \textit{Paper}\textsuperscript{\ref{code_dataset}} is an in-house dataset, each object denotes an individual academic paper that consists of a feature vector extracted from the textual content and structured attributes (e.g., {\sf affiliation}, {\sf venue}, and {\sf topic}). In {Tab. \ref{tab: Dataset}}, LID indicates local intrinsic dimensionality, and a larger LID value implies a ``harder'' dataset \cite{DPG}.

\vspace{0.1cm}
\noindent{\textbf{Compared methods.}} To verify the proposed NPGs' effectiveness with optimized edge selection and routing, we evaluate them alongside six existing state-of-the-art PGs.
\begin{itemize}[leftmargin=*]
  \item \textbf{HNSW} \cite{HNSW} is a hierarchical PG used widely in various fields; it is optimized by hardware or learning \cite{HM_ANN,LiZAH20}.
  \item \textbf{NSW} \cite{NSW} is the precursor of HNSW, and is a single-layer PG constructed by incrementally inserting data.
  \item \textbf{KGraph} \cite{NNDescent} is an approximate KNNG. It only considers the distance factor in edge selection.
  \item \textbf{DPG} \cite{DPG} maximizes the angle between neighbors on the basis of KGraph to alleviate redundant calculation.
  \item \textbf{NSG} \cite{NSG} ensures the routing path's monotonicity by approximating a Monotonic Search Network (MSNET) \cite{NSG}. 
  \item \textbf{NSSG} \cite{NSSG} is similar to DPG; it adjusts the angle between neighbors to adapt to different data characteristics.
  \item \textbf{NPG\_nsw} and \textbf{NPG\_kgraph} are two proposed NPGs in this paper that add our edge selection and routing atop NSW \cite{NSW} and KGraph \cite{NNDescent}, respectively ({Sec. \ref{sec: algorithm optimization}}).
\end{itemize}

We compare our hybrid query methods with six existing ones that have been used in many high-tech companies.

\begin{itemize}[leftmargin=*]
  \item \textbf{ADBV} \cite{ADBV} is a cost-based method proposed by Alibaba. It optimizes PQ \cite{PQ} for vector similarity search. 
  \item \textbf{Milvus} \cite{Milvus,Milvus_sigmod2021} divides the object set through frequently used attributes, and deploys ADBV \cite{ADBV} on each subset.
  \item \textbf{Vearch} \cite{Jingdong_lib,Jingdong_paper} is developed by Jingdong, which implements the hybrid query working off Strategy B.
  \item \textbf{NGT} \cite{NGT} is a vector similarity search library released by Yahoo Japan, which answers a hybrid query to conduct attribute filtering atop the candidates recalled by NGT.
  \item \textbf{Faiss} \cite{Faiss} is a library developed by Facebook, which answer a hybrid query based on IVFPQ (it uses k-means to cluster PQ codes into groups \cite{ADBV}) and Strategy A.
  \item \textbf{SPTAG} \cite{SPTAG} is a PG-based vector similarity search library from Microsoft and answers hybrid queries on Strategy B.
  \item \textbf{NHQ-NPG\_nsw} and \textbf{NHQ-NPG\_kgraph} are our hybrid query methods based on the NHQ integrating two NPGs.
\end{itemize}

\vspace{0.1cm}
\noindent{\textbf{Metrics.}} For index build performance, we record the index build time, peak memory overhead, index size, and graph quality ({Def. \ref{definition:graph_quality}}). We evaluate the search efficiency, accuracy and peak memory overhead to demonstrate search performance. In terms of vector similarity search, the search efficiency can be measured by \textit{queries per second} (\textit{QPS}) and \textit{Speedup}; \textit{QPS} is the ratio of the number of queries ($\# q$) to the search time ($t$), i.e., ${\# q}/{t}$; \textit{Speedup} is defined as ${|\mathcal{S}|}/{NDC}$, where $|\mathcal{S}|$ is the object set's size and is also the total number of distance calculations of the linear scan for a query, and \textit{NDC} is the number of distance calculations of searching on a PG. We use the \textit{Recall} rate to  evaluate the search accuracy, which is measured by {Eq. \ref{recall}}. As for hybrid query, \textit{QPS} is more suitable for evaluating search efficiency, because different methods' calculation cost is distinguishing. Unlike vector similarity search, attribute constraints are added to the \textit{Recall} rate formula in {Eq. \ref{recall}} for hybrid query processing, i.e., the elements in $\mathcal{D}$ and $\mathcal{G}$ have exactly the same attributes as the query object. In addition, \textit{selectivity} is used to evaluate each method's scalability. It is defined as $1-{|\mathcal{P}|}/{|\mathcal{S}|}$ in \cite{ADBV}, where $|\mathcal{P}|$ is the number of objects that match the given attributes, and $|\mathcal{S}|$ is the total number of objects.

\vspace{0.1cm}
\noindent{\textbf{Implementation setup.}} Most methods' codes are publicly accessible online, otherwise we implement the corresponding methods according to their papers. Given that all the compared approaches have parallel versions and SIMD, prefetching instructions' optimizations in their index construction codes, we build all the indexes in parallel with 64 threads and turn on these time-saving optimizations. However, considering that not all algorithms support the parallelization of a single query, we mainly use a single thread to perform search, which is a mainstream setting in related work \cite{NSG,NSSG}. {In addition, we implement a parallel version of our hybrid query method for comparison with two existing methods that parallelize a query (\textbf{Sec. \ref{sec: hybrid query}}).}

All codes are written in C++, and are compiled by g++ 6.5. All experiments are conducted on a Linux server with an Intel(R) Xeon(R) Gold 6248R CPU at 3.00GHz, and a 755G memory. We report the average results of all indicators by performing three repeated trials.

\vspace{0.1cm}
\noindent{\textbf{Parameters.}} Because doing parameters’ adjustment in the entire dataset may cause overfitting \cite{NSG}, we randomly sample a certain percentage of data from the entire dataset to form a validation dataset and search for the optimal values of all the adjustable parameters of each PG on each validation dataset. {In our experiment, all the optimal configurations are given in our open source library\textsuperscript{\ref{code_dataset}}.}

\begin{figure}
   \setlength{\belowcaptionskip}{0cm}
  \centering
  \tiny
  \stackunder[0.5pt]{\includegraphics[scale=0.27]{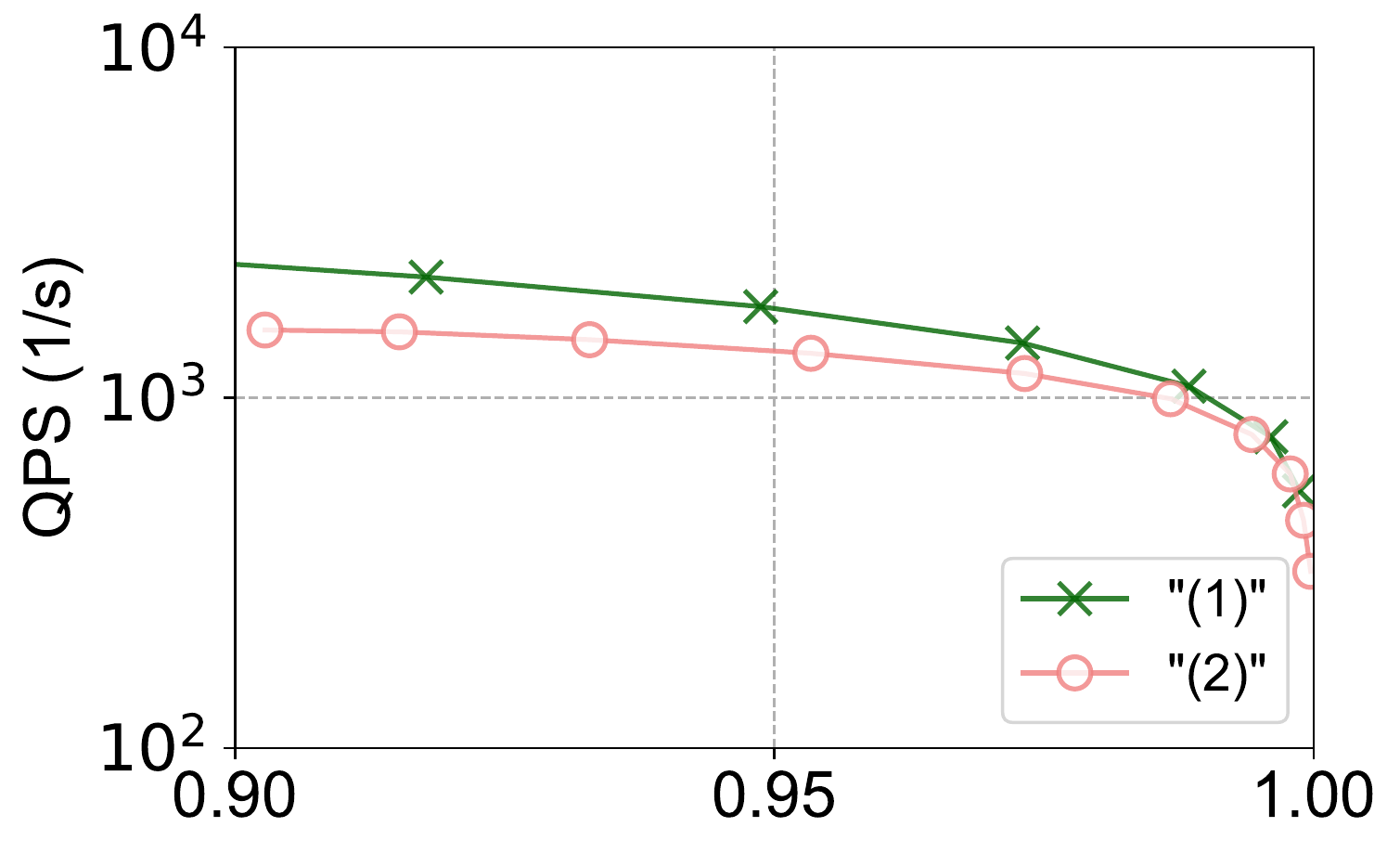}}{(a) Recall@10 (HNSW, SIFT1M)}\hspace{2.5mm}
  \stackunder[0.5pt]{\includegraphics[scale=0.27]{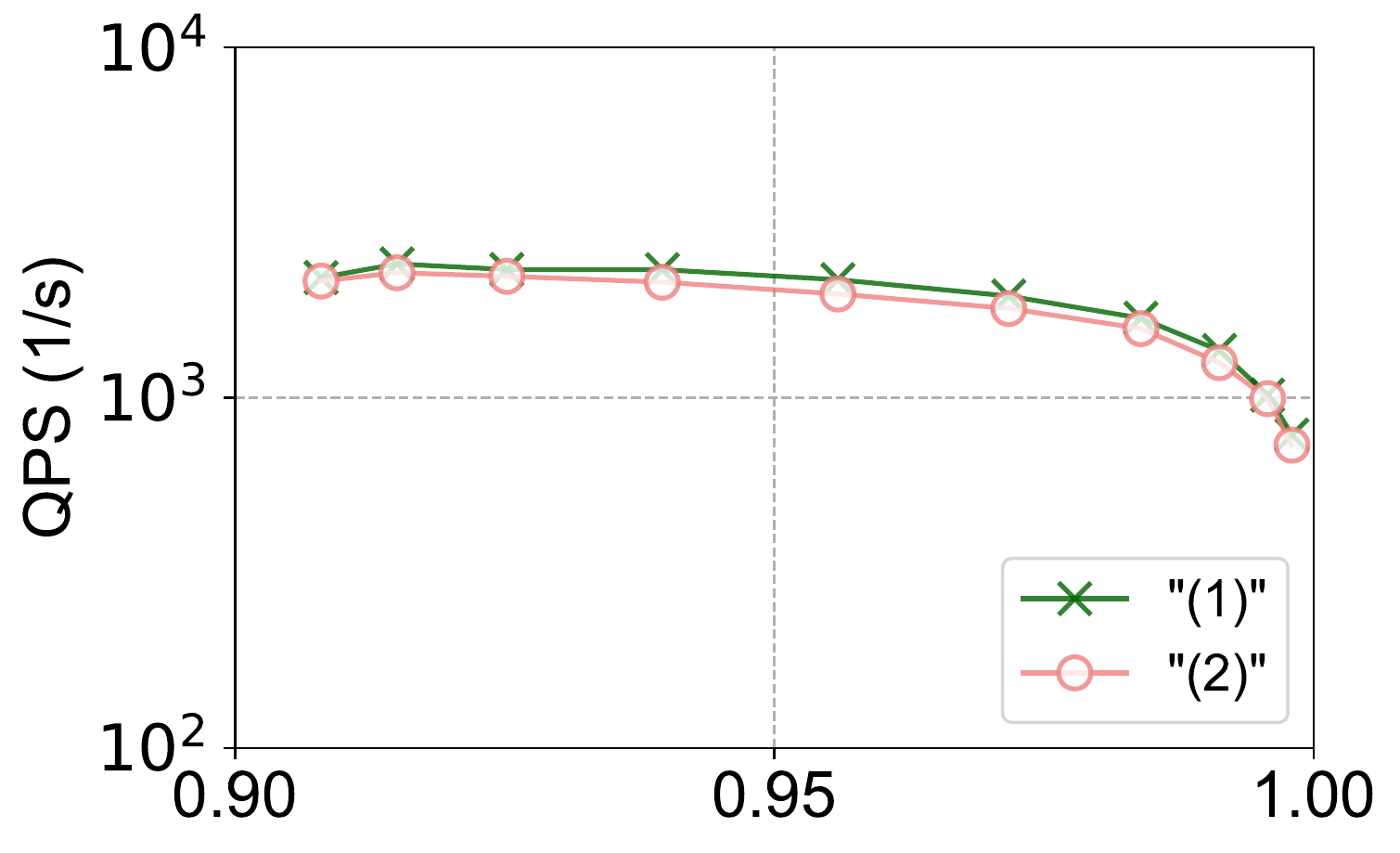}}{(b) Recall@10 (NSG, SIFT1M)}
  \caption{Comparison of two implementations in Strategy B.}
  \label{fig: strategy B}
\end{figure}

\begin{figure}
   \setlength{\belowcaptionskip}{0cm}
  \tiny
  \stackunder[0.5pt]{\includegraphics[scale=0.19]{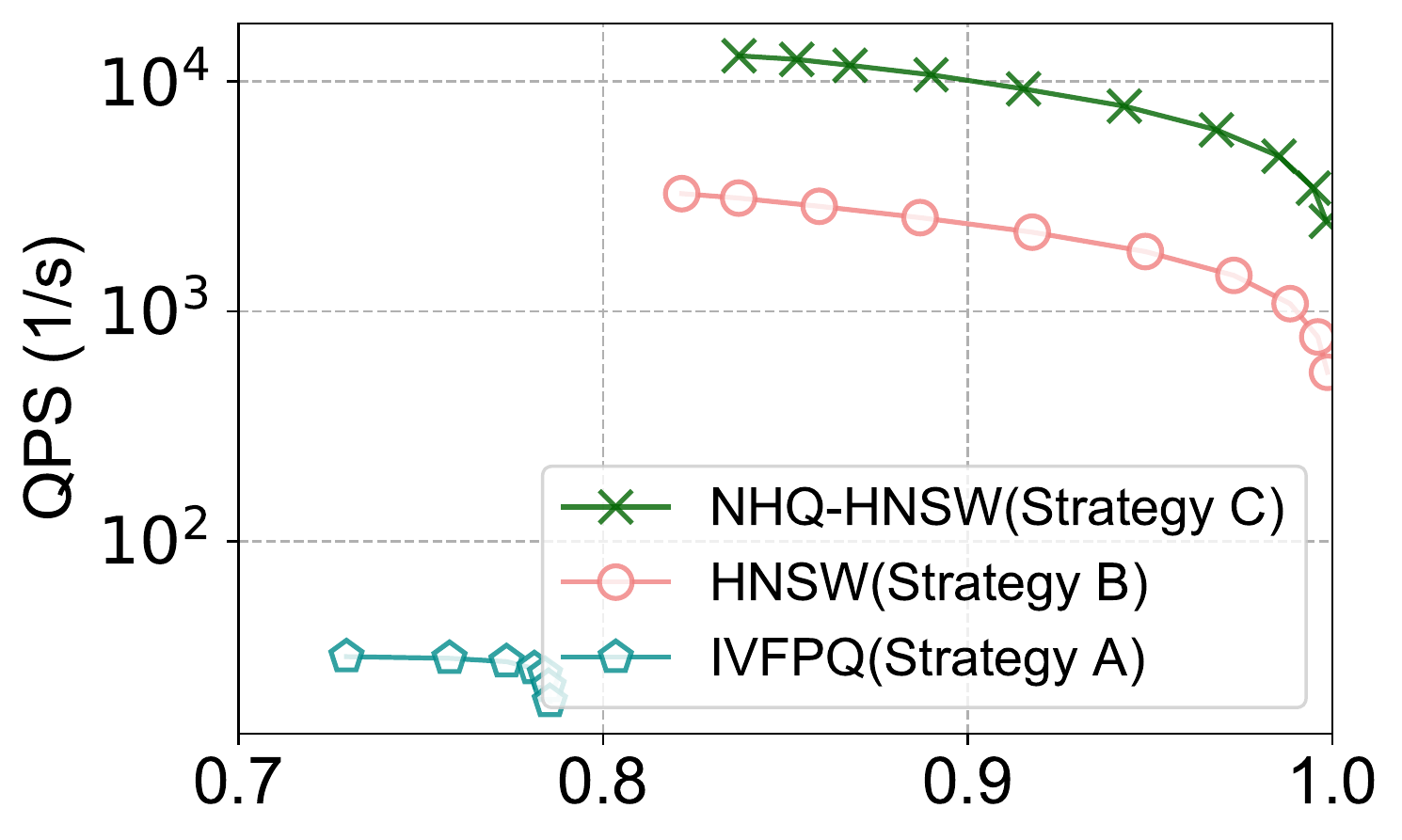}}{(a) Recall@10 (HNSW, SIFT1M)}\hspace{0.8mm}
  \stackunder[0.5pt]{\includegraphics[scale=0.19]{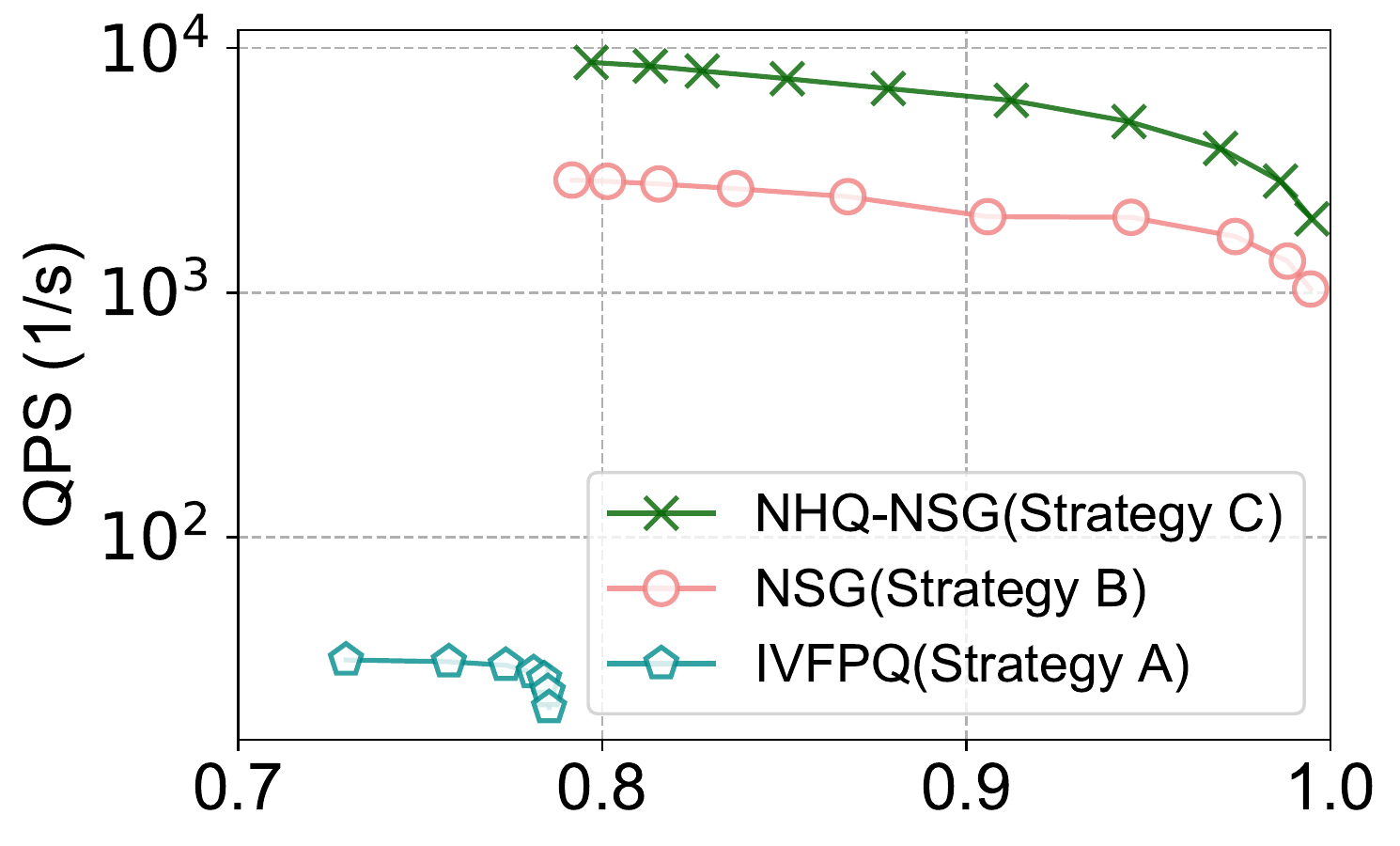}}{(b) Recall@10 (NSG, SIFT1M)}\hspace{0.4mm}
  \stackunder[0.5pt]{\includegraphics[scale=0.19]{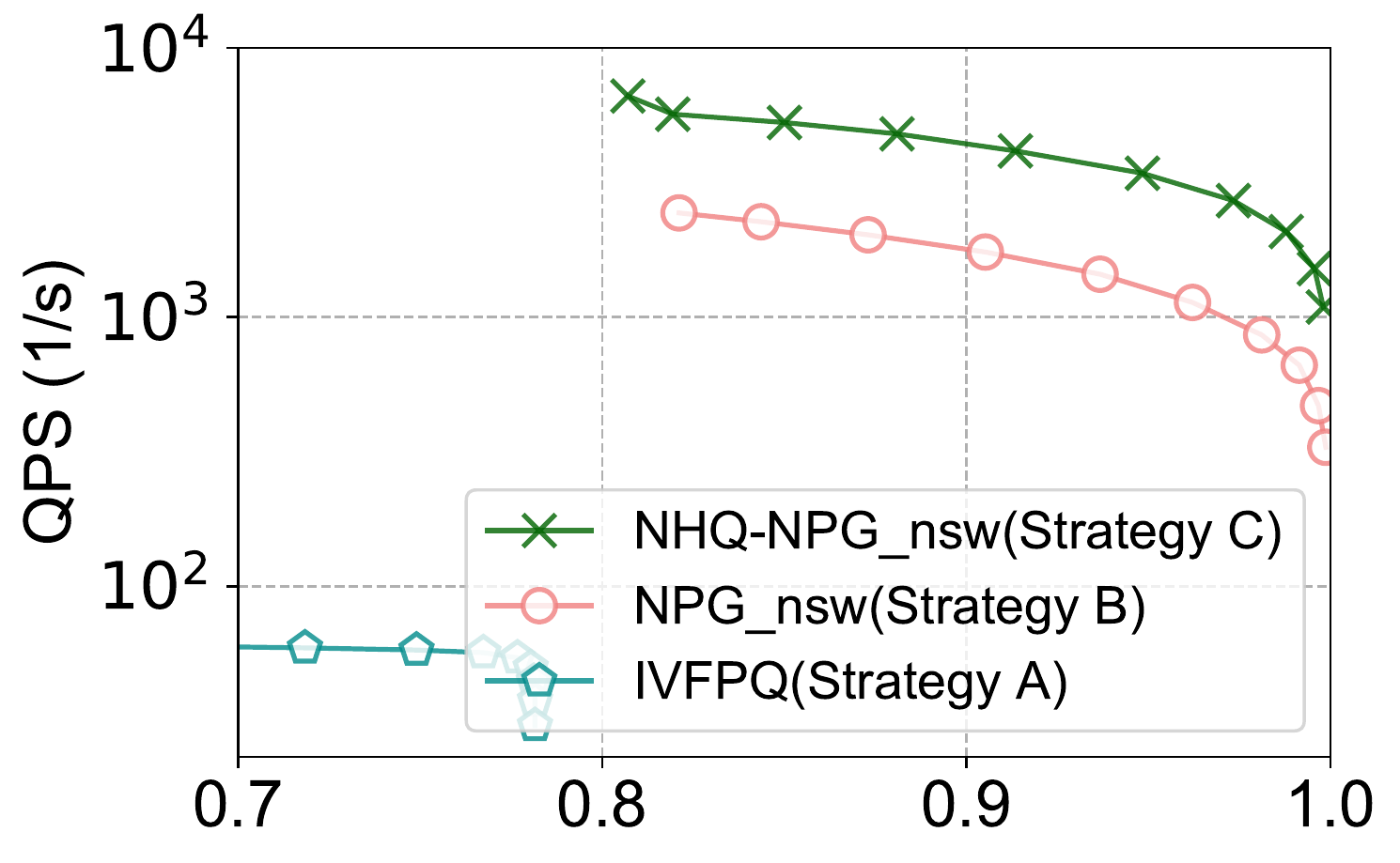}}{(c) Recall@10 (NPG\_nsw, SIFT1M)}
  \newline
  \stackunder[0.5pt]{\includegraphics[scale=0.19]{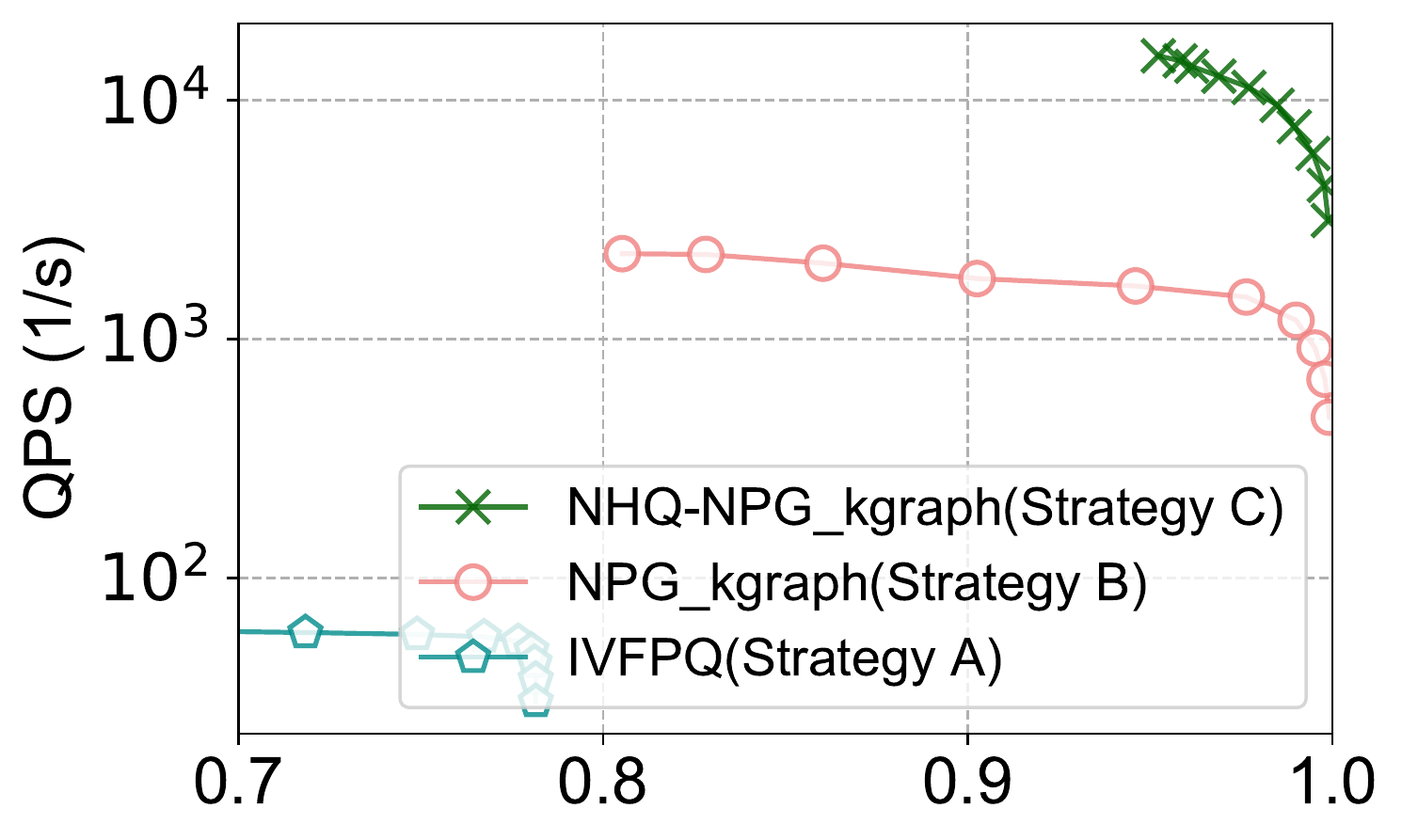}}{(d) Recall@10 (NPG\_kgraph, SIFT1M)}
  \stackunder[0.5pt]{\includegraphics[scale=0.19]{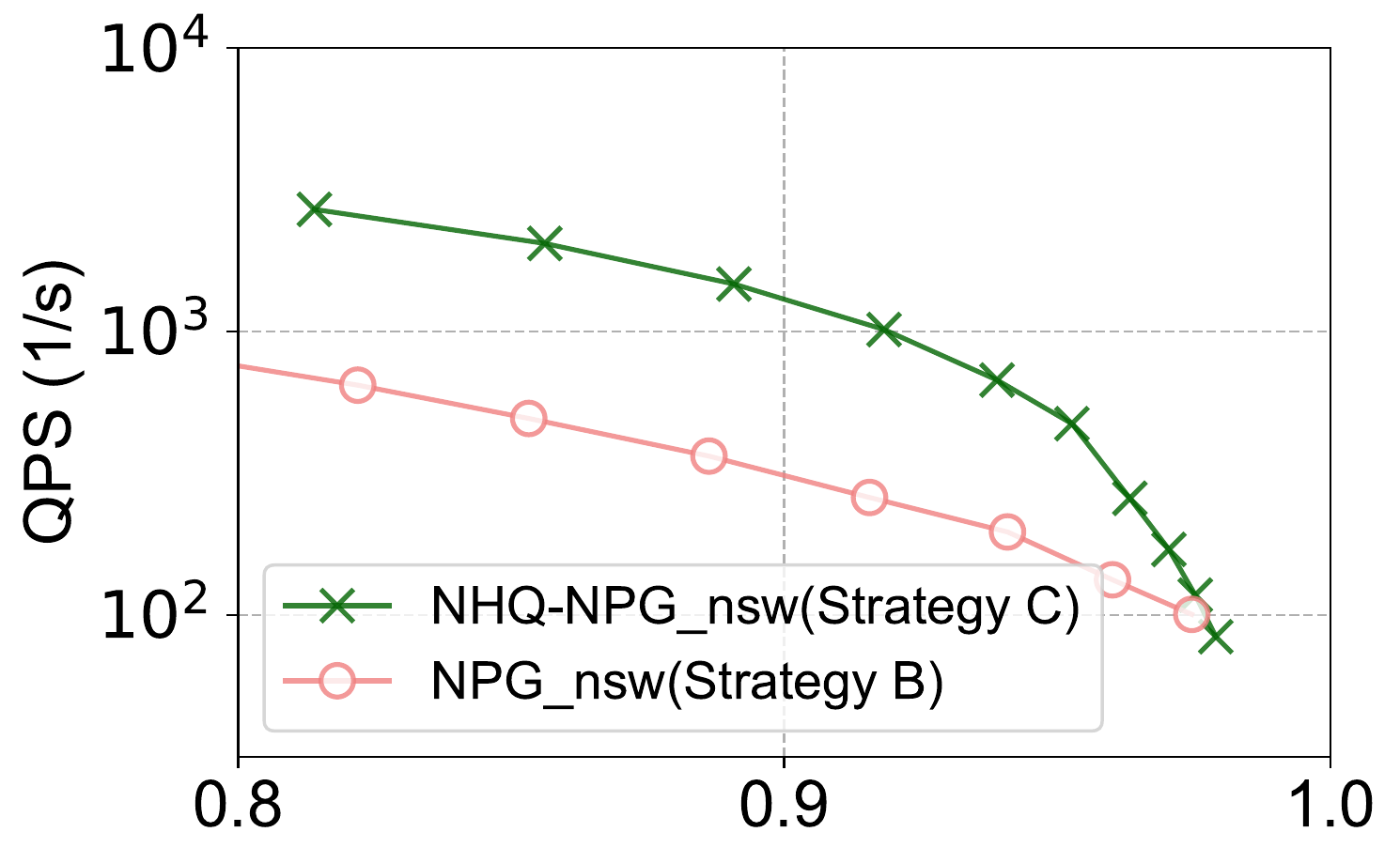}}{(e) Recall@10 (NPG\_nsw, GloVe)}
  \stackunder[0.5pt]{\includegraphics[scale=0.19]{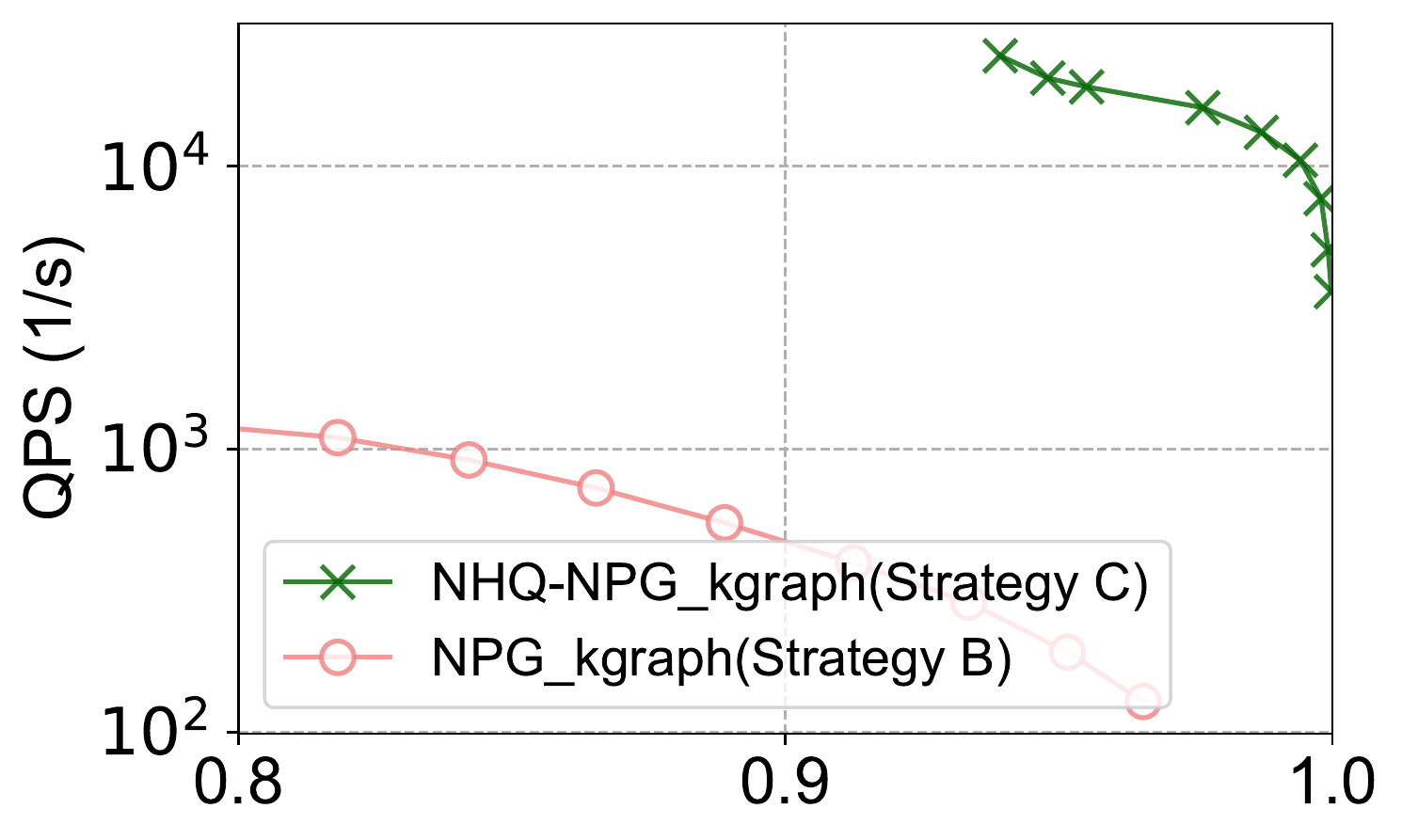}}{(f) Recall@10 (NPG\_kgraph, GloVe)}
  \caption{Hybrid Querying performance comparison of different strategies.}
  \label{fig: hq_validation_nhq}
\end{figure}

\subsection{Validation of NHQ framework.}
\label{sec: NHQ validation}
To verify the capabilities and universality of our NHQ (Strategy C), we implement the hybrid queries working off the ``decomposition-assembly'' model with several PGs (including HNSW, NSG, and our two NPGs) based on Strategy B. We also deploy IVFPQ \cite{PQ} in Strategy A following \cite{ADBV,Milvus_sigmod2021} because Strategy A does not support PG (\textbf{L4} in {Sec. \ref{sec:limitations}}). Meanwhile, we deploy HNSW and NSG on NHQ to form NHQ-HNSW and NHQ-NSG, respectively.

\setlength{\textfloatsep}{0cm}
\setlength{\floatsep}{0cm}
\begin{table*}[th!]
\setlength{\abovecaptionskip}{0cm}
\setlength{\belowcaptionskip}{-0.3cm}
\setstretch{0.8}
\fontsize{6.5pt}{3.3mm}\selectfont
    \centering
    \caption{Index build time, peak memory overhead, index size, and graph quality of different PG-based vector similarity search (the bold values are the best).}
    \label{tab: Index Construction}
    \setlength{\tabcolsep}{0.0085\linewidth}{
    \begin{tabular}{l|l|l|l|l|l|l|l|l|l|l|l|l|l|l|l|l}
    \hline
    \multirow{2}*{\textbf{Algorithm}} & \multicolumn{4}{c|}{\textbf{Build Time (s)}} & \multicolumn{4}{c|}{\textbf{Memory Overhead (GB)}} & \multicolumn{4}{c|}{\textbf{Index Size (MB)}} & \multicolumn{4}{c}{\textbf{Graph Quality}} \\
    \cline{2-17}
    ~ & SIFT1M & GIST1M & GloVe & Crawl & SIFT1M & GIST1M & GloVe & Crawl & SIFT1M & GIST1M & GloVe & Crawl & SIFT1M & GIST1M & GloVe & Crawl \\
    \hline
    \hline
    {HNSW} & 108 & 519 & 150 & 406 & 3.00 & 15.74 & 3.30 & 13.07 & 198 & 149 & 280 & 492 & 0.875 & 0.643 & 0.642 & 0.672 \\
    \hline
    {NSW} & 34 & 294 & \textbf{27} & 224 & 2.88 & 15.73 & 3.57 & 12.85 & 160 & 149 & 371 & 416 & 0.831 & 0.515 & 0.481 & 0.598 \\
    \hline
    {KGraph} & 31 & 228 & 101 & \textbf{71} & 4.61 & 8.41 & 7.34 & \textbf{8.46} & 504 & 465 & 686 & 774 & \textbf{0.998} & \textbf{0.992} & \textbf{0.942} & 0.929  \\
    \hline
    {DPG} & 90 & 519 & 150 & 406 & 5.51 & 9.04 & 6.88 & 12.72 & 640 & 741 & 861 & 1,383 & 0.995 & 0.988 & 0.875 & \textbf{0.971}\\
    \hline
    {NSG} & 195 & 1,906 & 656 & 1,392 & 4.80 & 17.03 & 12.07 & 24.39 & 100 & 58 & 68 & 107 & 0.547 & 0.363 & 0.479 & 0.280 \\
    \hline
    {NSSG} & 329 & 1,163 & 1,618 & 2,982 & 9.53 & 16.63 & 8.80 & 11.71 & 80 & 101 & 74 & 111 & 0.578 & 0.393 & 0.405 & 0.348 \\
    \hline
    {NPG\_nsw} & \textbf{27} & 285 & 105 & 231 & \textbf{2.87} & 15.72 & \textbf{3.29} & 12.83 & 155 & 145 & 275 & 408 & 0.621 & 0.418 & 0.460 & 0.446 \\
    \hline
    {NPG\_kgraph} & 37 & \textbf{172} & 215 & 320 & 4.75 & \textbf{7.16} & 9.16 & 15.58 & \textbf{95} & \textbf{55} & \textbf{59} & \textbf{91} & 0.541 & 0.398 & 0.431 & 0.273 \\
    \hline
    \end{tabular}
    }\vspace{-0.5cm}
\end{table*}

\textbf{Analysis.} {We first evaluate two different implementations of Strategy B (i.e., ``(1)'' and ``(2)'' discussed in {Sec. \ref{Implementation Strategies}}) under the same condition, which implements ``(1)'' and ``(2)'' on HNSW and NSG. As {Fig. \ref{fig: strategy B}} shows, ``(1)'' obtains a better \textit{QPS} vs \textit{Recall} trade-off than ``(2)''. A possible explanation is that current PGs are built with feature vectors rather than attributes, prematurely filtering those objects that mismatch the attributes may impair the vector similarity search's performance. To the best of our knowledge, existing hybrid query methods based on Strategy B deploy ``(2)'' rather than ``(1)''. Therefore, we follow this implementation of Strategy B (i.e., ``(2)'') when comparing to other strategies.} As {
Fig. \ref{fig: hq_validation_nhq}} shows, NHQ (Strategy C) beats others on different PGs, and maintains stable \textit{QPS} on datasets with various LIDs (such as (d) and (f)); however, the hybrid queries based on Strategy B drops significantly on datasets with a higher LID. Because of IVFPQ's limitations, the hybrid queries based on Strategy A are difficult to reach high accuracy.

\begin{figure}
  \setlength{\belowcaptionskip}{0cm}
  \centering
  \tiny
  \stackunder[0.5pt]{\includegraphics[scale=0.19]{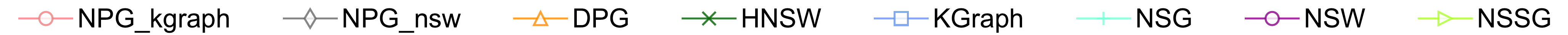}}{}
  \stackunder[0.5pt]{\includegraphics[scale=0.19]{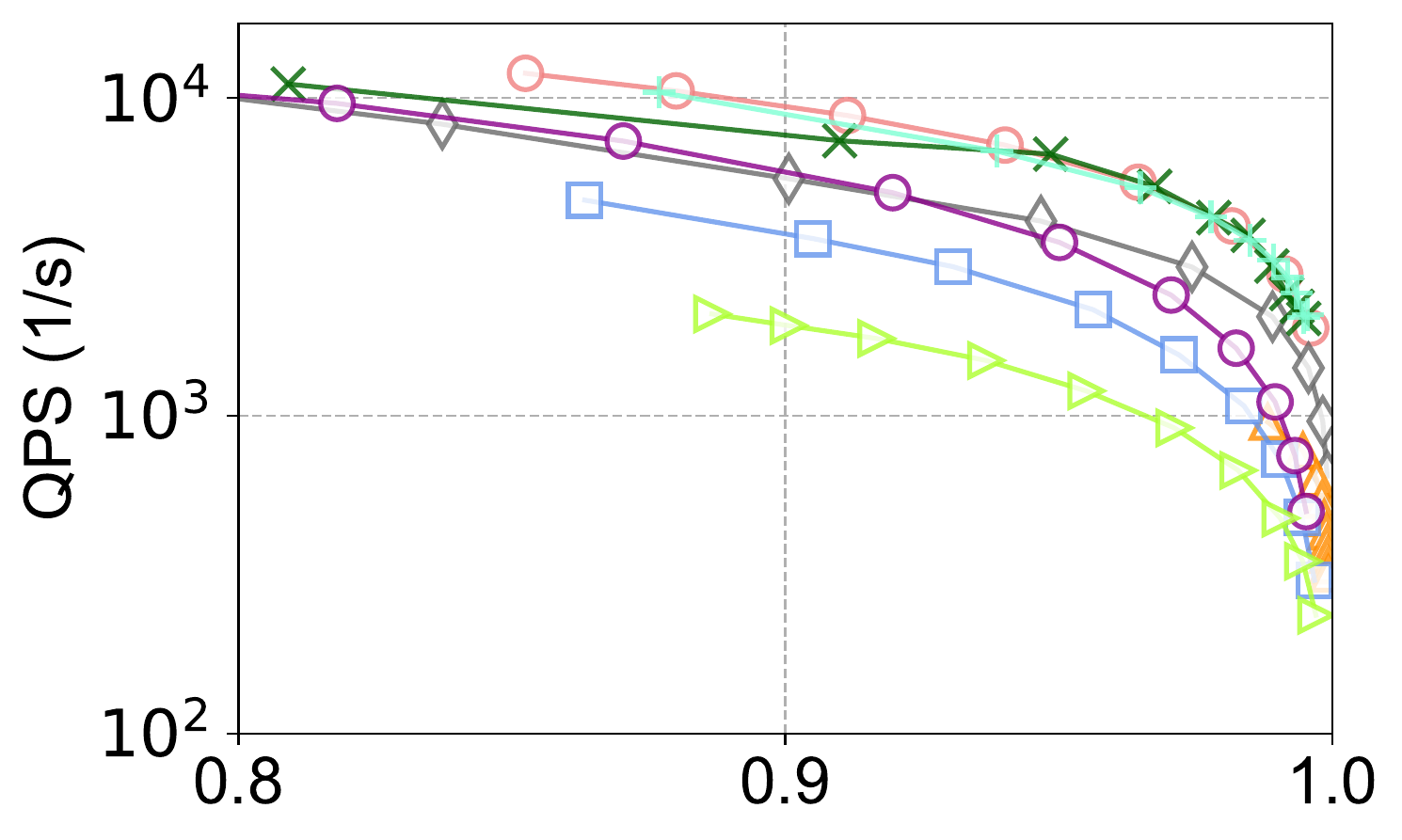}}{(a) Recall@10 (SIFT1M)}
  \stackunder[0.5pt]{\includegraphics[scale=0.19]{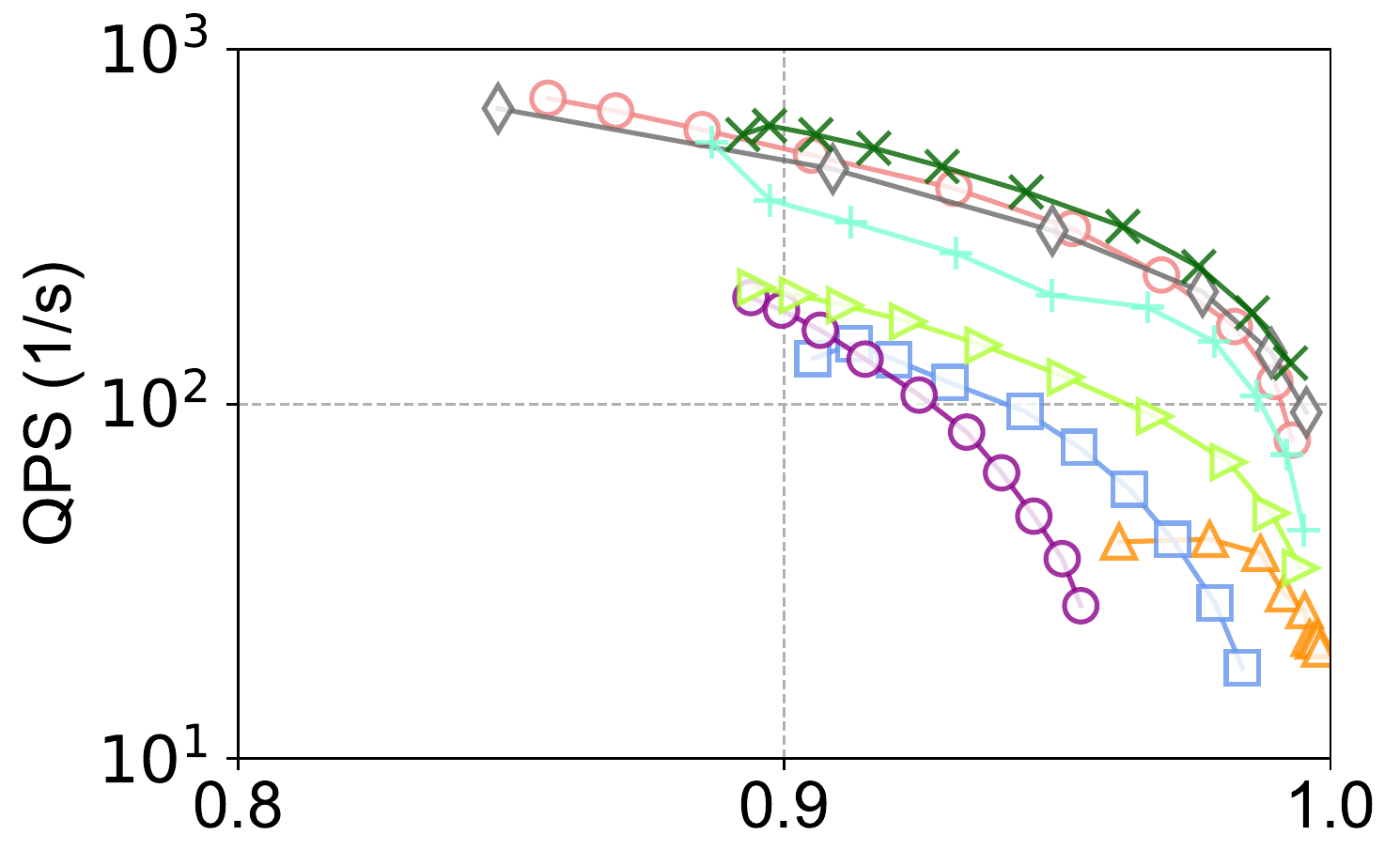}}{(b) Recall@10 (GIST1M)}
  \stackunder[0.5pt]{\includegraphics[scale=0.19]{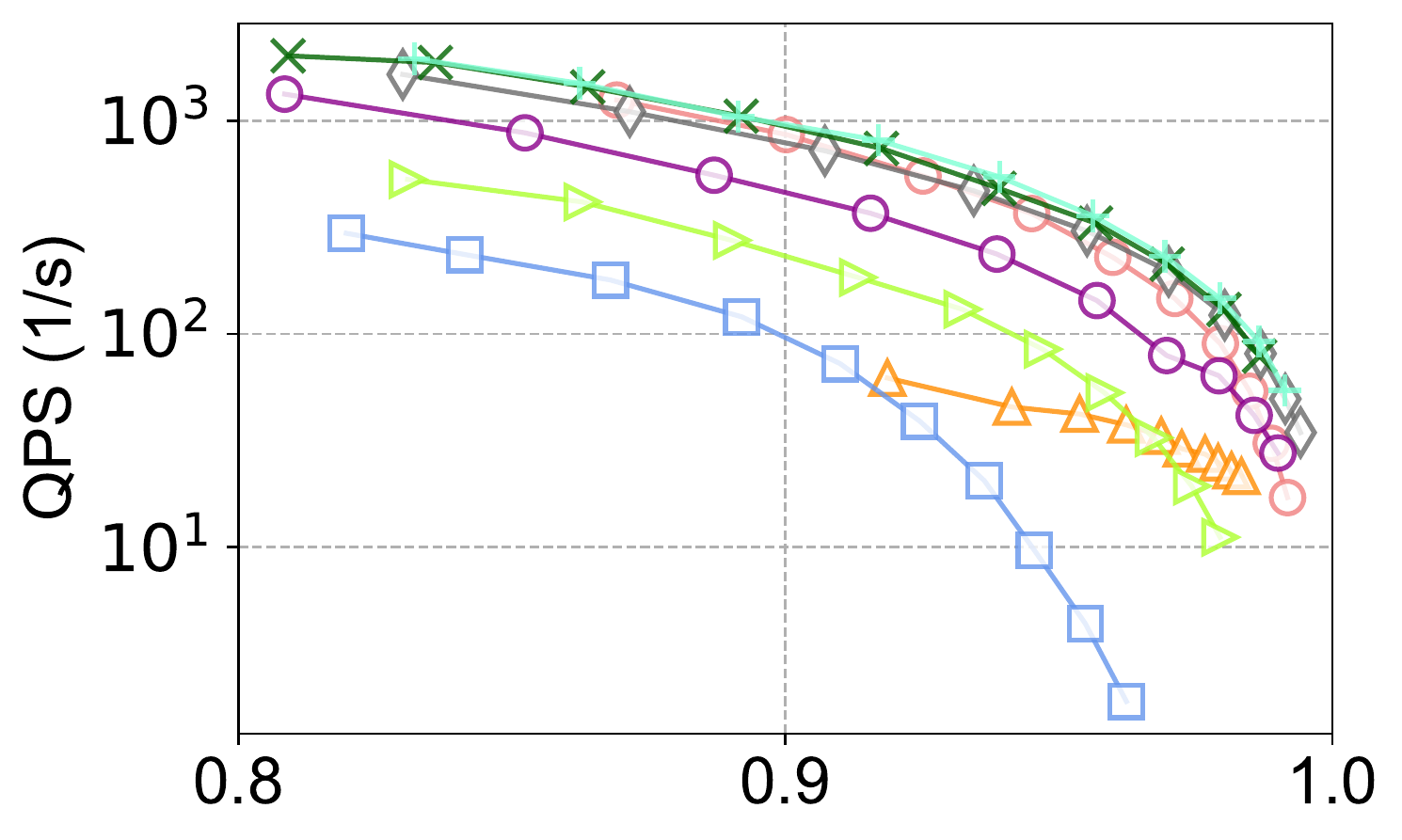}}{(c) Recall@10 (GloVe)}
  \newline
  \stackunder[0.5pt]{\includegraphics[scale=0.19]{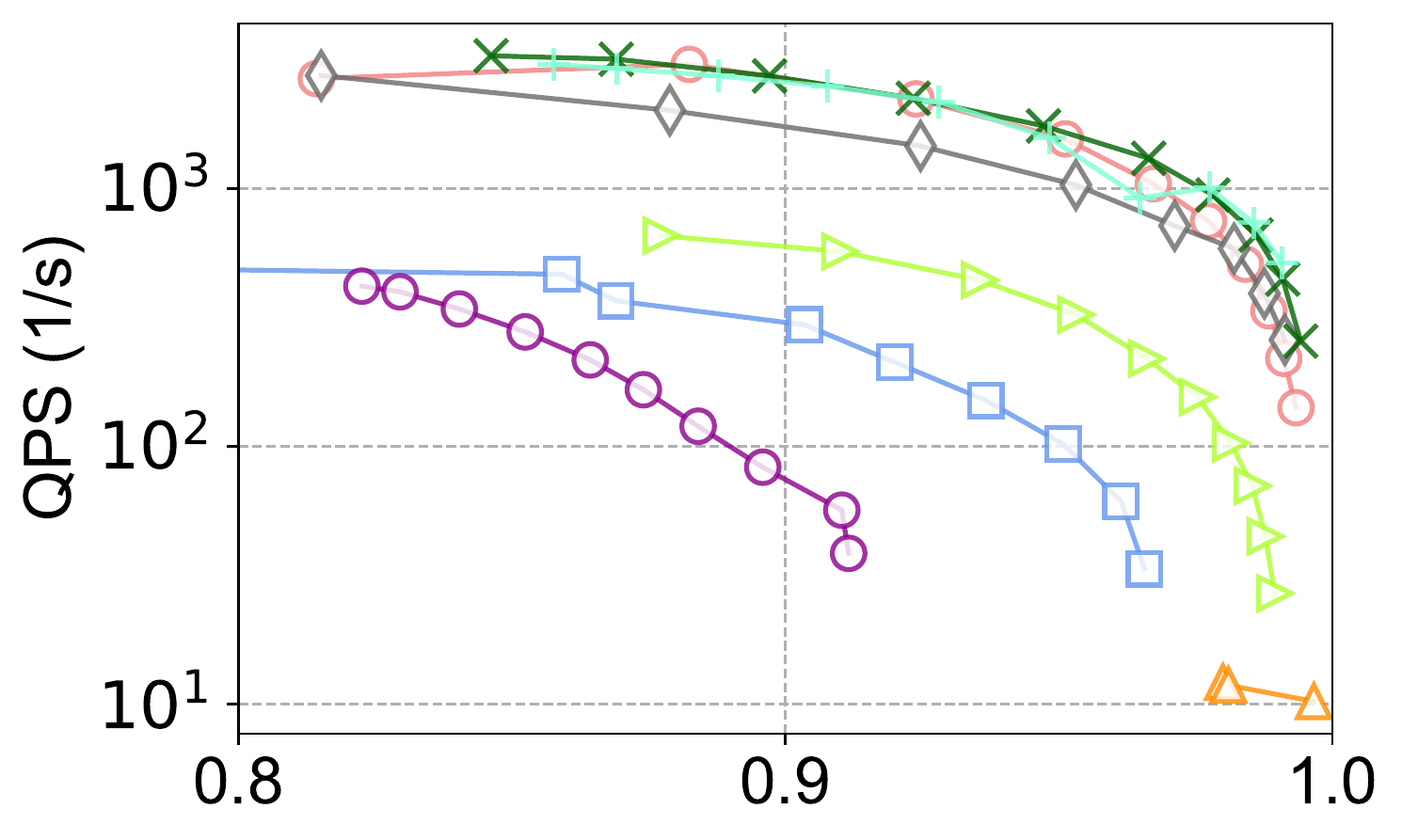}}{(d) Recall@10 (Crawl)}
  \stackunder[0.5pt]{\includegraphics[scale=0.19]{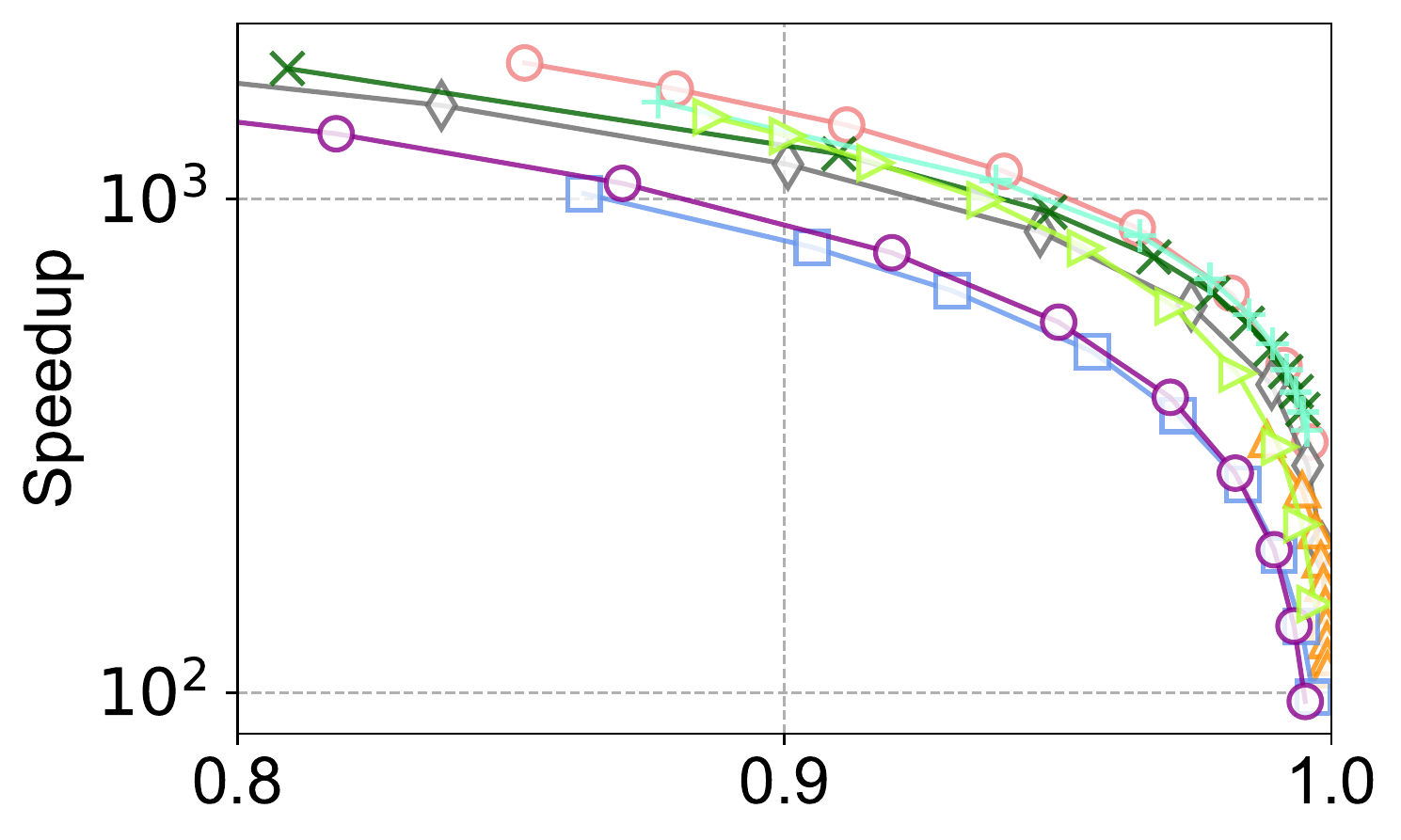}}{(e) Recall@10 (SIFT1M)}
  \stackunder[0.5pt]{\includegraphics[scale=0.19]{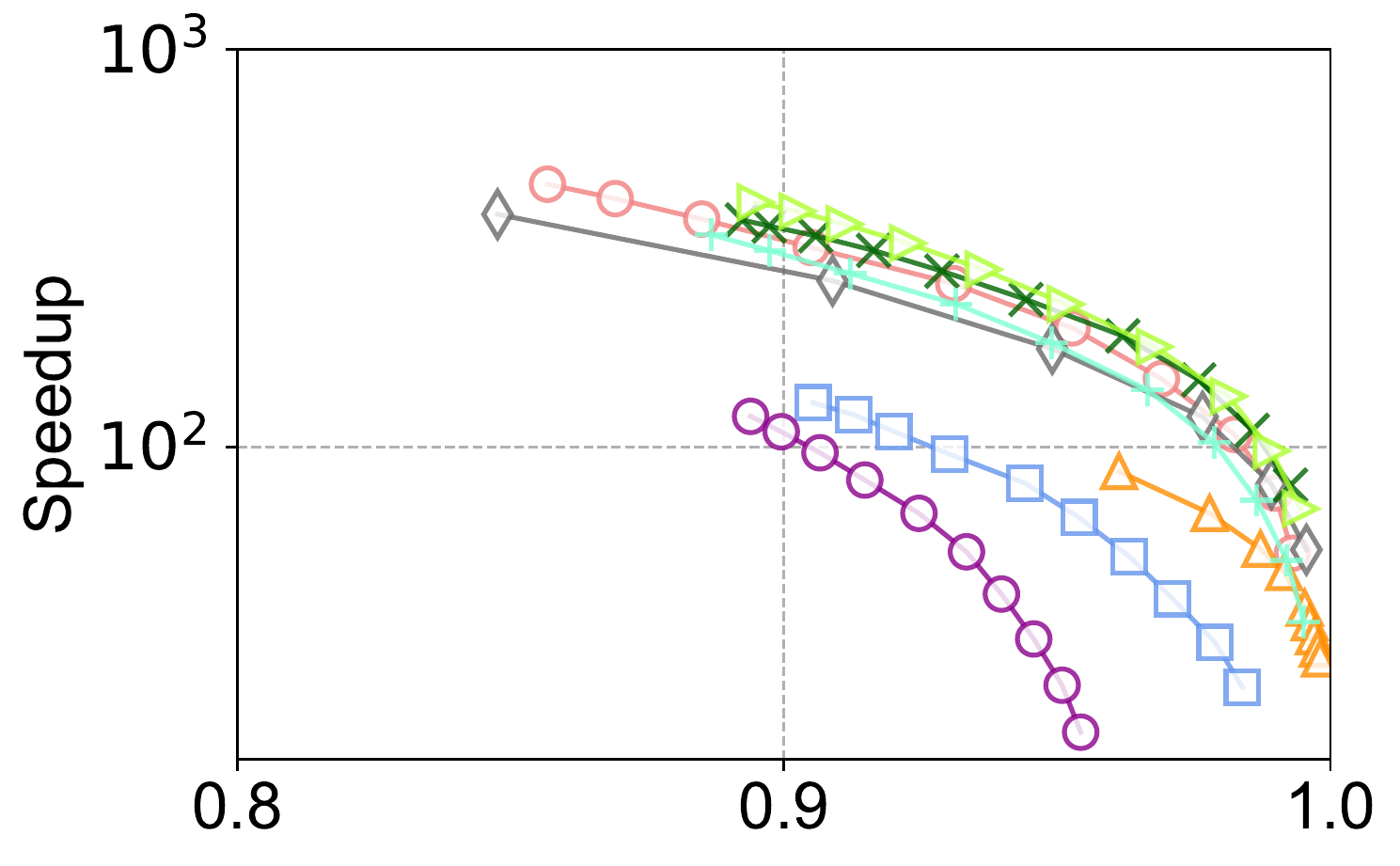}}{(f) Recall@10 (GIST1M)}
  \newline
  \stackunder[0.5pt]{\includegraphics[scale=0.19]{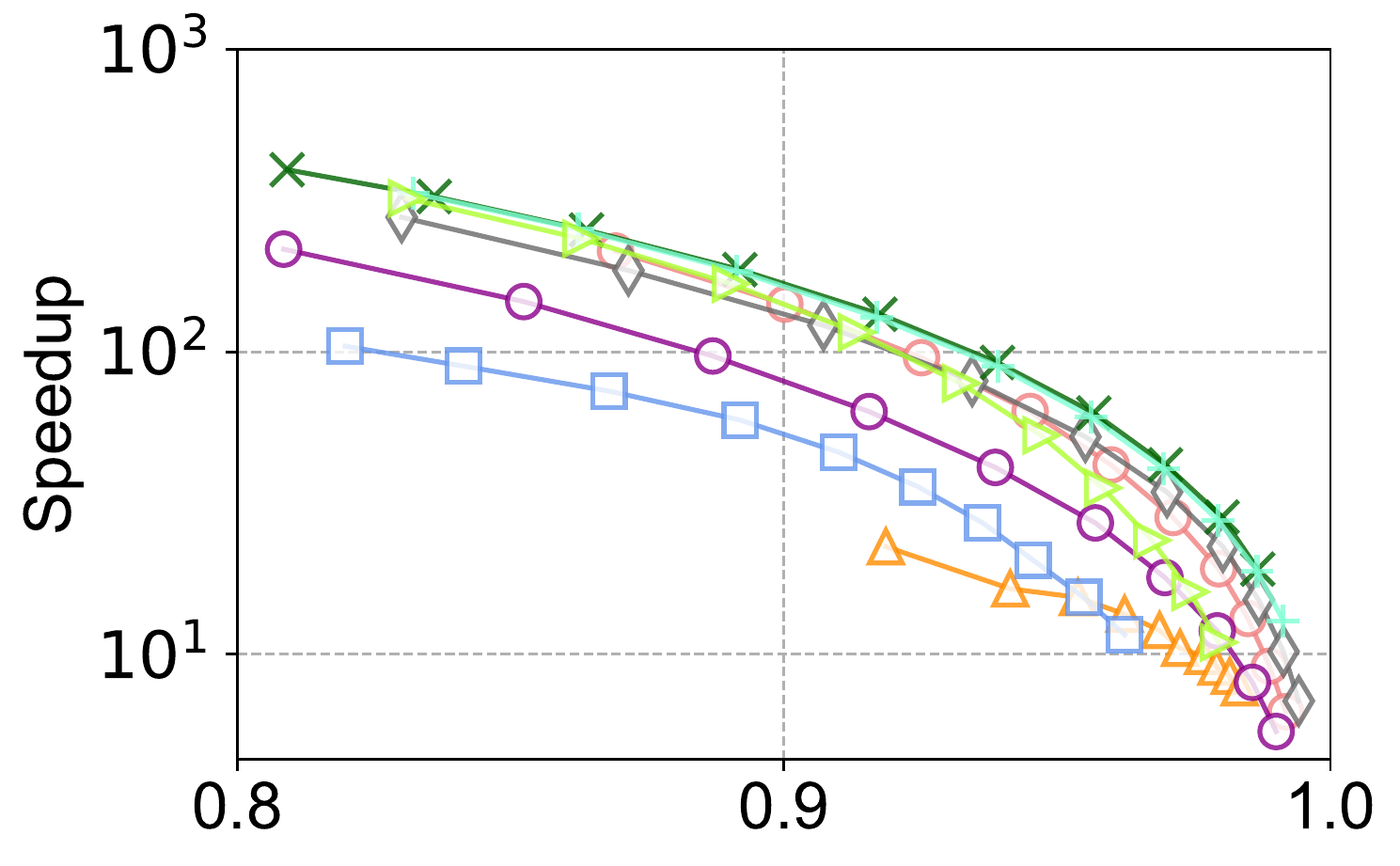}}{(g) Recall@10 (GloVe)}
  \stackunder[0.5pt]{\includegraphics[scale=0.19]{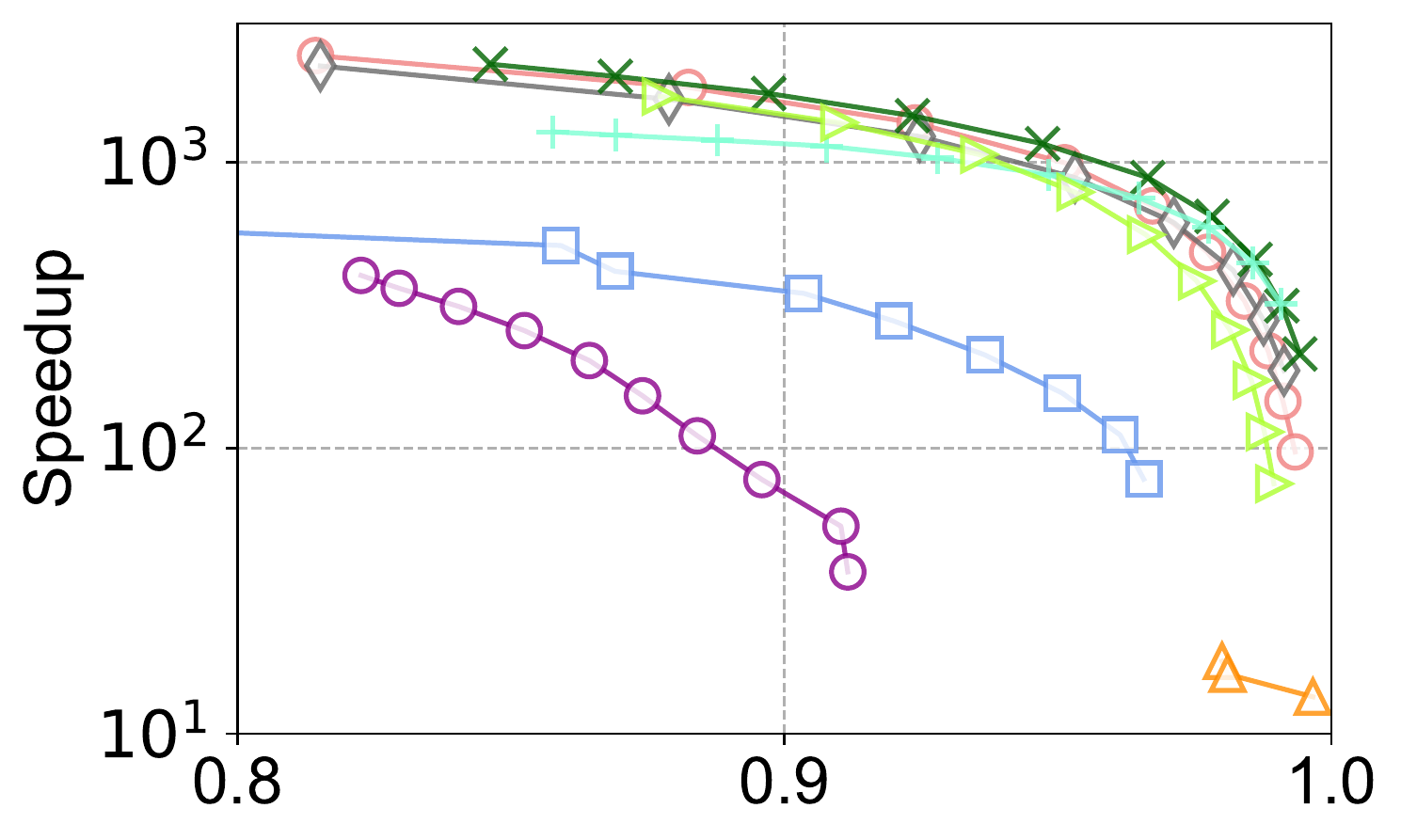}}{(h) Recall@10 (Crawl)}
  \stackunder[0.5pt]{\includegraphics[scale=0.19]{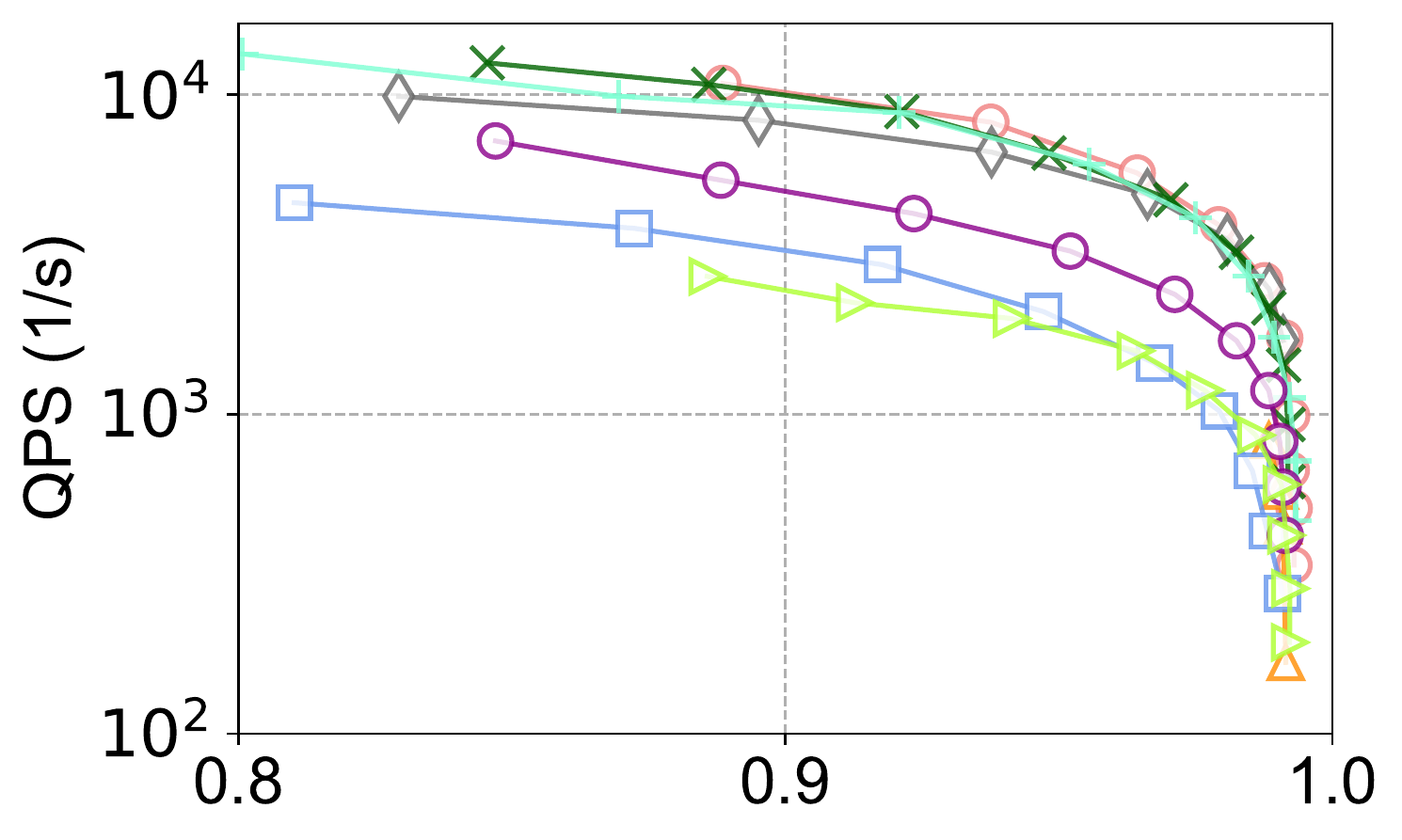}}{(i) Recall@1 (SIFT1M)}
  \caption{\textit{QPS} and \textit{Speedup} vs \textit{Recall}. Here, (a)–(d) are the \textit{QPS} vs \textit{Recall@10} of different PG-based algorithms; (e)–(h) are the \textit{Speedup} vs \textit{Recall@10}; (i) is the \textit{QPS} vs \textit{Recall@1} (the top right is better).}
  \label{fig: real_search_speedup}
\end{figure}

\subsection{Performance of NPG}
\label{sec: NPG evaluation}

\noindent{\textbf{Index building.}} As {Tab. \ref{tab: Index Construction}} shows, different PGs have different index build time, memory overhead, index size on different datasets; while graph quality mainly is determined by PG. Specifically, NSW, KGraph, and our two NPGs have higher index building efficiency; and KGraph and NPGs show lower memory overhead; the smallest index falls on NPG\_kgraph; and KGraph and DPG achieve higher graph quality. In general, the higher the dataset's LID, the longer it takes to build a PG on it, and the larger the memory overhead during index building. Overall, compared to state-of-the-art PGs (such as HNSW and NSG), the index building of our NPGs is more efficient, and takes up less memory. Additionally, NPGs significantly reduce the index size compared to their precursors (i.e., NSW and KGraph).

\textbf{Analysis.} The characteristics of the dataset will affect build performance. For example, the build time and memory overhead on GIST1M are higher than that on SIFT1M, this phenomenon is universal. NSW and KGraph only consider D1 ({Sec. \ref{sec: edge selection}}) when selecting edges, so their build time is smaller. An additional consideration of D2 is that it will demand extra time (e.g., NSSG). NPGs still have a low build time although they also take into account D2, which demonstrates that our edge selection is more efficient. In addition, our NPGs have a smaller index than the original PGs (e.g., KGraph) because our edge selection cuts off redundant neighbors in the same direction. Meanwhile, the diversity of neighbors makes NPGs link to not only the nearest neighbors, but also some other directions' ones farther away, which will improve query performance.

\noindent{\textbf{Search performance.}} As {Tab. \ref{tab: search_memory_overhead}} and {Fig. \ref{fig: real_search_speedup}} show, NPGs obtain state-of-the-art accuracy vs efficiency trade-off with lower memory overhead on most datasets. As the dataset's LID increases, the search efficiency of each PG decreases; for example, from SIFT1M to GIST1M ((a) and (b) in {Fig. \ref{fig: real_search_speedup}}), \textit{QPS} decreases by about one order of magnitude. In general, the fewer the number of distance calculations (which means a bigger \textit{Speedup}), the larger the \textit{QPS}; whereas the differences in some PGs' routing implementation lead to noncompliance with this phenomenon (e.g., NSSG). On the same dataset, for different \textit{Recall} requirements, such as \textit{Recall@1} ((i)) and \textit{Recall@10} ((a)), each PG's search performance ranking is almost the same, especially NPG\_kgraph, which shows robust scalability. Compared with HNSW and NSG, our NPGs occupy less memory during the search. Considering the graph quality in {Tab. \ref{tab: Index Construction}}, we can see that the optimal search performance does not require the highest graph quality, but a proper value ($0.2\sim0.7$ in different datasets) as explained in {Example \ref{example:distance_factor}}.

\setlength{\textfloatsep}{0cm}
\setlength{\floatsep}{0cm}
\begin{table}[t!]
\setlength{\abovecaptionskip}{0cm}
\setstretch{0.8}
\fontsize{6.5pt}{3.3mm}\selectfont
    \centering
    \caption{Peak memory overhead during search on PGs.}
    \label{tab: search_memory_overhead}
    \setlength{\tabcolsep}{0.0435\linewidth}{
    \begin{tabular}{l|l|l|l|l}
    \hline
    \multirow{2}*{\textbf{Algorithm}} & \multicolumn{4}{c}{\textbf{Memory Overhead (MB)}} \\
    \cline{2-5}
    ~ & SIFT1M & GIST1M & GloVe & Crawl \\
    \hline
    \hline
    {HNSW} & 1,733 & 11,526 & 1,838 & 9,125  \\
    \hline
    {NSG} & 1,609 & \textbf{7,564} & 1,958 & 14,799 \\
    \hline
    {NPG\_nsw} & 1,691 & 11,522 & 1,834 & 9,041 \\
    \hline
    {NPG\_kgraph} & \textbf{1,285} & 7,585 & \textbf{1,477} & \textbf{5,506} \\
    \hline
    \end{tabular}
    }
\end{table}


\setlength{\textfloatsep}{0cm}
\setlength{\floatsep}{0cm}
\begin{table*}[th!]
\setlength{\abovecaptionskip}{0cm}
\setlength{\belowcaptionskip}{-0.3cm}
\setstretch{0.8}
\fontsize{6.5pt}{3.3mm}\selectfont
    \centering
    \caption{Index build time and peak memory overhead of different hybrid query methods on eight real-world datasets (the bold items are the best).}
    \label{tab: Index Construction hybrid query}
    \setlength{\tabcolsep}{0.00835\linewidth}{
    \begin{tabular}{l|l|l|l|l|l|l|l|l|l|l|l|l|l|l|l|l}
    \hline
    \multirow{2}*{\textbf{Algorithm}} & \multicolumn{8}{c|}{\textbf{Build Time (s)}} & \multicolumn{8}{c}{\textbf{Memory Overhead (MB)}} \\
    \cline{2-17}
    ~ & SIFT1M & GIST1M & GloVe & Crawl & UQ-V & Msong & Audio & Enron & SIFT1M & GIST1M & GloVe & Crawl & UQ-V & Msong & Audio & Enron \\
    \hline
    \hline
    {ADBV} & 1,614 & 1,812 & 1,838 & 2,895 & 1,598 & 1,627 & 478 & 611 & 857 & 5,117 & 1,287 & 5,042 & 1,356 & 3,538 & 220 & 1,867 \\
    \hline
    {Milvus} & 2,131 & 2,526 & 2,645 & 3,389 & 2,116 & 2,113 & 607 & 933 & \textbf{621} & \textbf{2,795} & \textbf{761} & \textbf{2,112} & \textbf{954} & \textbf{1,639} & 215 & 1,534 \\
    \hline
    {Vearch} & 108 & 519 & 150 & 406 & \textbf{21} & 47 & \textbf{1} & 43 & 3,052 & 16,104 & 3,365 & 13,348 & 4,890 & 8,133 & 235 & 2,378  \\
    \hline
    {NGT} & 27 & 1,124 & 806 & 2,745 & 31 & 73 & \textbf{1} & 20 & 1,427 & 7,870 & 1,513 & 5,598 & 2,417 & 3,701 & 111 & 1,072\\
    \hline
    {Faiss} & 1,591 & 1,721 & 1,838 & 2,791 & 1,598 & 1,627 & 478 & 611 & 790 & 5,117 & 1,287 & 5,042 & 1,356 & 3,538 & 220 & 1,867 \\
    \hline
    {SPTAG} & 456 & 2,690 & 609 & 1,555 & 614 & 1,574 & 27 & 276 & 2,821 & 12,503 & 2,491 & 8,611 & 4,293 & 9,031 & 376 & 2,082 \\
    \hline
    {NHQ-NPG\_nsw} & \textbf{17} & 121 & \textbf{83} & 406 & 40 & 207 & 2 & 23 & 2,575 & 11,920 & 2,441 & 9,310 & 3,944 & 5,915 & \textbf{207} & 1,639 \\
    \hline
    {NHQ-NPG\_kgraph} & 25 & \textbf{70} & 189 & \textbf{188} & 25 & \textbf{61} & \textbf{1} & \textbf{3} & 3,847 & 7,328 & 5,057 & 8,969 & 4,337 & 9,030 & 260 & \textbf{998} \\
    \hline
    \end{tabular}
    }\vspace{-0.4cm}
\end{table*}

\textbf{Analysis.} It is difficult to organize the neighborhood relationship of a dataset with high LID via a PG (graph quality is low), which degrades the search performance (the ``curse of dimensionality'' \cite{DPG}). For searching on a PG, a large proportion of the search time is used for calculating the distance between vectors \cite{SONG}, so there exists a positive proportional relationship for \textit{Speedup} and \textit{QPS}. HNSW divides neighbors with different distances into different subgraphs via a hierarchical structure to accelerate the search, which increases memory overhead. NPGs diversify the neighbors' distribution on a single-layer graph, and achieves better search performance on the basis of inheriting the original PG's (i.e., NSW and KGraph) low memory overhead.

\subsection{Evaluation of Hybrid Query Methods}
\label{sec: hybrid query}

\begin{figure}
  \setlength{\belowcaptionskip}{0cm}
  \centering
  \tiny
  \stackunder[0.5pt]{\includegraphics[scale=0.19]{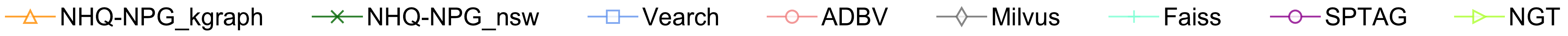}}{}
  \stackunder[0.5pt]{\includegraphics[scale=0.19]{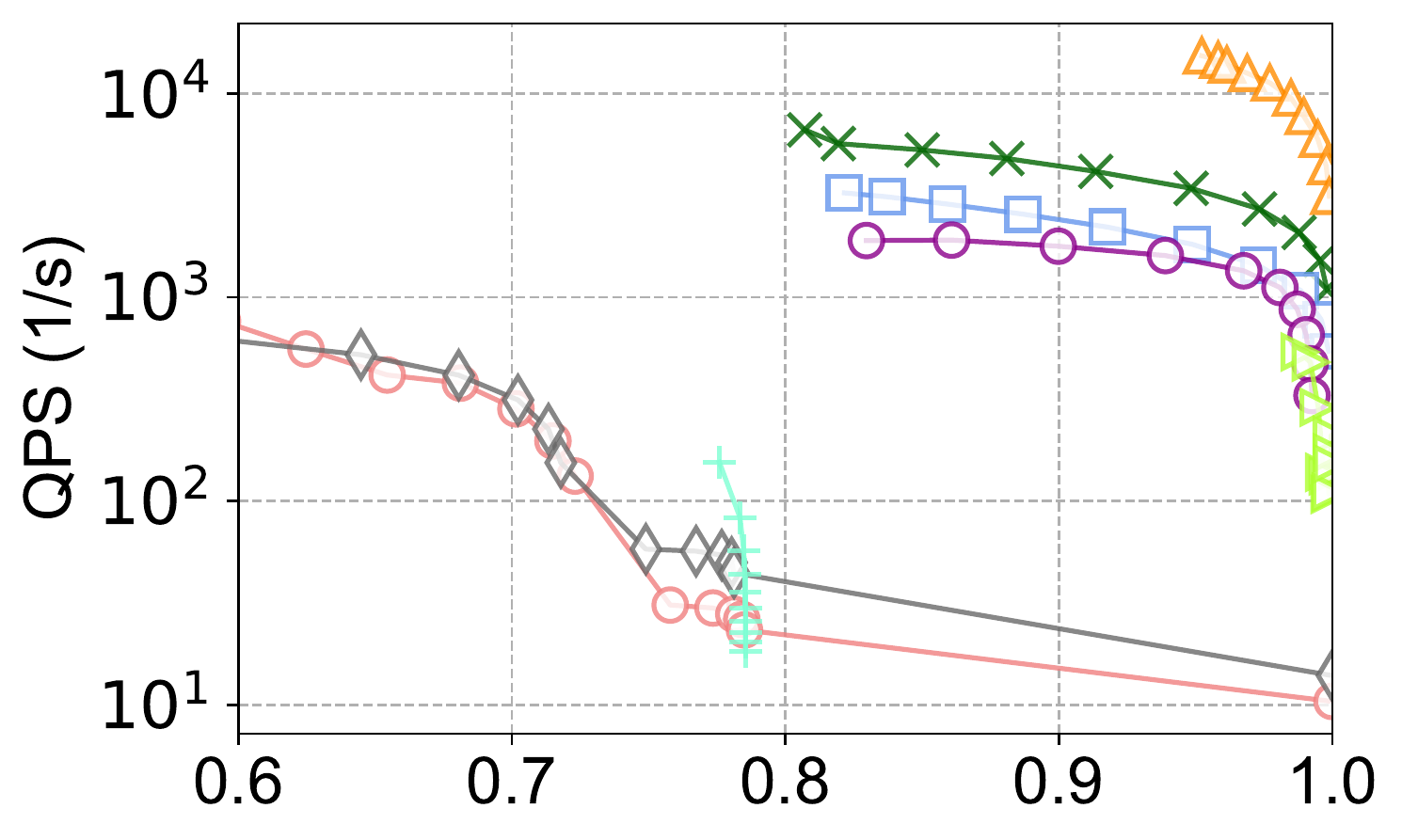}}{(a) Recall@10 (SIFT1M)}
  \stackunder[0.5pt]{\includegraphics[scale=0.19]{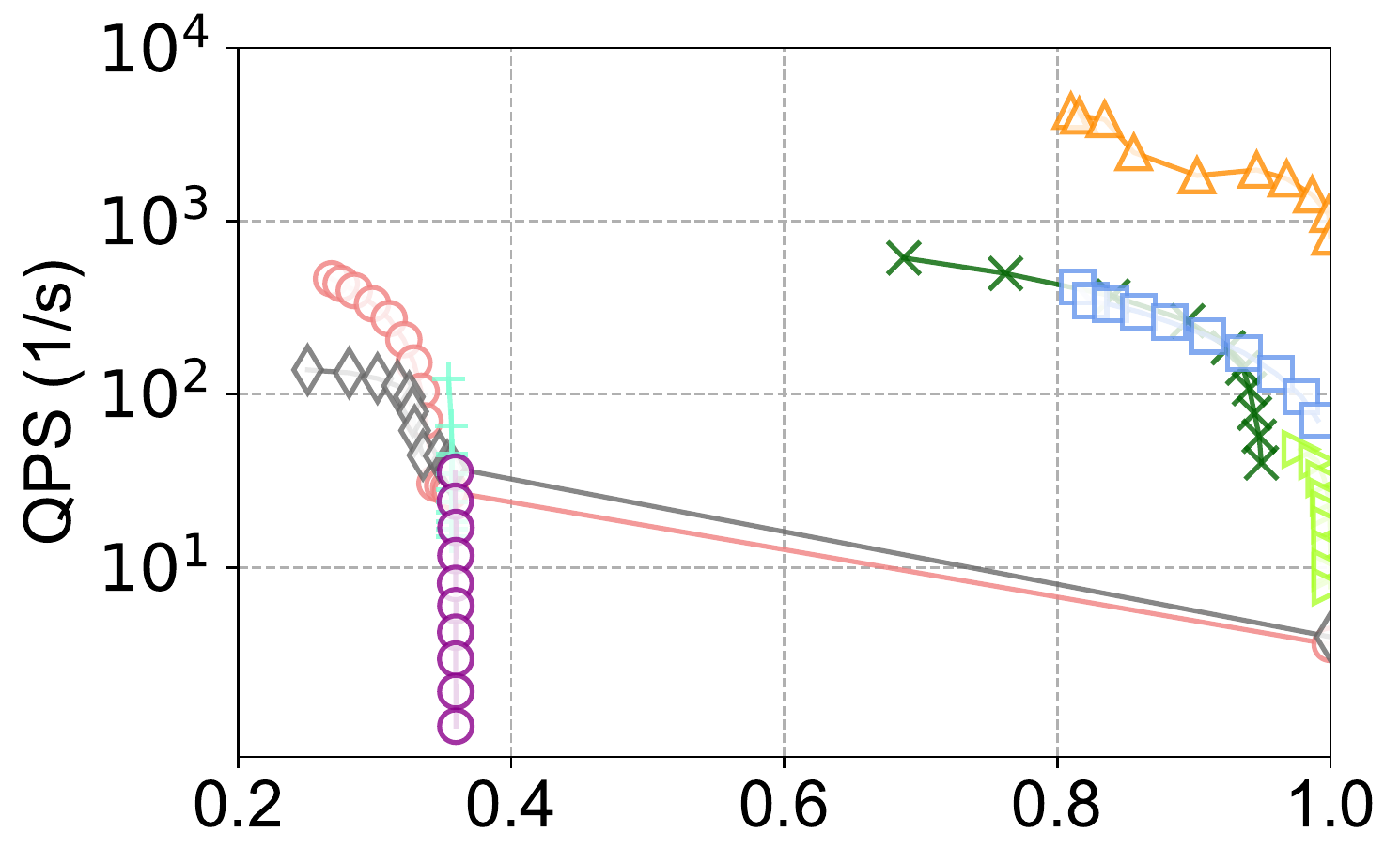}}{(b) Recall@10 (GIST1M)}
  \stackunder[0.5pt]{\includegraphics[scale=0.19]{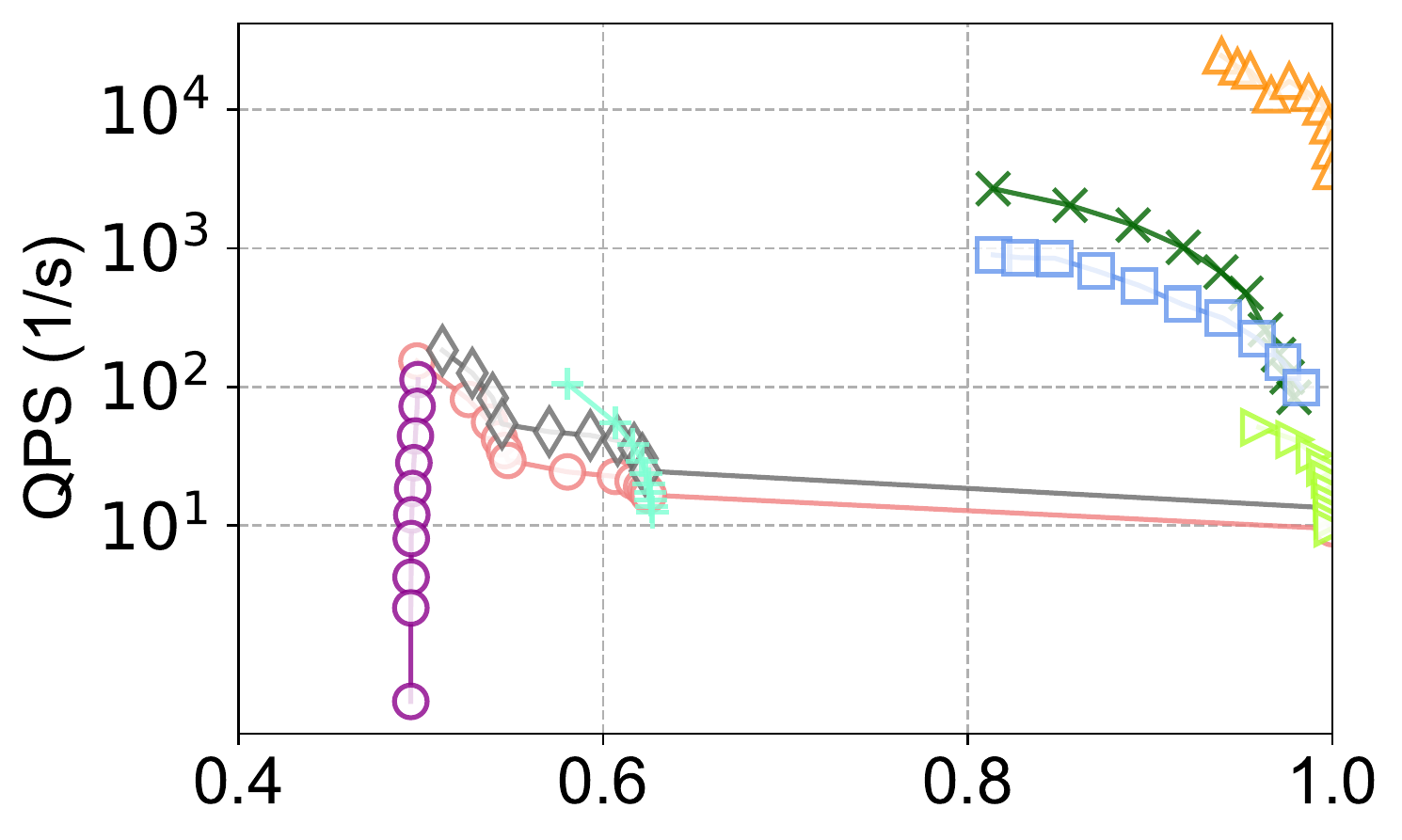}}{(c) Recall@10 (GloVe)}
  \newline
  \stackunder[0.5pt]{\includegraphics[scale=0.19]{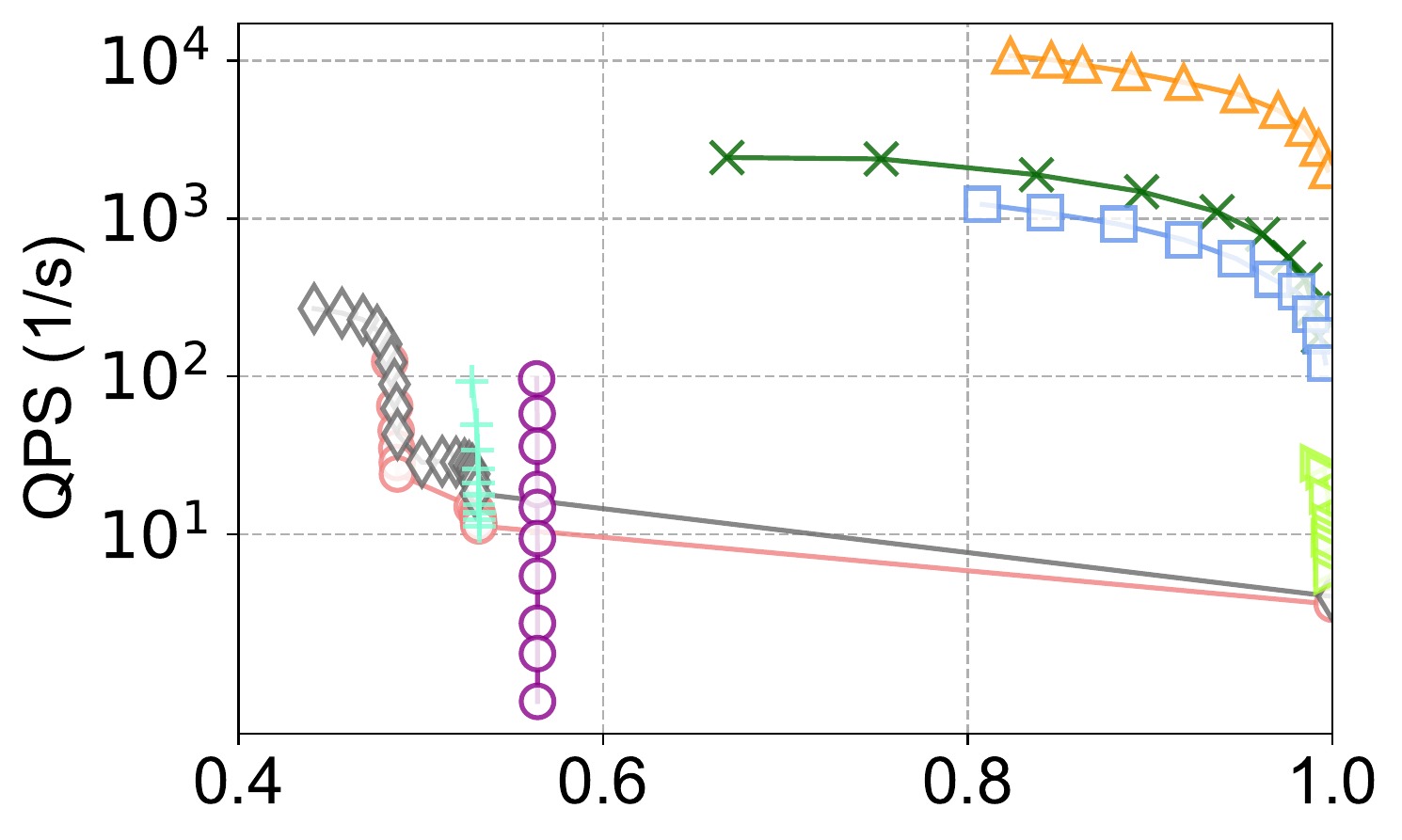}}{(d) Recall@10 (Crawl)}
  \stackunder[0.5pt]{\includegraphics[scale=0.19]{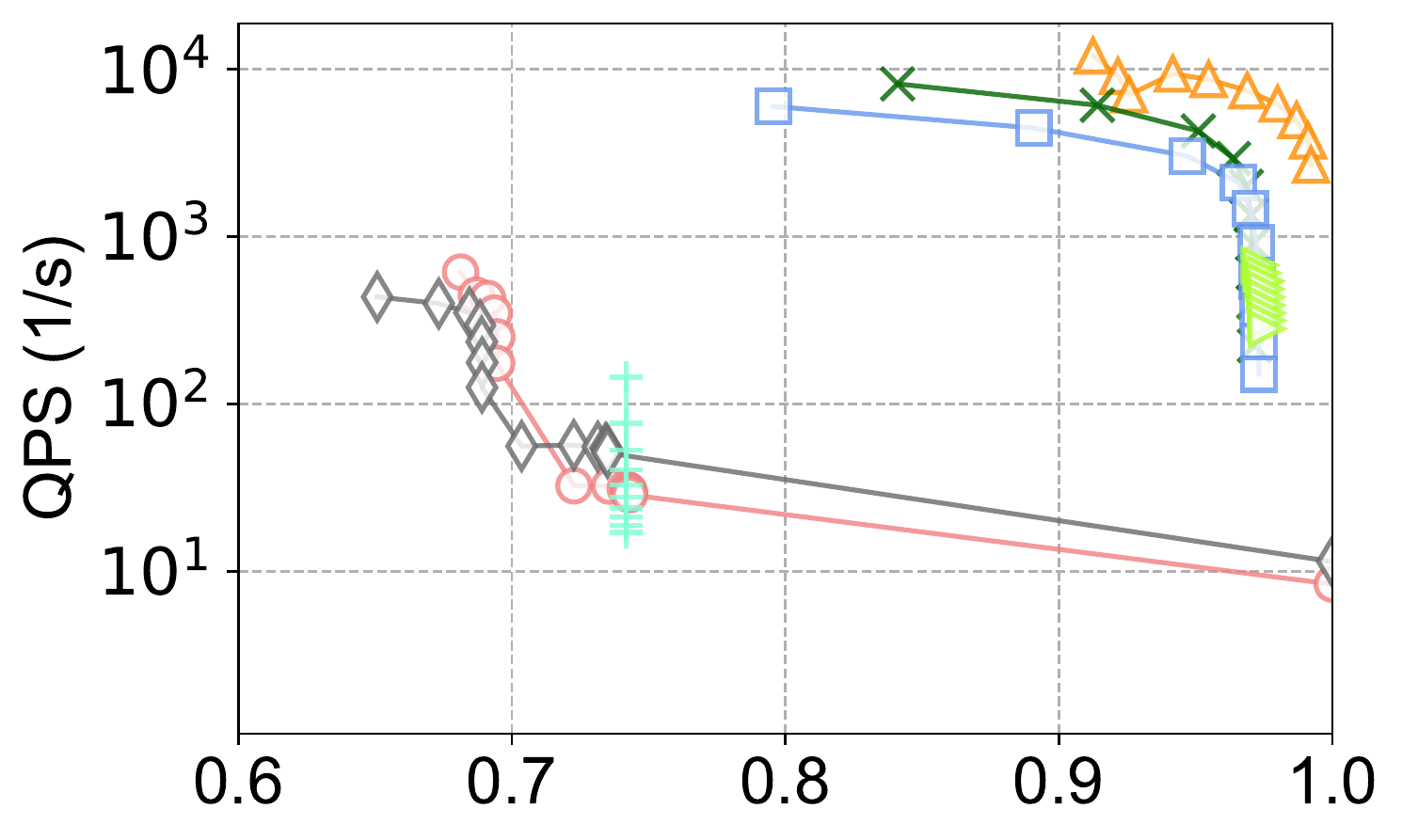}}{(e) Recall@10 (UQ-V)}
  \stackunder[0.5pt]{\includegraphics[scale=0.19]{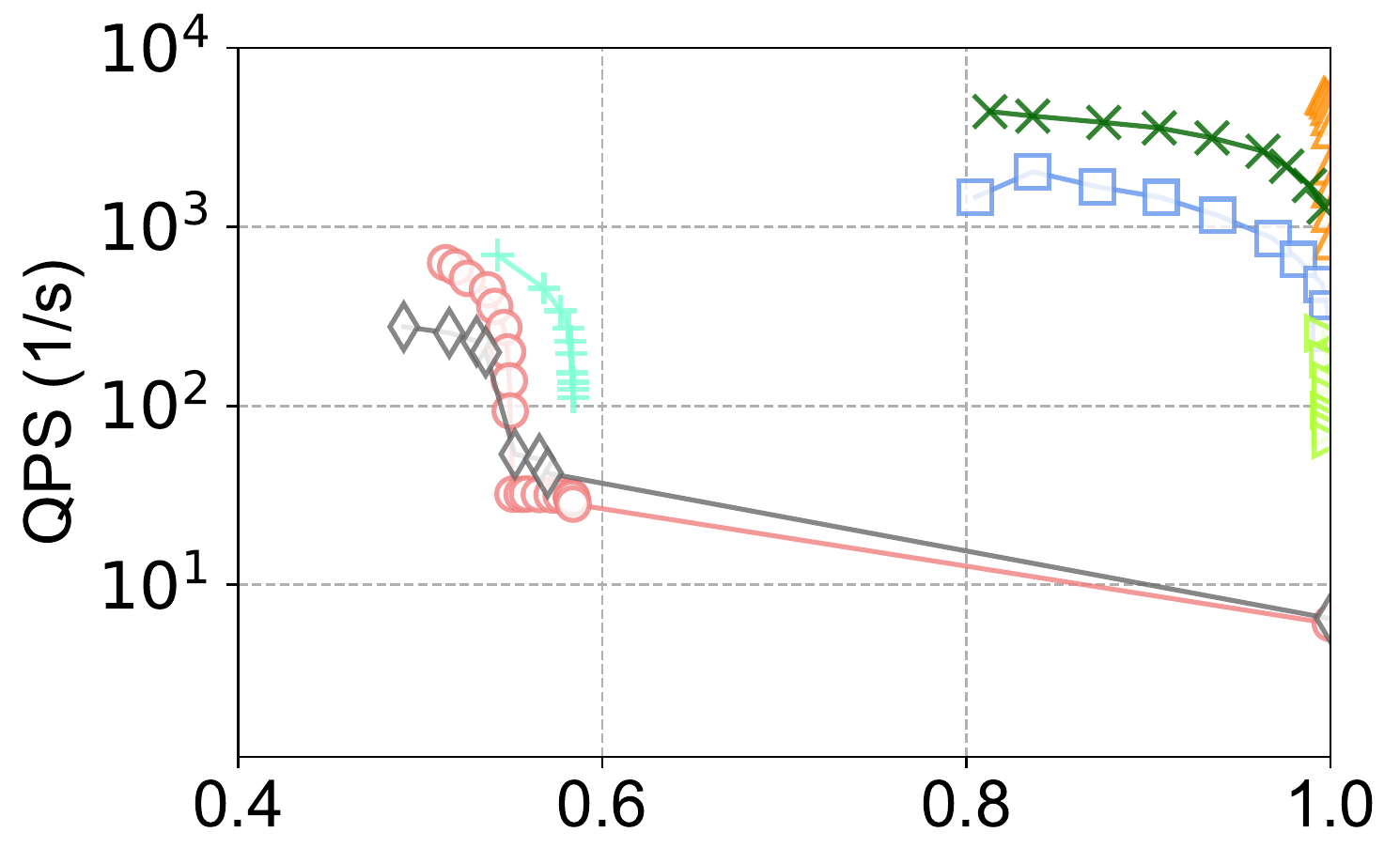}}{(f) Recall@10 (Msong)}
  \newline
  \stackunder[0.5pt]{\includegraphics[scale=0.19]{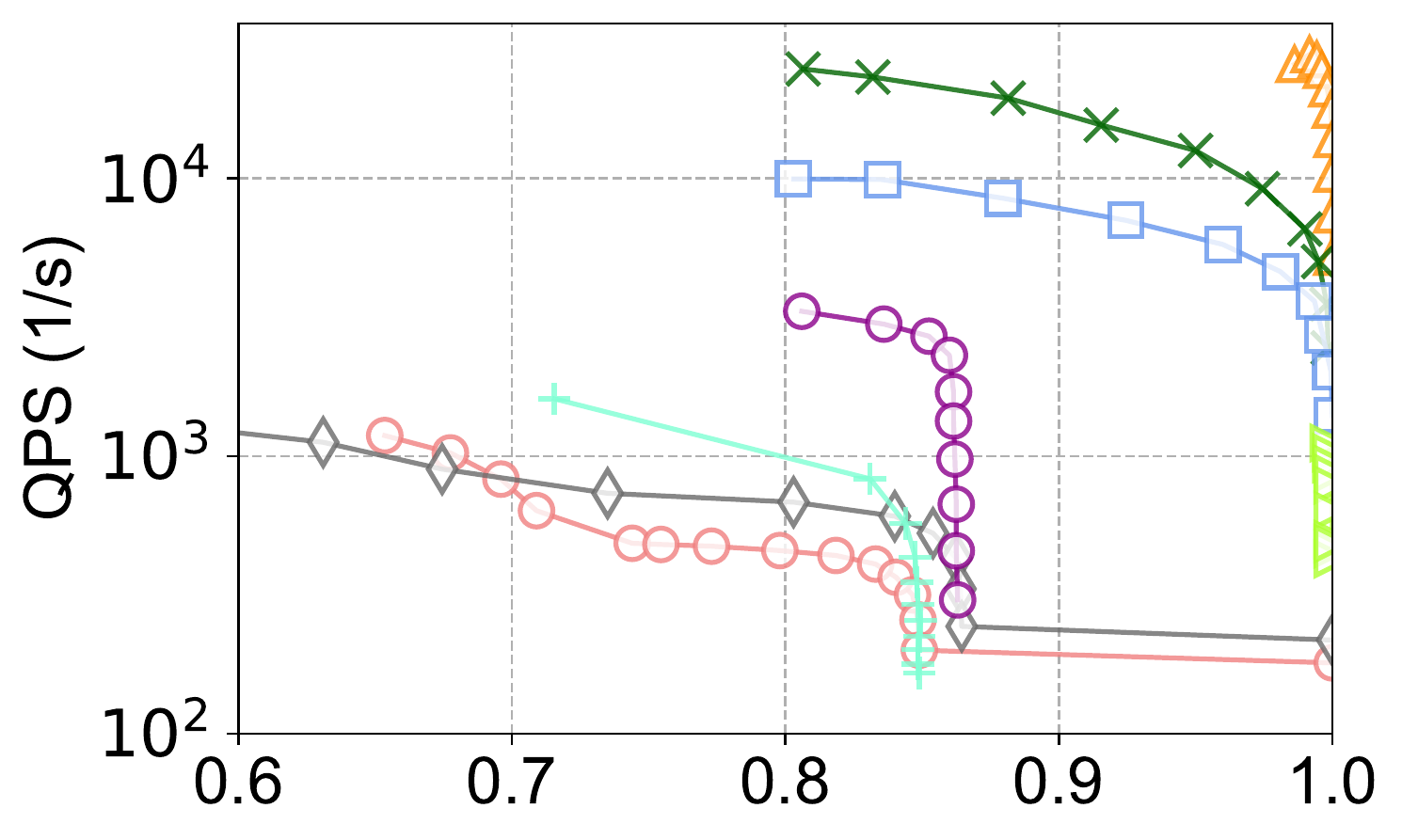}}{(g) Recall@10 (Audio)}
  \stackunder[0.5pt]{\includegraphics[scale=0.19]{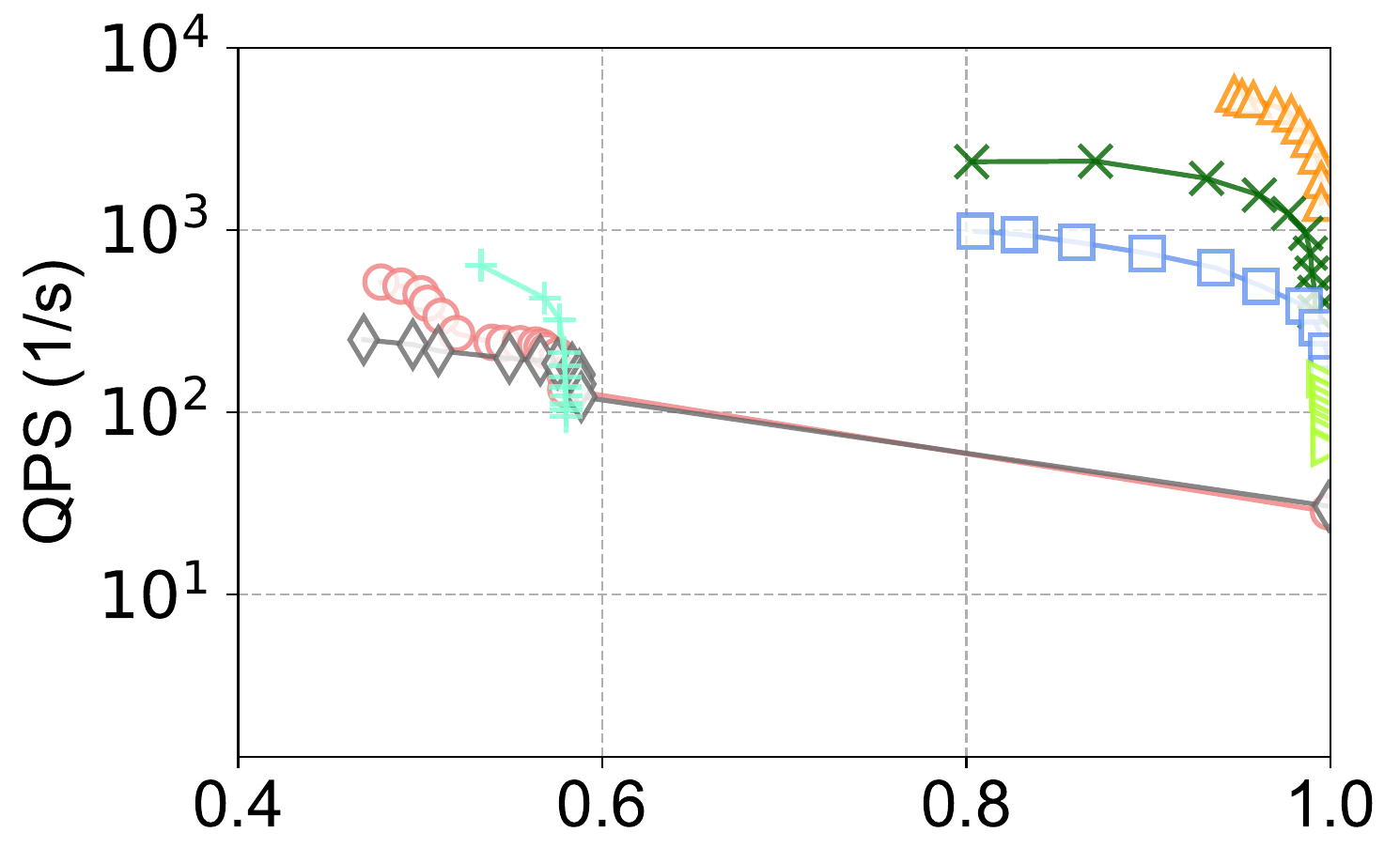}}{(h) Recall@10 (Enron)}
  \stackunder[0.5pt]{\includegraphics[scale=0.19]{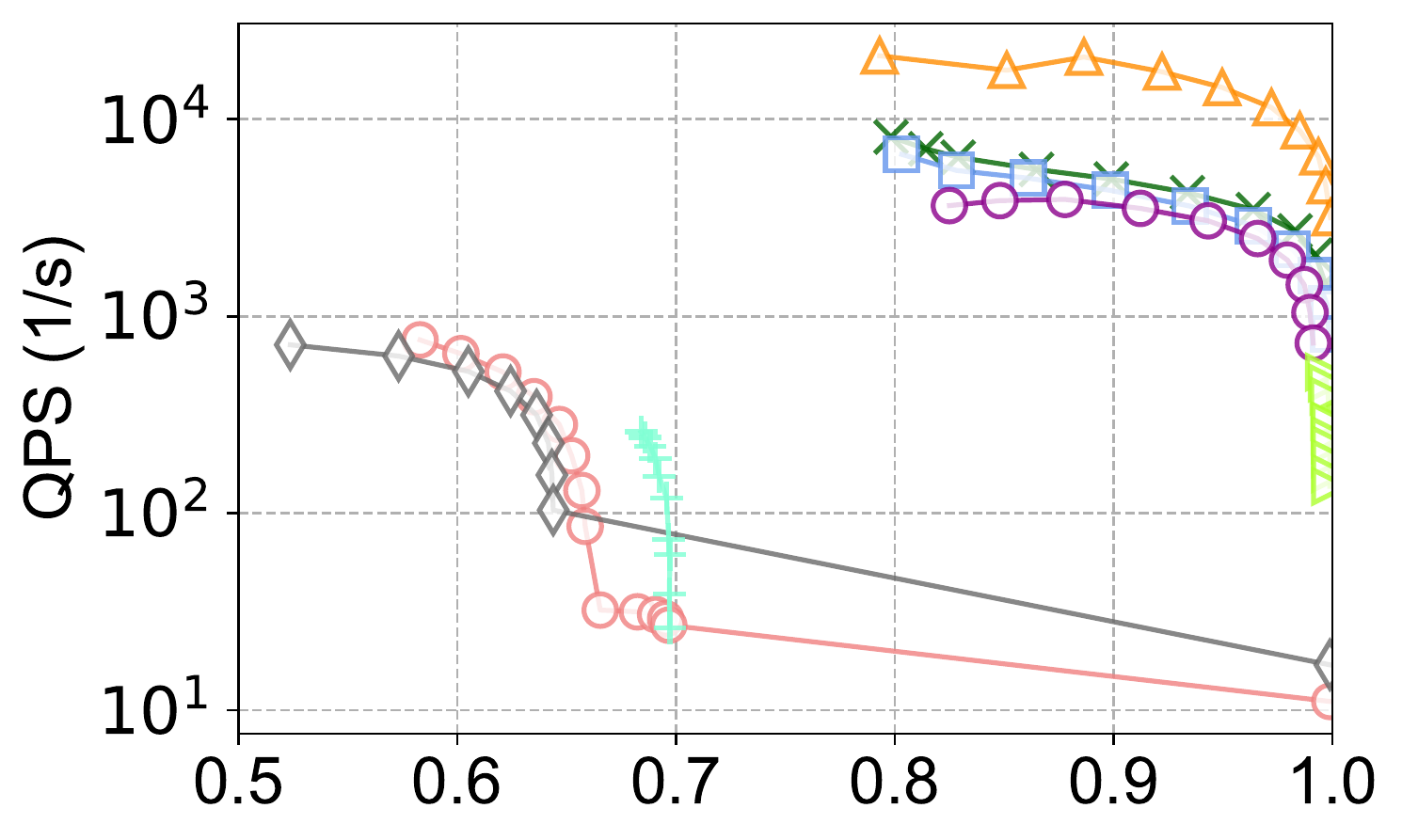}}{(i) Recall@1 (SIFT1M)}
  \caption{(a–h) are the queries per second (QPS) vs Recall@10 of different hybrid query methods; (i) is the QPS vs Recall@1. The attributes' dimension of all datasets is three.}
  \label{fig: hq_real_search_speedup}
\end{figure}

\noindent{\textbf{Index building.}} In {Tab. \ref{tab: Index Construction hybrid query}}, our approaches take the least time to complete index building on almost all datasets. As {Tab. \ref{tab: Index Construction hybrid query}} and {Tab. \ref{tab: hybrid_query_index_size}} show, although PQ-based methods (ADBV, Milvus, and Faiss) generally have smaller memory overhead and index size, they show the limited query performance in {Fig. \ref{fig: hq_real_search_speedup}}. If only PG-based methods are considered, ours are lower than other methods on most datasets (e.g., GIST1M). In general, the higher the dataset's LID, the longer it takes to complete index building, and the larger the memory overhead of building the index and the index size.

\textbf{Analysis.} Clustering on large-scale high-dimensional vectors results in low index building efficiency for PQ-based schemes (such as ADBV); however, these methods' memory overhead and index size are tiny when they comress into raw vectors. Milvus divides datasets according to attributes \cite{Milvus_sigmod2021}, and indexes are built independently on each subset, which reduces memory overhead. Milvus's and ADBV's index size far exceeds Faiss's, because they must generate more indexes for the cost model \cite{ADBV,Milvus_sigmod2021}. PG-based methods rely on raw feature vectors, which lead to a larger index size; nevertheless, our methods show the smallest index size among them, which is because our edge selection eliminates numerous redundant neighbors ({Example \ref{example:ORNG}}). Considering the significant advantages of our methods in hybrid query performance, more memory overhead during index building and a larger index size are worthwhile. Additionally, our methods reach the state-of-the-art hybrid query performance with the highest index building efficiency, which is the priority of most practical applications \cite{NSG}.

\setlength{\textfloatsep}{0cm}
\setlength{\floatsep}{0cm}
\begin{table}[t!]
\setlength{\abovecaptionskip}{0cm}
\setstretch{0.8}
\fontsize{6.5pt}{3.3mm}\selectfont
    \centering
    \caption{Index size of different hybrid query methods.}
    \label{tab: hybrid_query_index_size}
    \setlength{\tabcolsep}{0.008\linewidth}{
    \begin{tabular}{l|l|l|l|l|l|l|l|l}
    \hline
    \multirow{2}*{\textbf{Algorithm}} & \multicolumn{8}{c}{\textbf{Index Size (MB)}} \\
    \cline{2-9}
    ~ & SIFT1M & GIST1M & GloVe & Crawl & UQ-V & Msong & Audio & Enron \\
    \hline
    \hline
    {ADBV} & 113 & 343 & 130 & 292 & 148 & 200 & 11 & 75 \\
    \hline
    {Milvus} & 126 & 438 & 144 & 347 & 174 & 245 & 15 & 105 \\
    \hline
    {Vearch} & 691 & 3,903 & 736 & 2,832 & 1,141 & 1,904 & 50 & 526 \\
    \hline
    {NGT} & 672 & 3,939 & 665 & 2,688 & 1,171 & 1,803 & 49 & 530 \\
    \hline
    {Faiss} & \textbf{40} & \textbf{54} & \textbf{47} & \textbf{83} & \textbf{42} & \textbf{45} & \textbf{3} & \textbf{11} \\
    \hline
    {SPTAG} & 656 & 3,830 & 650 & 2,611 & 1,144 & 1,756 & 48 & 512 \\
    \hline
    {NHQ-NPG\_nsw} & 648 & 3,745 & 550 & 3,050 & 1,098 & 1,786 & 52 & 529 \\
    \hline
    {NHQ-NPG\_kgraph} & 561 & 3,709 & 491 & 2,346 & 1,041 & 1,678 & 42 & 500 \\
    \hline
    \end{tabular}
    }
\end{table}

\vspace{0.1cm}
\noindent{\textbf{Hybrid query performance.}} As {Fig. \ref{fig: hq_real_search_speedup}} shows, our methods are significantly better than existing methods w.r.t \textit{QPS} vs \textit{Recall} trade-off across all datasets. For example, when \textit{Recall@10} $> 0.99$, the \textit{QPS} of our methods is 2 to 3 orders of magnitude higher than others. Meanwhile, our methods' search performance is robustly scalable, which gives them a distinct advantage on harder datasets, since other methods (e.g., SPTAG) perform worse on these datasets. In addition, there is no \textit{Recall} data for Milvus and ADBV between (about) 0.8 and 0.99, as their PQ-based plans are can achieve only a limited \textit{Recall}; and to reach a high \textit{Recall} (\textit{Recall@10} =1), they execute a linear scan. NHQ-NPG\_kgraph easily obtain a high \textit{Recall} under high \textit{QPS}, so that its \textit{QPS} is too high to show in figure when \textit{Recall@10} $<$ (about) 0.9.

\textbf{Analysis.} Existing methods are based on Strategies A and B ({Fig. \ref{fig:implementation_strategies}}), which implement attribute filtering and vector similarity search separately, thereby limiting their ability to answer a hybrid query. In contrast, ours perform attribute filtering and vector similarity search concurrently by embedding feature vectors and attributes into a {composite index}. By jointly pruning, our methods avoid redundant computation and improve the query efficiency.

\begin{figure*}
   \setlength{\belowcaptionskip}{0cm}
  \tiny
  \stackunder[0.5pt]{\includegraphics[scale=0.23]{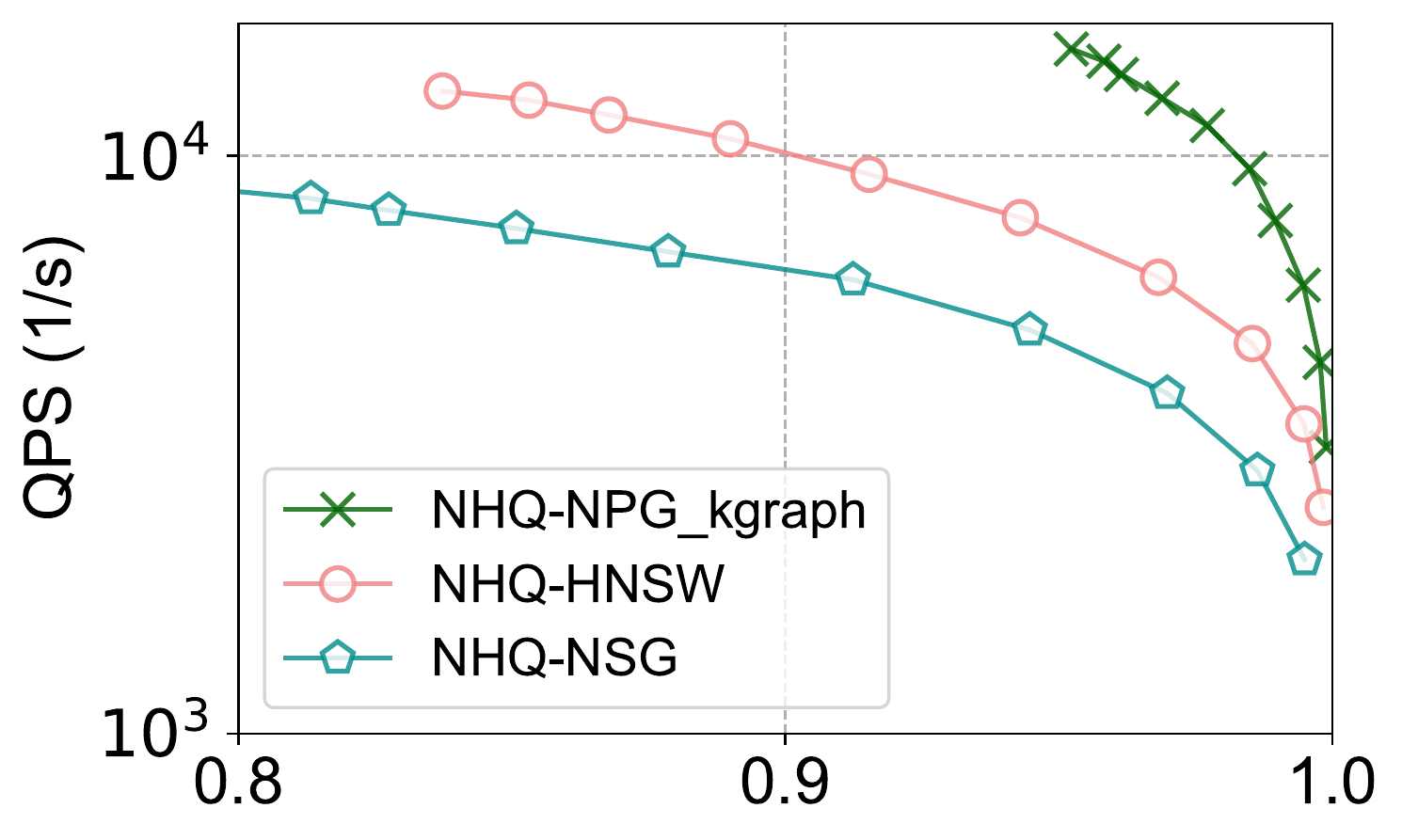}}{(a) Recall@10 (SIFT1M)}\hspace{1.4mm}
  \stackunder[0.5pt]{\includegraphics[scale=0.23]{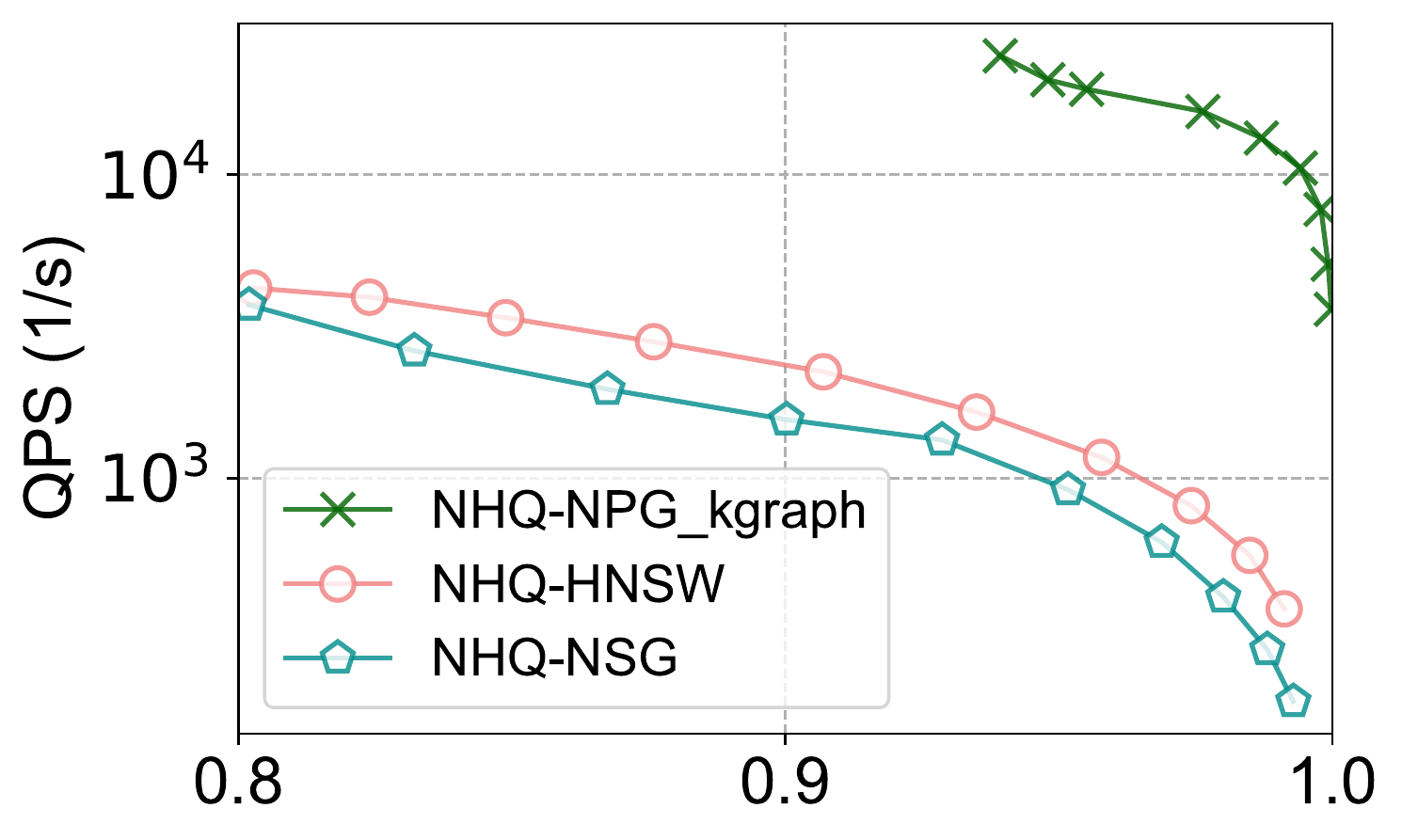}}{(b) Recall@10 (GloVe)}\hspace{1.4mm}
  \stackunder[0.5pt]{\includegraphics[scale=0.23]{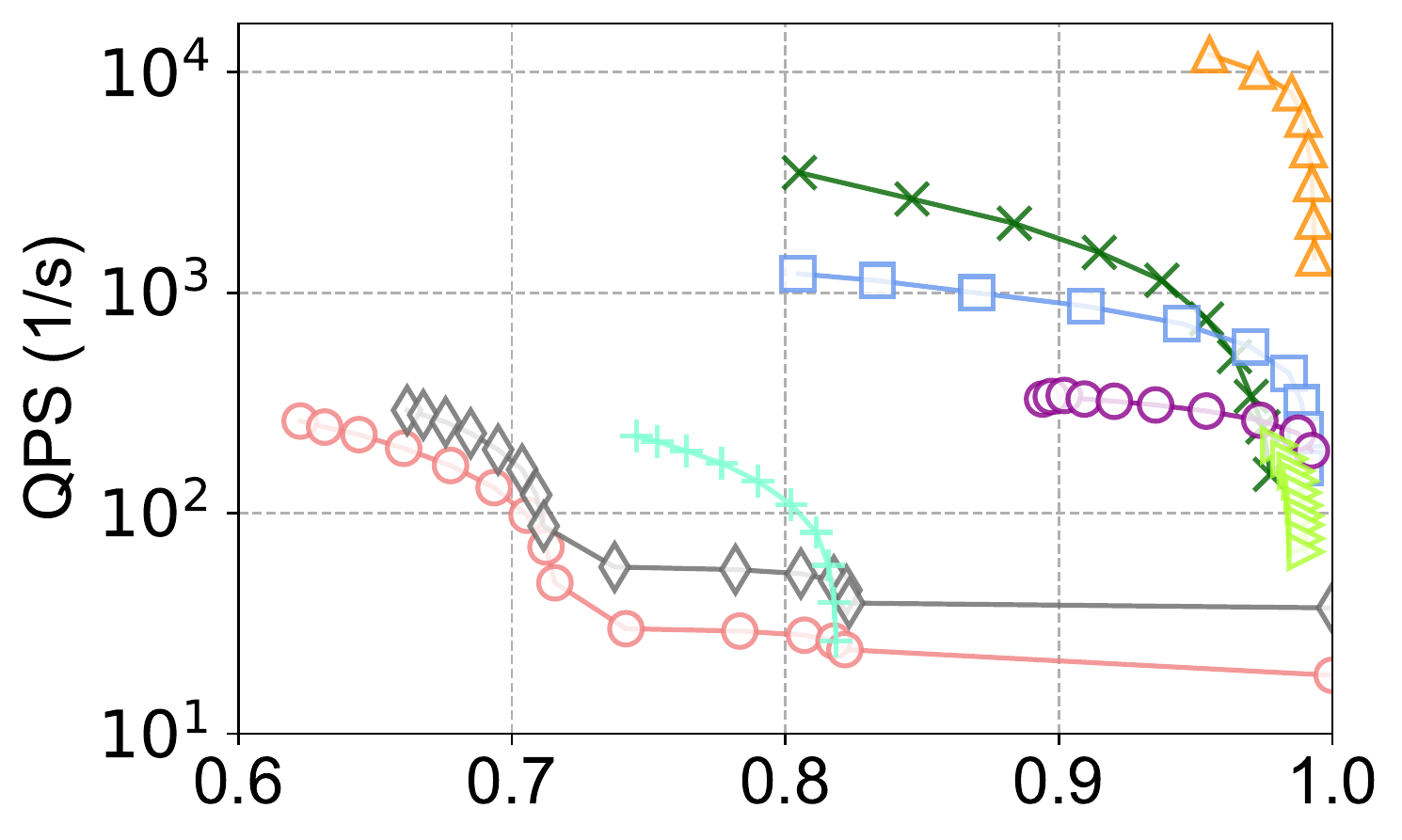}}{(c) Recall@10 ($m$=6)}\hspace{1.4mm}
  \stackunder[0.5pt]{\includegraphics[scale=0.23]{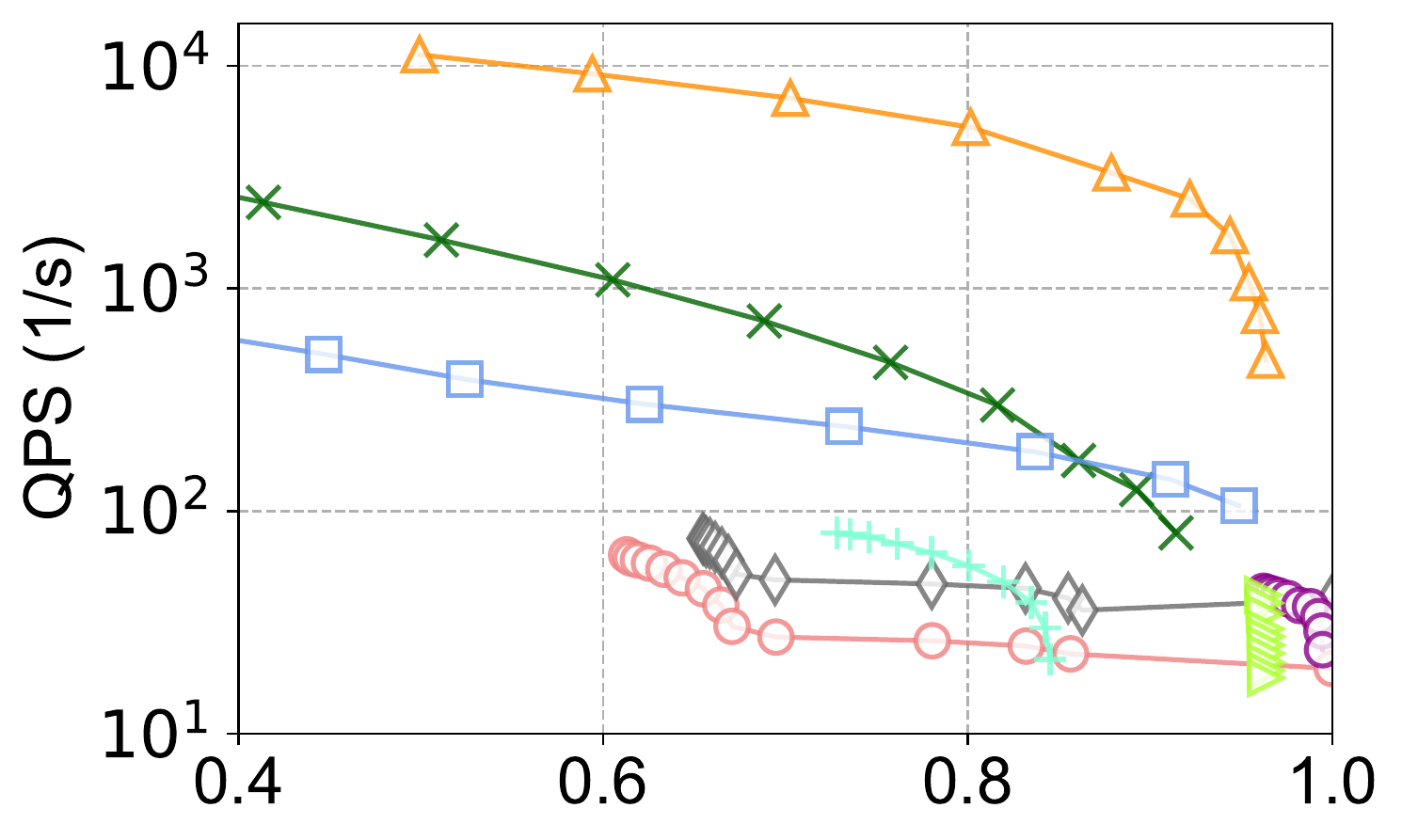}}{(d) Recall@10 ($m$=9)}\hspace{1.4mm}
  \stackunder[0.5pt]{\includegraphics[scale=0.23]{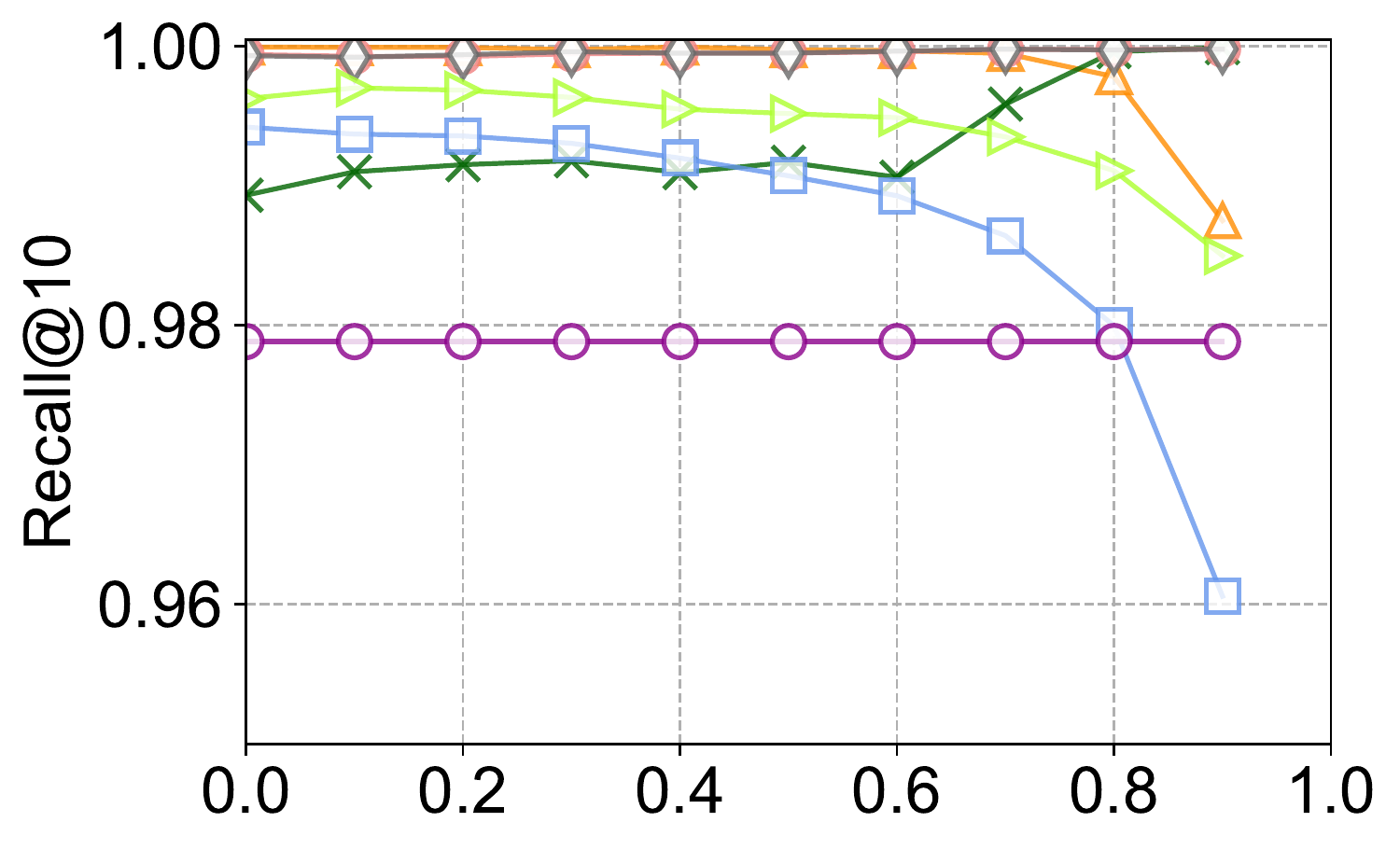}}{(e) $selectivity$}
  \newline
  \stackunder[0.5pt]{\includegraphics[scale=0.23]{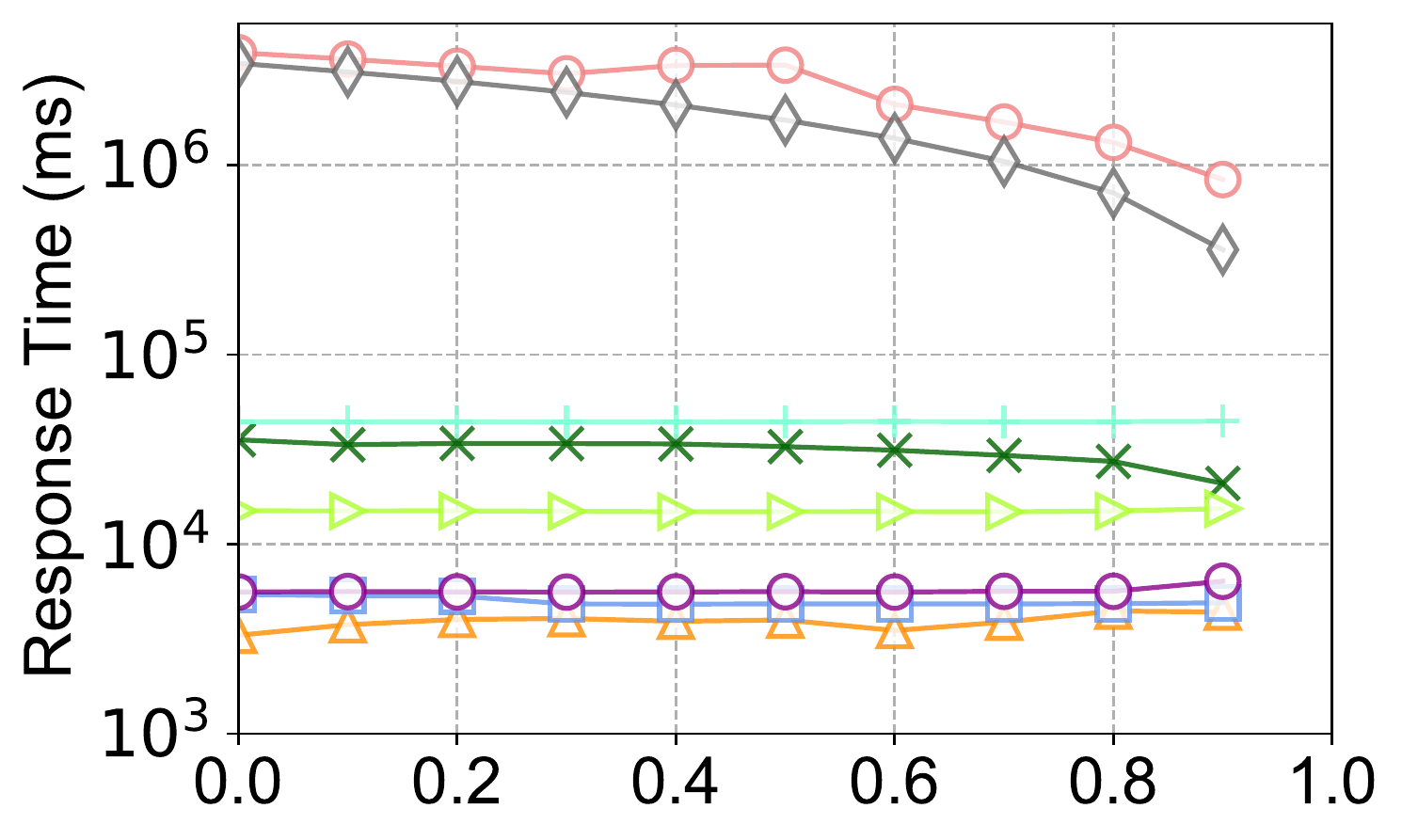}}{(f) $selectivity$}\hspace{1.4mm}
  \stackunder[0.5pt]{\includegraphics[scale=0.23]{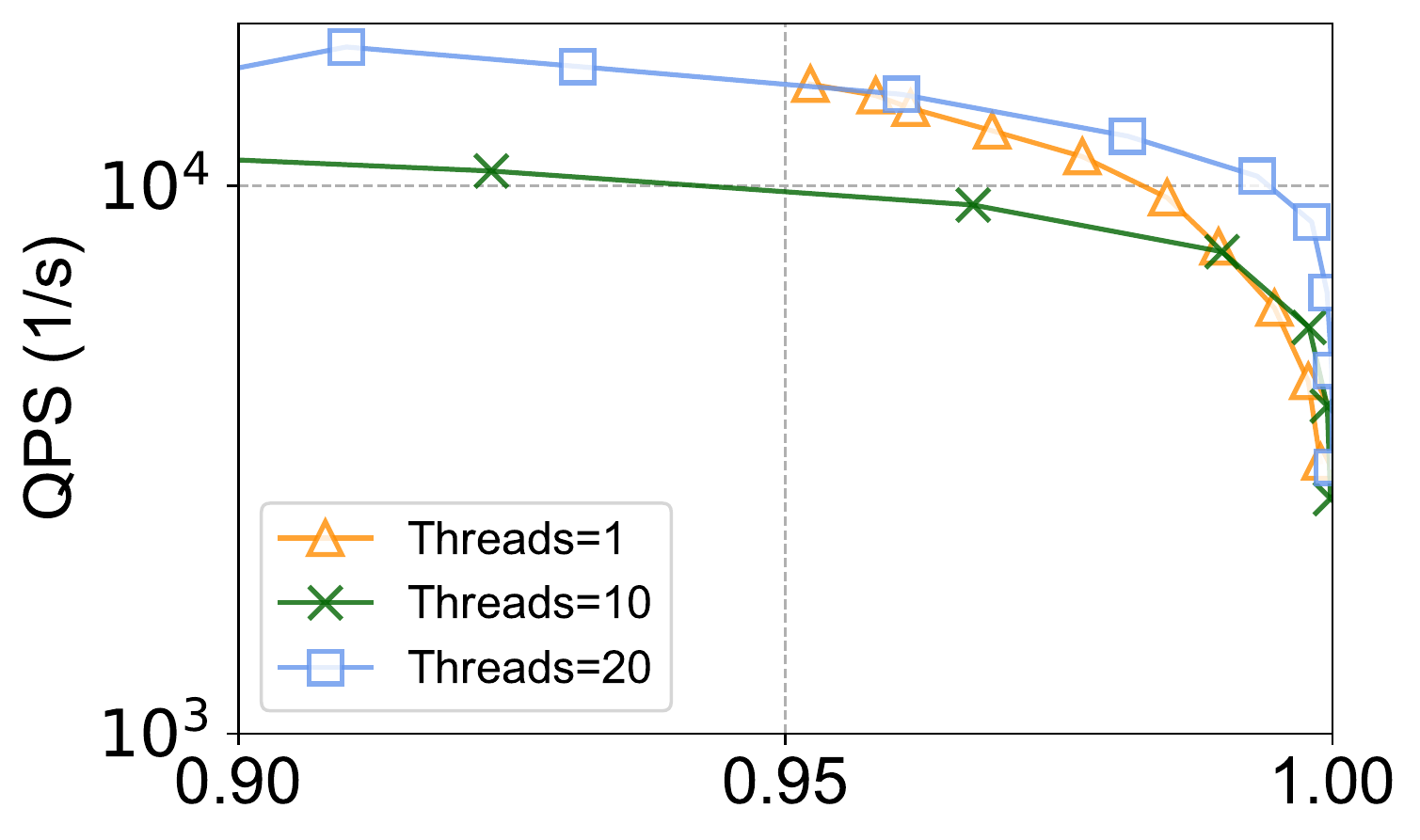}}{(g) Recall@10 (NHQ-NPG\_kgraph, SIFT1M)}\hspace{1.4mm}
  \stackunder[0.5pt]{\includegraphics[scale=0.23]{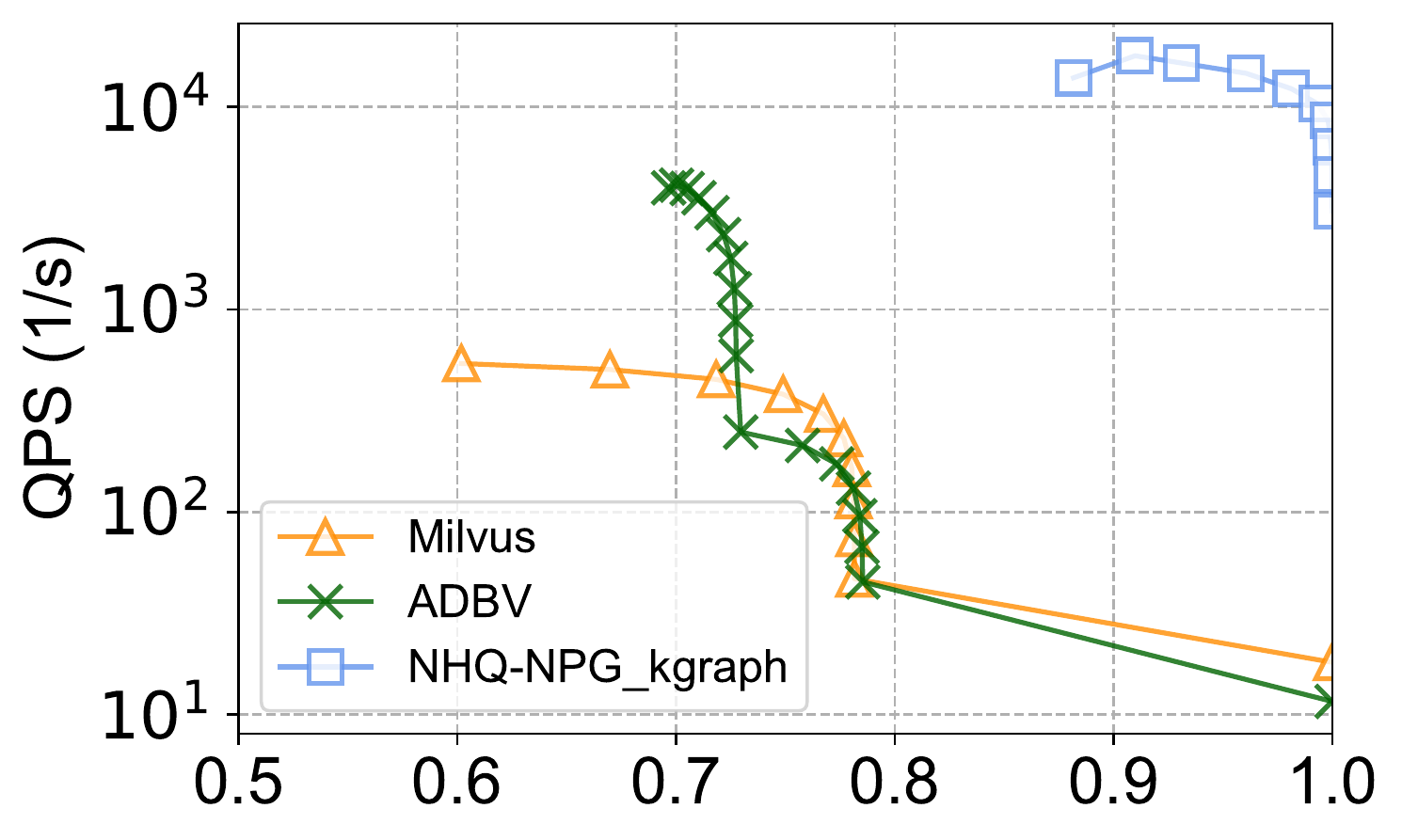}}{(h) Recall@10 (Threads = 20, SIFT1M)}\hspace{1.4mm}
  \stackunder[0.5pt]{\includegraphics[scale=0.23]{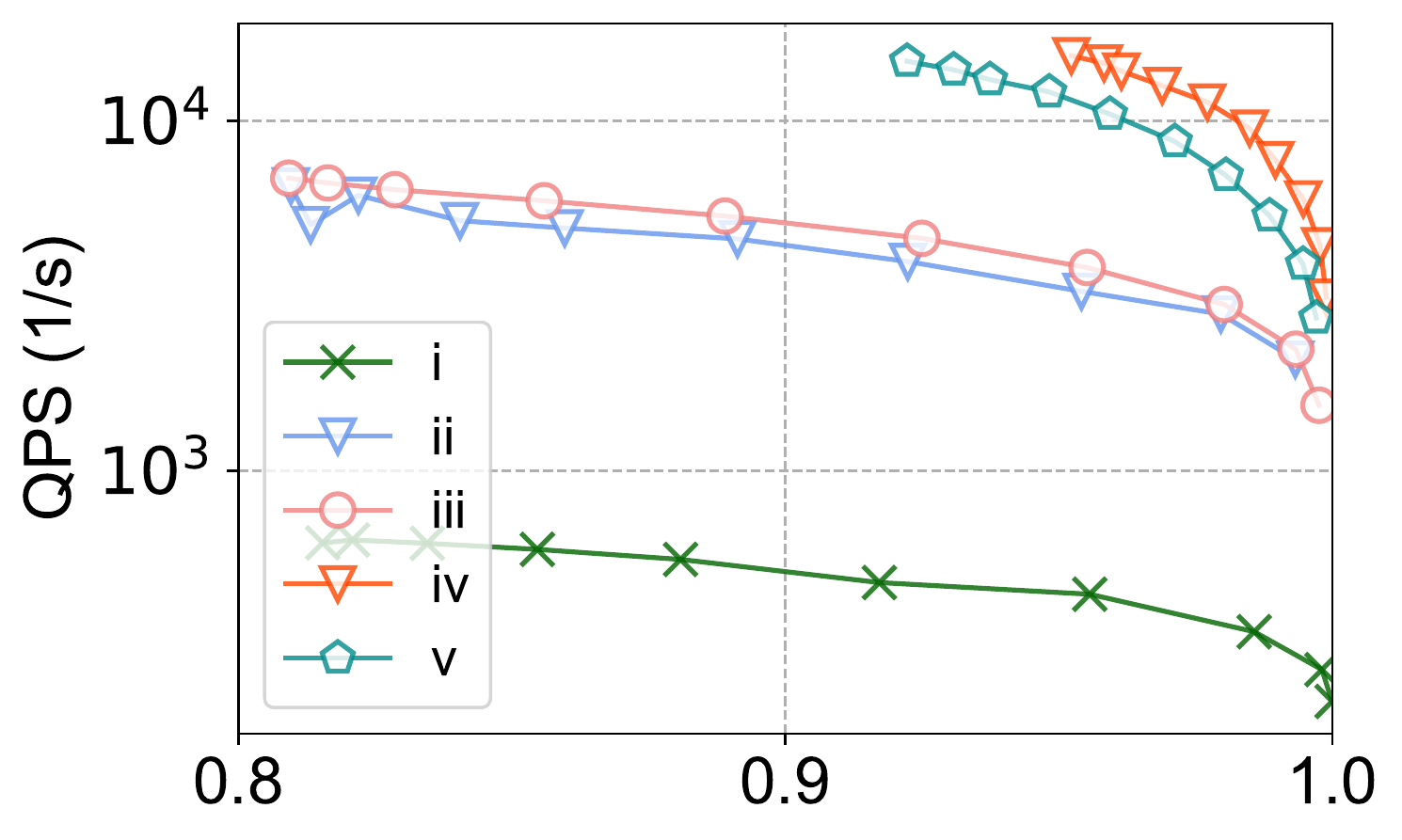}}{(i) Recall@10 (SIFT1M)}\hspace{1.4mm}
  \stackunder[0.5pt]{\includegraphics[scale=0.23]{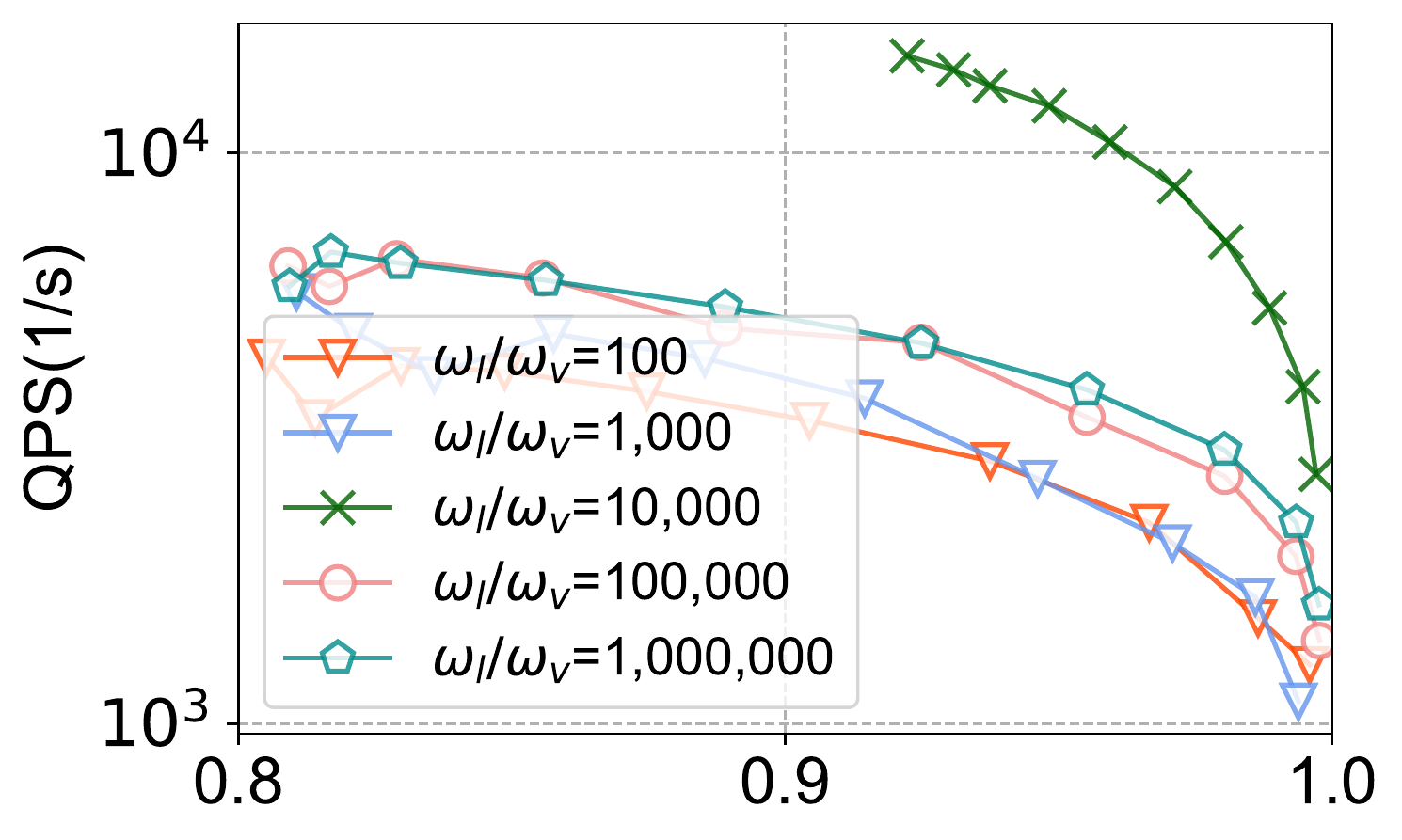}}{(j) Recall@10 (SIFT1M)}
  \caption{(a–b) Effect of hybrid query with different PGs on NHQ. (c–d) Performance under different attribute dimensions, refer to {Fig. \ref{fig: hq_real_search_speedup}} for the legend. (e–f) Performance under different \textit{selectivity}, refer to {Fig. \ref{fig: hq_real_search_speedup}} for the legend. (g–h) Performance under different number of threads. (i–j) Performance under different $\omega _{\nu}$ and $\omega _{\ell}$. In (i), curve {\romannumeral1}: $\omega _{\nu}={\chi}/({\delta + \chi})$, $\omega _{\ell}={\delta}/({\delta + \chi})$; curve {\romannumeral2} modifies $\chi$ by {Eq. \ref{equation:harmonic_adjust}} following curve {\romannumeral1}; curve {\romannumeral3}: $\omega _{\nu} = \delta _{max} ^{-1}$, $\omega _{\ell} = \chi _{max} ^{-1}$; curve {\romannumeral4}: $\omega _{\nu} = 1$, $\omega _{\ell} = {\delta}/{\chi _{max}}$; and curve {\romannumeral5}: ${\omega _{\ell}}/{\omega _{\nu}}=10,000$.}
  \label{fig: hq_other}\vspace{-0.5cm}
\end{figure*}

\vspace{0.1cm}
\noindent{\textbf{Effect of different PGs under NHQ.}} We implement HNSW \cite{HNSW}, NSG \cite{NSG}, and our NPG\_kgraph on NHQ, thus obtaining NHQ-HNSW, NHQ-NSG, and NHQ-NPG\_kgraph. We compare their performance on SIFT1M (with low LID) and GloVe (with high LID). As {Fig. \ref{fig: hq_other}(a) and (b)} illustrate, NHQ-NPG\_kgraph outperforms the others by a large margin, and this superiority becomes even more significant on the harder dataset. We may attribute this to NPGs' edge selection and joint pruning's optimization for NHQ-NPG\_kgraph, which further improve performance.

\vspace{0.1cm}
\noindent{\textbf{Different recall requirements.}} In {Fig. \ref{fig: hq_real_search_speedup}}(i) and (a), as the number of results increases, the \textit{QPS} of different methods degenerates. Note that the interval between our methods and others enlarges, which shows our methods' superiority when more targets need to be recalled.

\textbf{Analysis.} For existing methods, to improve the \textit{Recall}, we must obtain numerous candidates (intermediate results that satisfy one query constraint). For example, when $k = 10$ in \textit{Recall@$k$}, the number of candidates must reach $10 \times \sim 1000\times$ larger than $k$, which undermines the query efficiency. In contrast, our methods directly return the final results without intermediate candidates.

\vspace{0.1cm}
\noindent{\textbf{Dimension of attributes ($m$).}} Attributes' dimension corresponds to the number of attributes in a set of attributes. As shown in {Fig. \ref{fig: hq_other}(c) and (d)}, as the dimension increases, each method's query performance decreases by varying degrees. Our methods still retain the advantage in a higher attribute dimension by a large margin.


\vspace{0.1cm}
\noindent{\textbf{Selectivity.}} A larger \textit{selectivity} means fewer objects match the attributes constraint, which increases the difficulty in answering a hybrid query. In {Fig. \ref{fig: hq_other}(e) and (f)}, \textit{Recall@10} decreases under a high \textit{selectivity} for most methods. However, NHQ-NPG\_nsw appears to excel when there is large \textit{selectivity}. {One explanation is that, when the construction strategy inserts objects incrementally, it may link better with vertices possessing the same attributes but are farther away for feature vectors. This improves NHQ-NPG\_nsw’s ability to search objects with matched attributes, especially such objects are few (that is, high \textit{selectivity}).} NHQ-NPG\_kgraph reaches the highest \textit{Recall} with the least search time for different \textit{selectivity}. In addition, Faiss is not shown in {Fig. \ref{fig: hq_other}}(e) because its \textit{Recall@10} is only about 0.7.

\vspace{0.1cm}
\noindent{\textbf{Number of threads.}} {We implement a parallel version of hybrid queries on NHQ-NPG\_kgraph, which divides an object set into multiple subsets and builds a NHQ-NPG\_kgraph on each subset, so that we can search in parallel on these NHQ-NPG\_kgraphs. As shown in {Fig. \ref{fig: hq_other}}(g), more threads in the higher \textit{Recall} area (\textit{Recall@10 $>$ 0.98}) will improve query performance, while a single thread is more suitable for hybrid queries in the lower \textit{Recall} area (\textit{Recall@10 $<$ 0.95}). We explain that to reach a higher \textit{Recall} rate, we need to visit more vertices, where parallel search is more advantageous; but for a lower \textit{Recall} rate, with a small number of vertices visited, the percentage of time consumed to frequently create and start a thread will increase (for the total query time), and search with multi-threads will degrade query performance. Nevertheless, in {Fig. \ref{fig: hq_other}}(h), NHQ-NPG\_kgraph under a parallel implementation shows better hybrid query performance than Milvus and ADBV.}

\subsection{Parameter Sensitivity}
\label{sec: parameter sensitivity}
For NHQ, $\omega _{\nu}$ and $\omega _{\ell}$ in {Eq. \ref{hybrid_distance}} are a pair of parameters that regulate the weights of $\delta(\nu(e_i),\nu(e_j))$ (abbr. as $\delta $) and $\chi (\ell(e_i),\ell(e_j))$ (abbr. as $\chi$) in $\mathit{\Gamma} (e_i,e_j)$ (abbr. as $\mathit{\Gamma}$), which impacts the query performance. In the following, based on experimental observations, we evaluate the different settings' effects on performance. Because of space constraints, we only show the execution results of NHQ-NPG\_kgraph on SIFT1M, considering that other datasets and methods working off NHQ show similar phenomena.

In general, we treat feature vectors and attributes equally; that is, $\delta $ and $\chi$ have the same contribution to $\mathit{\Gamma}$. We set $\omega _{\nu}={\chi}/({\delta + \chi})$ and $\omega _{\ell}={\delta}/({\delta + \chi})$, so $\mathit{\Gamma}={2 \cdot \delta \cdot \chi}/({\delta + \chi})$ is the harmonic mean of $\delta$ and $\chi$. As {Fig. \ref{fig: hq_other}}(i) (\textbf{curve \romannumeral1}) shows, this setting does not perform well for hybrid queries. We found that $\chi = 0$ is a common occurrence (two objects possessing the same attributes), and then $\mathit{\Gamma}=0$ no matter what value $\delta$ is, which contradicts the same contribution of the two. To resolve this, we adjust $\chi$ as
\vspace{-0.1cm}
\begin{equation}
  \label{equation:harmonic_adjust}
  \setlength{\nulldelimiterspace}{0pt}
  \chi= \left\{\begin{IEEEeqnarraybox}[\relax][c]{l's}
    1, &for $\chi = 0$ \\
    \chi \cdot c, &for $\chi \neq 0$
  \end{IEEEeqnarraybox}\right.\quad ,
\end{equation}
where $c$ is a given constant, and it holds that $c>1$. This alleviates the defect caused by $\chi = 0$, and it significantly improves the hybrid query performance (\textbf{curve {\romannumeral2}}, $c=100$). Nevertheless, a suitable $c$ is difficult to obtain, so we turn to the other two options to deal with the fact that $\delta$ and $\chi$ are in different value spaces. One option is to map both $\delta$ and $\chi$ to $[0,1]$ by setting $\omega _{\nu} = \delta _{max} ^{-1}$ and $\omega _{\ell} = \chi _{max} ^{-1}$, where $\delta _{max}$ and $\chi _{max}$ are the max value of $\delta$ and $\chi$ on a given object set $\mathcal{S} $. We further promote query performance under this setting (\textbf{curve \romannumeral3}). However, it needs to compute $\delta _{max}$ of $\mathcal{S} $ in advance, which is not easy in a real-world scenario because $\mathcal{S}$ is dynamic. The other option is to map $\chi$ to $\delta$'s value space, i.e., $\omega _{\nu} = 1$ and $\omega _{\ell} = {\delta}/{\chi _{max}}$, where $\chi _{max}=m$, $m$ is the dimension of attributes, and it can be obtained easily. As curve {\textbf{\romannumeral4}} depicts, this approach achieves the best performance in all our settings. Additionally, we also study a naive setting method, where $\omega _{\nu}$ and $\omega _{\ell}$ are fixed as constants. We traverse multiple possible ${\omega _{\ell}}/{\omega _{\nu}}$ values by grid search. {Fig. \ref{fig: hq_other}}(j) illustrates that different ${\omega _{\ell}}/{\omega _{\nu}}$ leads to a huge difference in hybrid query performance, and the optimal value falls to 10,000 (\textbf{curve {\romannumeral5}} in {Fig. \ref{fig: hq_other}}(i)), which makes it close to the median of all $\delta$ on $\mathcal{S}$. 
Thus, we chose the settings of \textbf{curve {\romannumeral4}} for all other experiments.

\subsection{Use Case Study}
\label{sec: use case}
We deployed NHQ-NPG\_kgraph into an academic expert finding system \cite{liujun} to provide hybrid query processing services for academic papers with attributes on the \textit{Paper} dataset. An important component in such system is the semantically similar papers retrieval that supports attribute filtering. Given a user input hybrid query with a descriptive text as the unstructured constraint and paper's some attributes as the structured constraints, expert finding aims at extracting the top-$k$ experts from top-$m$ papers that have the semantically similar vectors to the given query text as well as satisfying the given attribute constraints. Because we extract the experts from the returned papers, the results of hybrid queries would directly affect the retrieved experts. As comparison, we also implemented two existing methods for retrieving papers (i.e., Vearch and HNSW).

\setlength{\textfloatsep}{0cm}
\setlength{\floatsep}{0cm}
\begin{table}[t!]
\setlength{\abovecaptionskip}{0cm}
\setstretch{0.8}
\fontsize{6.5pt}{3.3mm}\selectfont
    \centering
    \caption{Experts output by the academic expert finding engine under different query methods (the bold items are correct).}
    \label{tab: experts_output}
    \setlength{\tabcolsep}{0.045\linewidth}{
    \begin{tabular}{l|l|l}
    \hline
    \textbf{HNSW} & \textbf{Vearch} & \textbf{NHQ-NPG\_kgraph} \\
    \hline
    \hline
    \textbf{John Collomosse} $\surd$ & \textbf{John Collomosse} $\surd$ & \textbf{John Collomosse} $\surd$ \\
    \hline
    Mayu Iwata $\times $ & \textbf{Rahul Duggal} $\surd$ & \textbf{Rahul Duggal} $\surd$ \\
    \hline
    Yanhao Zhang $\times $ & - & \textbf{Hosnieh Sattar} $\surd$ \\
    \hline
    Takuma Yamaguchi $\times $ & - & \textbf{Thanh-Toan Do} $\surd$ \\
    \hline
    \textbf{Rahul Duggal} $\surd$ & - & \textbf{Gregory Zelinsky} $\surd$ \\
    \hline
    \end{tabular}
    }
\end{table}

\textbf{Analysis.} {Tab. \ref{tab: experts_output}} shows the top-5 experts who have published papers on {\sf CVPR} for a hybrid query having the abstract of \cite{CollomosseBJ19} as the unstructured constraint and {\sf CVPR} as an attribute constraint under a given query time. Our method obtains the best results in compared methods. HNSW only returns the experts who are relevant to the query text without considering the {\sf CVPR} constraint. Although Vearch adds the {\sf CVPR} constraints, the experts are insufficient after attribute filtering. However, the query latency will surge when increasing the number of candidate experts.

\section{Related Work}
\label{related_work}
To provide a better sense of how our approach compares to other efforts in vector similarity search and hybrid query processing, here we briefly discuss related work by topic.

\vspace{0.1cm}
\noindent{\textbf{Vector similarity search.}} Vector similarity search—or approximate nearest neighbor search (ANNS)—is a task to obtain results similar to a given query through a well-designed vector index \cite{Milvus_sigmod2021,graph_survey_vldb2021}. According to the different indexes, we can divide existing methods into four categories: tree-based \cite{Silpa-AnanH08,AroraSK018,MujaL14}; hashing-based \cite{GongWOX20,HuangFZFN15,LiZSWT020}; quantization-based \cite{PQ,AndreKS15,ScaNN}; and PG-based \cite{HNSW,NSSG,graph_survey_vldb2021,DiskANN,HM_ANN}. Recently, PGs have yielded state-of-the-art \textit{Speedup} vs \textit{Recall} trade-off \cite{DPG,NSG}, which has sparked substantial interest. So far, researchers have proposed dozens of PGs using different optimizations. However, these works are full-fledged solutions for content retrieval of unstructured data, and they cannot deal effectively with attribute filtering \cite{ADBV}.

\vspace{0.1cm}
\noindent{\textbf{Hybrid query processing.}} Hybrid queries entails advanced query processing with structured and unstructured constraints \cite{Milvus_sigmod2021}, which requires not only feature vectors' similarity, but also attributes' consistency \cite{Wangwei_tutorial,Qin000W21}. Increasingly, this technique is indispensable to many applications \cite{ADBV}. Existing hybrid query methods \cite{XuLWX20,ADBV,Milvus_sigmod2021,Jingdong_paper} perform vector similarity search and attribute filtering separately along Strategies A and B in {Fig. \ref{fig:implementation_strategies}}, which leads to some limitations ({L1}–{L4} in \textbf{Sec. \ref{sec:intro}}). By contrast, our methods nimbly and adeptly address issues without these limitations.

\section{Conclusion}
\label{conclusion}
In this paper, we tackle hybrid queries with structured and unstructured query constraints. We design a NHQ framework that is tailored for hybrid queries, which is friendly to most current PGs. We present two NPGs with optimized edge selection and routing strategies that obtain better performance compared to existing PGs. We deploy our proposed NPGs in NHQ to form two NPG-based hybrid query methods that are $10\times$ faster than the state-of-the-art competitors. Interesting future and ongoing work includes applying our proposed framework to multimodal search to cope with more complex query requirements.

\section*{Acknowledgments}
The National Natural Science Foundation of China (Number 62072149) supported this work. Additional funding was provided by the Primary Research \& Development Plan of Zhejiang Province (Number 2021C03156 and 2021C02004). This work was also supported by the Fundamental Research Funds for the Provincial Universities of Zhejiang (Number GK219909299001-006).

\bibliographystyle{IEEEtran}
\bibliography{IEEEabrv, mybibli.bib}

\end{document}